\documentclass[oldversion]{aa} 
\usepackage{graphicx}
\usepackage{aalongtable}
\usepackage{txfonts}
\usepackage{rotating}
\usepackage{lscape}
\newcommand{\gsim}{\;\lower.6ex\hbox{$\sim$}\kern-7.75pt\raise.65ex\hbox{$>$}\;}
\newcommand{\lsim}{\;\lower.6ex\hbox{$\sim$}\kern-7.75pt\raise.65ex\hbox{$<$}\;}

\begin{document}

\title{Chemistry of multiple stellar populations in the mono-metallic,
in situ, bulge globular cluster NGC~6388\thanks{Based on observations collected at  ESO telescopes under
programmes 073.D-0211 and  073.D-0760,  381.D-0329,  095.D-0834, and
099.D-0047.}
\thanks{Full Tables A.1, A.2, A.3, and A.5 are only available at the CDS via anonymous ftp to
cdsarc.u-strasbg.fr (130.79.128.5) or via http://cdsarc.
u-strasbg.fr/viz-bin/cat/J/A+A/??/??.}
 }

\author{
Eugenio Carretta\inst{1}
\and
Angela Bragaglia\inst{1}
}

\authorrunning{Carretta and Bragaglia}
\titlerunning{Detailed abundances of NGC~6388}

\offprints{E. Carretta, eugenio.carretta@inaf.it}

\institute{
INAF-Osservatorio di Astrofisica e Scienza dello Spazio di Bologna, Via Gobetti
 93/3, I-40129 Bologna, Italy}

\date{}

\abstract{We present the homogeneous abundance analysis for a combined sample of
185 giants in the bulge globular cluster (GC) NGC~6388. Our results are used to
describe the multiple stellar populations and differences or analogies with
bulge field stars. Proton-capture elements indicate that a single class of
first-generation polluters is sufficient to reproduce both the extreme and
intermediate parts of the anti-correlations among light elements O, Na, Mg, and
Al, which is at odds with our previous results based on a much smaller sample.
The abundance pattern of other species in NGC~6388 closely tracks the trends
observed in bulge field stars. In particular, the $\alpha$-elements, including
Si, rule out an accreted origin for NGC~6388, confirming our previous results
based on iron-peak elements, chemo-dynamical analysis, and the age-metallicity
relation. The neutron-capture elements are generally uniform, although the
[Zr/Fe] ratio shows an intrinsic scatter, correlated to Na and Al abundances.
Instead, we do not find enhancement in neutron-capture elements for stars whose
photometric properties would classify NGC~6388 as a type II GC. Together with
the homogeneity in [Fe/H] we found in a previous paper, this indicates we need
to better understand the criteria to separate classes of GCs, coupling
photometry, and spectroscopy. These results are based on  abundances of 22
species (O, Na, Mg, Al, Si, Ca, Ti, Sc, V, Cr, Mn, Fe, Co, Ni, Zn, Y, Zr, Ba,
La, Ce, Nd, and Eu) from UVES spectra  sampling proton-, $\alpha-$,
neutron-capture elements, and Fe-peak elements. For 12 species, we also obtain
abundances in a large number of giants (up to 150) from GIRAFFE spectra.   
}

\maketitle

\section{Introduction}

The high-mass, high-metallicity globular cluster (GC) NGC~6388, located in the
bulge of the Milky Way (MW) had been poorly studied using high resolution
spectroscopy, despite its relevance. The small apparent angular diameter and the
large contamination by field stars made it difficult to observe a large number
of cluster members (see, e.g. Wallerstein et al. 2007). This has been amended
with the use of multi-object spectrographs (such as FLAMES at the ESO VLT; see
Carretta et al. 2007a, Lanzoni et al. 2013) and large surveys such as APOGEE
(M\'esz\'aros et al. 2020: M20).  Additionally, a more efficient choice of candidate
members can be provided by the $Gaia$ mission (see, i.e. the
astrometric analysis in Vasiliev and Baumgardt 2021).

This paper presents all the data from our series on NGC~6388 `reloaded', aimed
at providing a detailed description of the chemical properties of
multiple stellar population in this GC. We started by using the existing data in
the ESO archive and added new observations purposely designed to complete
the set of setups needed to characterise the chemical composition.  In
particular, we sought to include all the light elements involved in the
(anti-)correlations typical of multiple stellar populations detected in GCs
(Carretta et al. 2009a,b, 2010a, Gratton et al. 2012a, 2019, Bastian and Lardo
2018). 

In Carretta and Bragaglia (2018), we analysed UVES spectra of 24 giants (adding
17 new stars to our original set of seven in Carretta et al. 2007a). The
resulting pattern of proton-capture elements seemed to imply the necessity of
two classes of polluters (the stars  that produced the characterising set of
light elements in GCs) to explain the anti-correlations between O and Na, Mg, and
Al. 

The next step (Carretta and Bragaglia 2019) was to  discuss the abundance of Mg,
Ca, and Sc for the full set of 185 stars observed with both UVES and  GIRAFFE.
By comparing the abundance ratios in NGC~6388 to those of field stars of similar
metallicity, we detected no significant differences for Ca and Sc (at variance
with NGC~2808 and a few other massive GCs: see Carretta and Bragaglia 2021),
whereas lighter proton-capture species, such as Si, showed clear variations
correlated to Mg depletion and Al enhancement. Together with prediction from
stellar nucleosynthesis, these observations allowed us to pinpoint the range of
internal temperature for the first generation (FG) polluters in NGC~6388
($\sim100-150$ MK), with some difference between the cases of the massive  stars
and asymptotic giant branch (AGB) stars. 

We then concentrated on the metal abundance of this cluster and presented 
(Carretta and Bragaglia 2022a) the atmospheric parameters for our new data set.
We also compared our results to literature results and addressed the issue of a metallicity
spread. The latter is predicted as a possible property of a separate class of
GCs (type II, see below), defined mainly on photometric ground. However, we did
not find any significant spread in [Fe/H], not even considering stars belonging
to the so-called anomalous red giants in the chromosome map, a diagnostic based
on Hubble Space Telescope UV photometry used to define the hypothesised class of
type II GCs (see Milone et al. 2017). Either an intrinsic spread in iron (and
neutron-capture elements, see below) is not a necessary condition to belong to
the so-called type II GCs, or NGC~6388 is not a type II GC, as claimed. 

Finally, in Carretta and Bragaglia (2022b), we addressed the recent claim that
NGC~6388 may have been accreted to the MW, based on the proposed under-abundance
of Sc, V, and Zn in four stars (Minelli et al. 2021a). We instead demonstrated
how our derived  abundances, for a much larger and significant sample of cluster
stars, are fully compatible with those of the MW bulge stars and clusters of
similar metallicity. Together with chemo-dynamical considerations and the
placement of NGC~6388 firmly on the in situ branch of the age-metallicity
diagram, our observations conclusively indicate an in situ origin for this GC.

\begin{figure}
\centering
\includegraphics[scale=0.40]{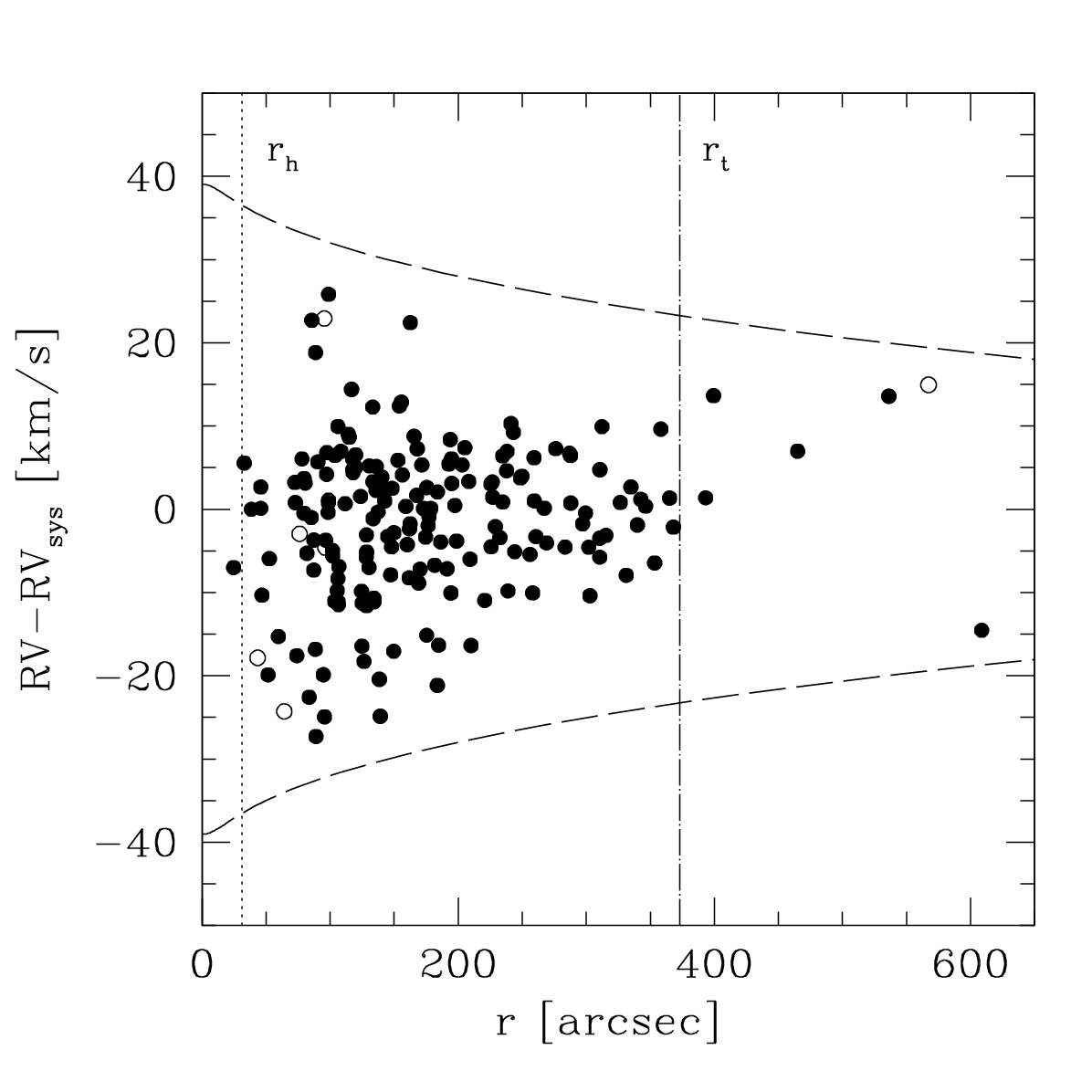}
\includegraphics[scale=0.40]{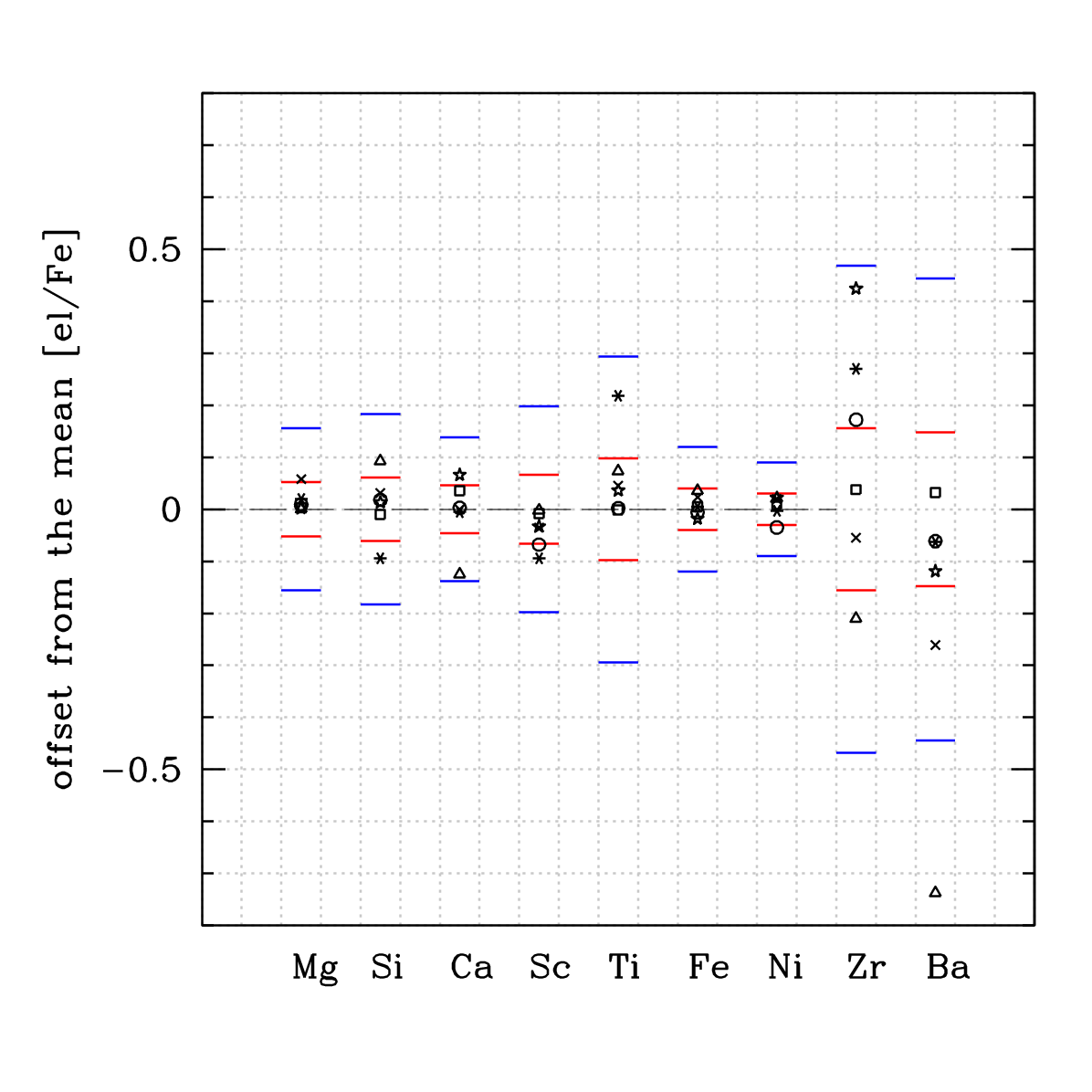}
\caption{ Checks on stars with dubious membership. Upper panel: Radial velocity
profile of NGC~6388 based on  our sample. The RV refers to the cluster systemic
velocity of  82.5 km~s$^{-1}$. The open circles indicate stars defined as
non-members in APOGEE DR17. The vertical lines indicate the half-mass and tidal
radii from Harris (2010). The dashed lines show the $\pm3\sigma$ velocity
dispersion profile of the best-fit King model as in Lanzoni et al. (2013). All
185 stars fall well inside its limits. Lower panel: For stars in common with
APOGEE DR17 and considered non-members there, we show the offsets from the
cluster mean elemental ratios ([el/Fe]). For each element, the red and blue
lines indicate the $1\sigma$ and $3\sigma$ limits, respectively. Only the Ba
value for star l63p075 is off by more than $3\sigma,$ and we conservatively
excluded this abundance from the computed average in
Table~\ref{t:meanabuUNIFIN}.}
\label{f:newfig1}
\end{figure}

In the present paper, we complete and show the analysis of the dataset,
presenting the abundances of 22 species in NGC~6388 from high resolution optical
spectroscopy, which permitted a full and homogeneous chemical characterisation of
this GC. We discuss our results regarding the multiple stellar population
phenomenon, and similarities and differences
with the underlying bulge stellar populations.

The paper is organised as follows. A brief summary of data selection and
observations is provided in Sect.~2, whereas the abundance analysis and error
budget are described in Sect.~3. Results for proton capture and the heavier
elements ($\alpha$-capture, Fe-peak, and neutron-capture elements) are discussed
in Sect.~4 and Sect.~5, respectively. A summary of the properties of NGC~6388 is
provided in Sect.~6.

\section{Summary of data selection and observations}

Details on the full procedure we followed to select member stars to be observed
in NGC~6388, heavily contaminated by bulge and disc stars, are discussed at
length in Carretta and Bragaglia (2022a). We exploited previously made
spectroscopic observations (mainly for kinematics) in NGC~6388 to select radial
velocity (RV) members. The large value of the systemic RV (80 km~s$^{-1}$,
Harris 2010) allowed us to individuate bona fide cluster members. Culling our
targets from the programmes 073.D-0760 (PI Catelan),  381.D-0329 (PI Lanzoni),
and 095.D-0834 (PI Henault-Brunet), we analysed good quality archive data (taken
with GIRAFFE high resolution setups HR13 and HR21 or with UVES) or acquired new 
observations using the same setups (programme 099.D-0047).

Our observing strategy was to obtain GIRAFFE HR13 spectra (to derive atmospheric
parameters and abundances of the light elements O, Na, Mg, Si) and HR21 spectra
(to derive Al abundances) for as many stars as possible in NGC~6388. Moreover,
new UVES/FLAMES spectra were acquired for 12 giants. The new observations were
performed from April to August 2017. The wavelength coverage is 6120-6405~\AA\
for HR13, 8484-9001~\AA\ for HR21, and 4800-6800~\AA\ for UVES/FLAMES. The median
S/N values are 93, 116, and 50, respectively.

Coordinates, magnitudes, original samples (or new observations), and RVs are
listed in Carretta and Bragaglia (2022a) for the 12 and 150 stars with UVES and 
GIRAFFE spectra, respectively. Similar data for 24 giants with previously analysed
UVES/FLAMES spectra can be found in Carretta and Bragaglia
(2018). 

We consider all stars in our sample to be good candidates as members of NGC~6388
based on the combination of RV and metallicity (presented in previous papers),
taking advantage of our derived very small spread in [Fe/H] (0.04 dex from 185
stars  as compared to 0.074 dex from 9 stars in M20).  As already discussed in
CB22a, for a few stars membership is more dubious because of their proper
motions (measured by Gaia) or because they fall outside the tidal radius (if we
take the value in Harris 2010, but not if we consider the much larger value in
Baumgardt's database, which would be outside the upper panel in
Fig.~\ref{f:newfig1}). Additionally, the referee noted that six of our targets
are not considered cluster members in APOGEE DR17. However, as shown in 
Fig.~\ref{f:newfig1} 
(upper panel) all stars in our sample fall well within the $3\sigma$ velocity
dispersion profile of the best-fit King model (for details, see Lanzoni et al.
2013, their Fig. 10 and Table 3). Additionally, if we consider the six candidate
non-members, we see that the abundances of elements not involved in the multiple
populations phenomenon are the same as the bulk of the cluster sample
(Fig.~\ref{f:newfig1},
lower panel). Furthermore, we find that four of them belong to the second
generation, based on their Na and O abundances (see Sect. 4), which decreases
the chance of  them being field interlopers. In the following, we treat all
stars as members, bearing in mind that further studies (e.g. better astrometric
data in future Gaia data releases) will help settle the matter.

\begin{figure}
\centering
\includegraphics[scale=0.40]{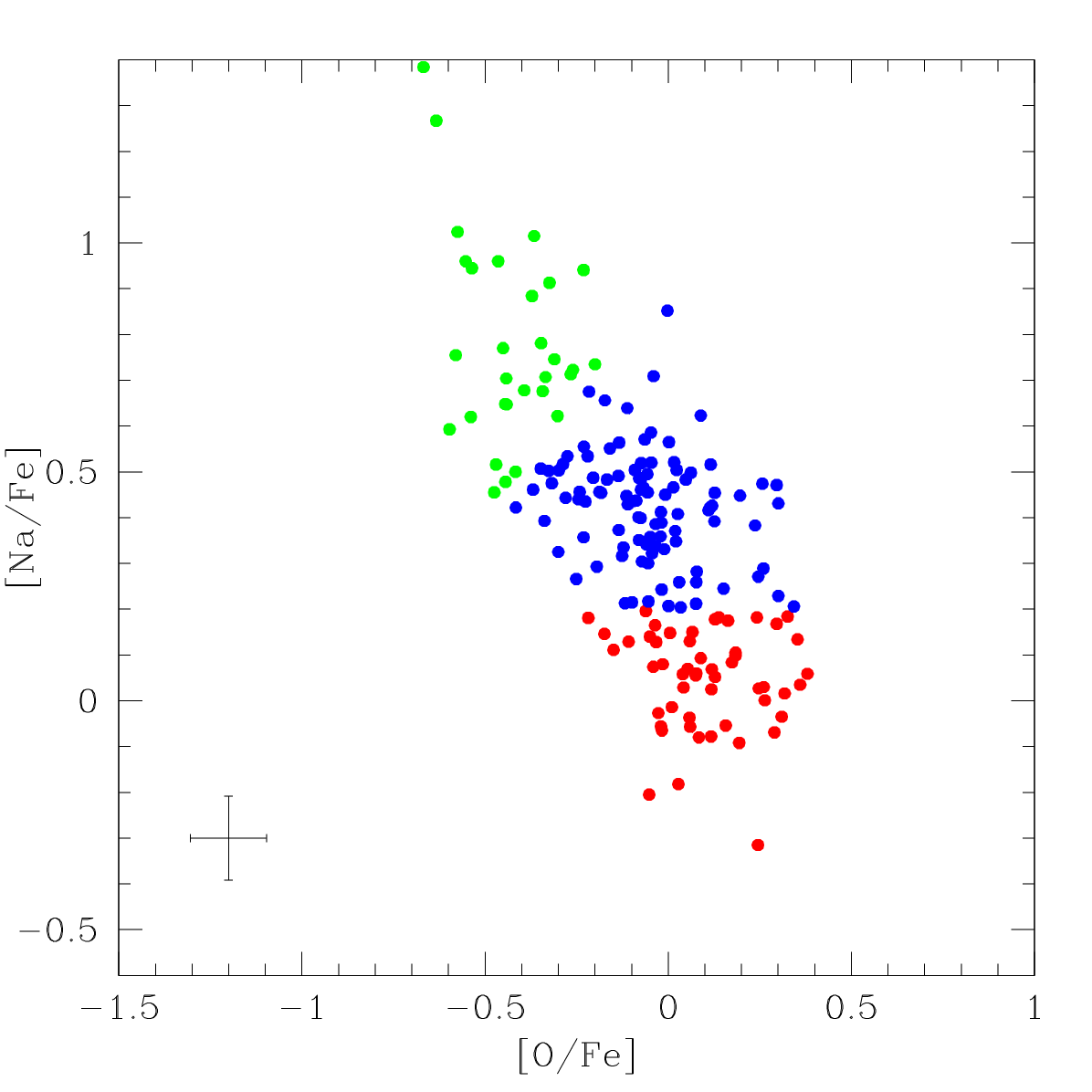}
\caption{[Na/Fe] abundance ratios as a function of [O/Fe] ratios in NGC~6388. 
Stars are colour-coded according their chemical composition following the
definitions in Carretta et al. (2009b): red for stars with primordial
composition (P), blue for those with intermediate composition (I), and green for stars
with heavily altered composition (E). Internal error bars are also reported.}
\label{f:solonao}
\end{figure}

\section{Abundance analysis and error budget}

The same procedure adopted for GCs previously analysed in our FLAMES survey (see
Carretta et al. 2006, 2009a,b) was
used here to derive atmospheric parameters and abundances. The metallicity of
NGC~6388 is reported and discussed in Carretta and Bragaglia (2022a). As usual,
our final effective temperatures T$_{\rm eff}$ were derived in two steps. First estimates
were obtained from the calibrations of Alonso et al. (1999, 2001). Afterwards,
these estimates were refined using a relation between the values from the
first step and $K$
magnitudes of the stars. Surface gravities were then obtained by using the above
temperatures, adopting distance modulus $(m-M)_V=16.14$ and reddening
$E(B-V)=0.37$ from Harris (2010), bolometric
corrections from Alonso et al. (1999), and masses of  0.90 $M_\odot$ and $M_{\rm
bol,\odot}=4.75$.

For the abundance analysis we used equivalent widths (EW) measured with the ROSA package (Gratton 1988), after correcting those measured on GIRAFFE spectra
to the system given by EWs from UVES spectra. The microturbulent velocities
$v_t$ were derived by minimising the slope of the relation between Fe~{\sc i}
abundances and expected line strength  (Magain 1984). Models with appropriate
atmospheric parameters whose abundances matched those derived from Fe~{\sc i}
lines were then interpolated within the Kurucz (1993) grid of model
atmospheres.  Adopted atmospheric parameters and derived abundances of Fe are
listed in Table 2 of Carretta and Bragaglia (2022a).
The absence of trends as a function of T$_{\rm eff}$ and the good agreement
between Fe~{\sc i} and Fe~{\sc ii} and between Ti~{\sc i} and Ti~{\sc ii}
are a good test of the reliability of the adopted scale of atmospheric
parameters.

\begin{table*}
\centering
\caption{Mean abundances in NGC~6388 for the present and previous samples.}
\begin{tabular}{lrcccrcccrccccc}
\hline
 Element             &stars &  avg &  rms & ref &stars &  avg&  rms & ref &stars &  avg &  rms & ref & Sun &ref\\
\hline

$[$Fe/H$]${\sc i}    & 12 & $-$0.509 & 0.039 & (1) &150& $-$0.488 & 0.040 & (2) & 35 & $-$0.480 & 0.045 & (3) & 7.54 & 1\\ 
$[$Fe/H$]${\sc ii}   & 12 & $-$0.491 & 0.039 & (1) &149& $-$0.478 & 0.053 & (2) & 35 & $-$0.415 & 0.081 & (3) & 7.49 & 1\\ 
$[$O/Fe$]${\sc i}    & 12 & $-$0.210 & 0.187 & (1) &148& $-$0.040 & 0.227 & (2) & 35 & $-$0.201 & 0.190 & (3) & 8.79 & 1\\ 
$[$Na/Fe$]${\sc i}   & 12 &   +0.544 & 0.334 & (1) &150&   +0.342 & 0.283 & (2) & 35 &   +0.514 & 0.230 & (3) & 6.21 & 1\\ 
$[$Mg/Fe$]${\sc i}   & 12 &   +0.206 & 0.043 & (1) &149&   +0.221 & 0.052 & (2) & 35 &   +0.212 & 0.049 & (3) & 7.43 & 1\\ 
$[$Al/Fe$]${\sc i}   & 12 &   +0.385 & 0.391 & (1) &125&   +0.464 & 0.303 & (2) & 35 &   +0.423 & 0.350 & (3) & 6.23 & 3\\
$[$Si/Fe$]${\sc i}   & 11 &   +0.342 & 0.029 & (1) &150&   +0.309 & 0.061 & (2) & 34 &   +0.350 & 0.063 & (3) & 7.53 & 1\\ 
$[$Ca/Fe$]$ {\sc i}  & 12 &   +0.082 & 0.047 & (1) &150&   +0.069 & 0.046 & (2) & 35 &   +0.062 & 0.049 & (3) & 6.27 & 1\\ 
$[$Sc/Fe$]${\sc ii}  & 12 & $-$0.016 & 0.059 & (1) &150& $-$0.018 & 0.067 & (2) & 35 & $-$0.027 & 0.084 & (3) & 3.13 & 1\\ 
$[$Ti/Fe$]${\sc i}   & 12 &   +0.292 & 0.050 & (1) &150&   +0.270 & 0.098 & (2) & 35 &   +0.295 & 0.086 & (3) & 5.00 & 1\\ 
$[$Ti/Fe$]${\sc ii}  & 12 &   +0.282 & 0.039 & (1) &   &          &       &     & 35 &   +0.244 & 0.086 & (3) & 5.07 & 1\\ 
$[$V/Fe$]${\sc i}    & 12 &   +0.228 & 0.103 & (1) &   &          &       &     & 35 &   +0.259 & 0.143 & (3) & 3.97 & 1\\ 
$[$Cr/Fe$]${\sc i}   & 12 & $-$0.138 & 0.039 & (1) &   &          &       &     & 35 & $-$0.104 & 0.081 & (3) & 5.67 & 1\\ 
$[$Mn/Fe$]${\sc i}   & 12 & $-$0.220 & 0.061 & (1) &   &          &       &     & 35 & $-$0.218 & 0.052 & (3) & 5.34 & 1\\ 
$[$Co/Fe$]${\sc i}   & 12 &   +0.043 & 0.069 & (1) &   &          &       &     & 35 &   +0.041 & 0.078 & (3) & 4.92 & 2\\ 
$[$Ni/Fe$]${\sc i}   & 12 &   +0.023 & 0.022 & (1) &150&   +0.034 & 0.030 & (2) & 35 &   +0.032 & 0.029 & (3) & 6.28 & 1\\ 
$[$Zn/Fe$]${\sc i}   &  9 &   +0.128 & 0.125 & (1) &   &          &       &     & 31 &   +0.102 & 0.239 & (3) & 4.59 & 1\\ 
$[$Y/Fe$]${\sc i}    & 11 & $-$0.323 & 0.183 & (1) &   &          &       &     & 32 & $-$0.259 & 0.253 & (3) & 2.24 & 2\\ 
$[$Zr/Fe$]${\sc i}   & 10 & $-$0.101 & 0.096 & (1) &138& $-$0.226 & 0.156 & (2) & 30 & $-$0.113 & 0.156 & (3) & 2.60 & 2\\ 
$[$Zr/Fe$]${\sc ii}  & 10 & $-$0.186 & 0.203 & (1) &   &          &       &     & 29 & $-$0.106 & 0.181 & (3) &      &  \\ 
$[$Ba/Fe$]${\sc ii}  & 11 & $-$0.084 & 0.154 & (1) &149& $-$0.016 & 0.135 & (2) & 33 & $-$0.115 & 0.140 & (3) & 2.22 & 3\\ 
$[$La/Fe$]${\sc ii}  & 12 &   +0.155 & 0.065 & (1) &   &          &       &     & 35 &   +0.191 & 0.141 & (3) & 1.22 & 2\\ 
$[$Ce/Fe$]${\sc ii}  & 11 & $-$0.295 & 0.059 & (1) &   &          &       &     & 34 & $-$0.299 & 0.142 & (3) & 1.63 & 3\\ 
$[$Nd/Fe$]${\sc ii}  & 11 &   +0.032 & 0.082 & (1) &   &          &       &     & 34 &   +0.032 & 0.091 & (3) & 1.50 & 2\\ 
$[$Eu/Fe$]${\sc ii}  & 11 &   +0.253 & 0.037 & (1) &   &          &       &     & 32 &   +0.248 & 0.063 & (3) & 0.55 & 3\\ 
\hline
\end{tabular}
\begin{list}{}{}
\item[(1)] This work; new UVES spectra.
\item[(2)] This work; GIRAFFE spectra.
\item[(3)] Total sample with UVES spectra (this work + Carretta et al. 2007 + Carretta and Bragaglia 2018).
\item References for solar: 1=Gratton et al. (2003); 2=Anders and Grevesse (1989); 3=R.Gratton (priv.comm.).] 
\item The average [Ba/Fe] abundance from GIRAFFE is computed by excluding
the star l63p075 (see text).
\end{list}
\label{t:meanabuUNIFIN}
\end{table*}

For the other elements, we proceeded as in our FLAMES survey. 
Oxygen abundances were obtained
from the forbidden [O~{\sc i}] line at 6300~\AA\ (more rarely also from the 
6363~\AA\ line) after cleaning the telluric lines as described in Carretta et
al. (2007a). To correct the Na abundances for non-LTE effects, we adopted the
prescriptions from Gratton et al. (1999). 
Finally, references for the hyperfine structure corrections applied to Sc, V,
Mn, and Co can be found in Gratton et al. (2003).

Average abundances are given in Table~\ref{t:meanabuUNIFIN} for the 12 stars
observed with UVES and the 150 stars with GIRAFFE spectra. We also list the
combination of the abundances of the present sample from UVES with those
previously analysed in Carretta and Bragaglia (2018) to show how good the
consistency is between the two studies that we merged together.
In this table, the neutral species are referred to as Fe~{\sc i}
abudances, whereas abundance ratios of ionised species are computed using  singly ionised iron. Solar reference abundances are 
reported in the penultimate column of Table~\ref{t:meanabuUNIFIN}, with their
sources listed in the last column.

Our procedure to estimate star to star errors due to uncertainties in the 
adopted atmospheric parameters and in $EW$ measurements is  described
in detail in Carretta et al. (2009a) for UVES and Carretta et al. (2009b) for
GIRAFFE. For the present work, the outcomes are summarised in
Table~\ref{t:sensitivityuUNIFIN} and Table~\ref{t:sensitivitymUNIFIN} for
abundances obtained from UVES and GIRAFFE spectra, respectively. For the sake of
completeness, we also report the values relative to iron in these tables (see
Carretta and Bragaglia 2022a, Tables 4 and 5). 

\begin{table*}
\centering
\caption[]{Sensitivities of abundance ratios to variations in the atmospheric
parameters and to errors in the equivalent widths and errors in abundances for
stars of NGC~6388 observed with UVES.}
\begin{tabular}{lrrrrrrrr}
\hline
Element     & Average   & T$_{\rm eff}$ & $\log g$ & [A/H]   & $v_t$    & EWs     & Total   & Total      \\
            & n. lines  &      (K)      &  (dex)   & (dex)   &kms$^{-1}$& (dex)   &Internal & Systematic \\
\hline        
Variation&              &  50           &   0.20   &  0.10   &  0.10    &         &         &            \\
Internal &              &   6           &   0.04   &  0.04   &  0.12    & 0.02   &         &            \\
Systematic&             &  17           &   0.06   &  0.02   &  0.04    &         &         &            \\
\hline
$[$Fe/H$]${\sc  i}& 100 &  $-$0.027     &   +0.041 &  +0.023 & $-$0.044 & 0.019   &0.058    &0.024         \\
$[$Fe/H$]${\sc ii}&  13 &  $-$0.119     &   +0.122 &  +0.040 & $-$0.030 & 0.052   &0.072    &0.055         \\
$[$O/Fe$]${\sc  i}&   2 &    +0.041     &   +0.038 &  +0.017 &   +0.039 & 0.134   &0.142    &0.058         \\
$[$Na/Fe$]${\sc i}&   4 &    +0.062     & $-$0.088 &  +0.017 &   +0.055 & 0.095   &0.117    &0.103         \\
$[$Mg/Fe$]${\sc i}&   4 &    +0.004     & $-$0.027 &$-$0.000 &   +0.018 & 0.095   &0.097    &0.016         \\
$[$Al/Fe$]${\sc i}&   2 &    +0.059     & $-$0.037 &$-$0.018 &   +0.017 & 0.134   &0.136    &0.115         \\
$[$Si/Fe$]${\sc i}&   8 &  $-$0.043     &   +0.018 &$-$0.002 &   +0.026 & 0.067   &0.074    &0.020         \\
$[$Ca/Fe$]${\sc i}&  16 &    +0.081     & $-$0.073 &$-$0.009 & $-$0.023 & 0.047   &0.058    &0.038         \\
$[$Sc/Fe$]${\sc ii}&  8 &    +0.094     & $-$0.034 &$-$0.006 & $-$0.024 & 0.067   &0.074    &0.038         \\
$[$Ti/Fe$]${\sc i}&  26 &    +0.104     & $-$0.045 &$-$0.005 & $-$0.035 & 0.037   &0.058    &0.042         \\
$[$Ti/Fe$]${\sc ii}& 11 &    +0.079     & $-$0.043 &$-$0.008 & $-$0.027 & 0.057   &0.067    &0.033         \\
$[$V/Fe$]${\sc i} &  12 &    +0.109     & $-$0.041 &  +0.003 & $-$0.039 & 0.055   &0.074    &0.051         \\
$[$Cr/Fe$]${\sc i}&  20 &    +0.067     & $-$0.043 &$-$0.005 &   +0.005 & 0.042   &0.044    &0.033         \\
$[$Mn/Fe$]${\sc i}&   7 &    +0.045     & $-$0.030 &  +0.009 & $-$0.011 & 0.071   &0.073    &0.025         \\
$[$Co/Fe$]${\sc i}&   5 &  $-$0.001     &   +0.006 &$-$0.002 & $-$0.009 & 0.085   &0.085    &0.020         \\
$[$Ni/Fe$]${\sc i}&  36 &  $-$0.005     &   +0.015 &  +0.003 &   +0.008 & 0.032   &0.033    &0.008         \\
$[$Zn/Fe$]${\sc i}&   1 &  $-$0.035     &   +0.006 &$-$0.005 & $-$0.013 & 0.189   &0.190    &0.044         \\
$[$Y/Fe$]${\sc i}&    2 &    +0.122     & $-$0.023 &$-$0.005 & $-$0.012 & 0.134   &0.135    &0.069         \\
$[$Zr/Fe$]${\sc i}&   5 &    +0.105     & $-$0.012 &$-$0.001 & $-$0.006 & 0.085   &0.086    &0.046         \\
$[$Zr/Fe$]${\sc ii}&  1 &    +0.089     & $-$0.034 &$-$0.003 &   +0.015 & 0.189   &0.190    &0.060         \\
$[$Ba/Fe$]${\sc ii}&  3 &    +0.126     & $-$0.062 &  +0.006 & $-$0.056 & 0.109   &0.130    &0.054         \\
$[$La/Fe$]${\sc ii}&  2 &    +0.128     & $-$0.043 &$-$0.002 & $-$0.029 & 0.134   &0.139    &0.050         \\
$[$Ce/Fe$]${\sc ii}&  1 &    +0.122     & $-$0.049 &$-$0.005 & $-$0.001 & 0.189   &0.190    &0.047         \\
$[$Nd/Fe$]${\sc ii}&  5 &    +0.122     & $-$0.045 &$-$0.004 & $-$0.026 & 0.085   &0.092    &0.049         \\
$[$Eu/Fe$]${\sc ii}&  2 &    +0.109     & $-$0.044 &$-$0.004 &   +0.013 & 0.134   &0.135    &0.041         \\

\hline
\end{tabular}
\label{t:sensitivityuUNIFIN}
\end{table*}

\begin{table*}
\centering
\caption[]{Sensitivities of abundance ratios to variations in the atmospheric
parameters and to errors in the equivalent widths and errors in abundances for
stars of NGC~6388 observed with GIRAFFE.}
\begin{tabular}{lrrrrrrrr}
\hline
Element     & Average   & T$_{\rm eff}$ & $\log g$ & [A/H]   & $v_t$    & EWs     & Total   & Total      \\
            & n. lines  &      (K)      &  (dex)   & (dex)   &kms$^{-1}$& (dex)   &Internal & Systematic \\
\hline        
Variation&              &  50           &   0.20   &  0.10   &  0.10    &         &         &            \\
Internal &              &   6           &   0.04   &  0.04   &  0.22    & 0.03    &         &            \\
Systematic&             &  57           &   0.06   &  0.02   &  0.01    &         &         &            \\
\hline
$[$Fe/H$]${\sc  i}& 20 &    +0.014     &   +0.032 &  +0.022 & $-$0.060 & 0.027   &0.067    &0.020        \\
$[$Fe/H$]${\sc ii}&  3 &  $-$0.079     &   +0.118 &  +0.040 & $-$0.018 & 0.070   &0.079    &0.097        \\
$[$O/Fe$]${\sc  i}&  2 &  $-$0.001     &   +0.049 &  +0.017 &   +0.057 & 0.086   &0.104    &0.025        \\
$[$Na/Fe$]${\sc i}&  2 &    +0.025     & $-$0.054 &$-$0.009 &   +0.031 & 0.086   &0.092    &0.040        \\
$[$Mg/Fe$]${\sc i}&  2 &  $-$0.015     & $-$0.009 &$-$0.012 &   +0.050 & 0.086   &0.100    &0.018        \\
$[$Al/Fe$]${\sc i}&  2 &    +0.016     & $-$0.054 &$-$0.020 &   +0.033 & 0.086   &0.093    &0.037        \\
$[$Si/Fe$]${\sc i}&  3 &  $-$0.056     &   +0.014 &  +0.001 &   +0.047 & 0.070   &0.085    &0.064        \\
$[$Ca/Fe$]${\sc i}&  5 &    +0.048     & $-$0.064 &$-$0.016 &   +0.007 & 0.055   &0.057    &0.058        \\
$[$Sc/Fe$]${\sc ii}& 2 &    +0.068     & $-$0.033 &$-$0.005 & $-$0.015 & 0.086   &0.088    &0.078        \\
$[$Ti/Fe$]${\sc i}&  3 &    +0.069     & $-$0.035 &$-$0.019 & $-$0.019 & 0.070   &0.074    &0.080        \\
$[$Ni/Fe$]${\sc i}& 11 &  $-$0.022     &   +0.019 &  +0.001 &   +0.035 & 0.037   &0.051    &0.026        \\
$[$Zr/Fe$]${\sc i}&  4 &    +0.084     & $-$0.020 &$-$0.019 &   +0.037 & 0.061   &0.073    &0.097        \\
$[$Ba/Fe$]${\sc ii}& 1 &    +0.090     & $-$0.067 &  +0.005 & $-$0.060 & 0.122   &0.137    &0.105        \\
                                                                           
\hline
\end{tabular}
\label{t:sensitivitymUNIFIN}
\end{table*}

We varied one parameter at a time by the amount listed in the first row and 
repeated the abundance analysis for all stars. The averages provide the
sensitivities of abundance ratios to changes in the atmospheric parameters 
(i.e. $\Delta{\rm [el/Fe]}/\Delta {\rm (par)}$) and are listed in the main body
of the tables. Star-to-star (internal) errors and  systematic errors in each
parameter are in the second and third rows. The typical internal errors in
abundances due to the measurements of EWs (0.02 dex and 0.03 dex for UVES and
GIRAFFE, respectively) are estimated as the average rms scatter in Fe
abundance divided by the square root of the typical number of measured Fe lines
(100 and 20 lines from UVES and GIRAFFE spectra, respectively). 
Finally, the  total internal and
systematic errors in the derived abundances are obtained by summing in
quadrature the contributions of individual error sources, weighted according to
the errors relative to each parameter.

Abundances for individual stars are listed in the Appendix A. For tables
containing elements from the whole UVES+GIRAFFE sample, only an excerpt is
provided as a guidance of the content. The complete tables can be found at CDS,
Strasbourg.

\section{Proton-capture elements}

Sample and observations in NGC~6388 were purposely tailored to measure the
largest set of elements involved in proton-capture reactions resulting in the
network of correlations and anti-correlations observed in GC stars.
Starting from the lightest species (O, Na, Table~A.1) up to the heavier Ca
(Table~A.2) and Sc (Table~A.3), almost all the interested elements are sampled (potassium is still missing).
We also show how, in NGC~6388, these elements define the classical set of 
anti-correlations and correlations typical of massive GCs, with a couple of 
notable exceptions that we discuss below.

\subsection{The Na-O anti-correlation}

The main chemical signature of multiple populations in any GC is evident in
Figure~\ref{f:solonao}, where we show the classical anti-correlation
between Na and O abundances in 183 stars of NGC~6388. Stars are grouped (and
colour-coded) according to their chemical compositions following the PIE scheme
as defined in Carretta et al. (2009b). 
The stars with primordial composition  (P, in red in Figure~\ref{f:solonao}) are
those with the lowest Na abundances, comprised between 
[Na/Fe]$_{\rm min}$\footnote{Conservatively, we excluded from this estimate 
the three stars with the lowest Na values, which seem to be outliers separated
from the bulk of stars along the anti-correlation.} ($=-$0.1 dex) and 
[Na/Fe]$_{\rm min}+0.3$ dex. This choice
guarantees the interception of almost the totality of stars with pristine
composition, since 0.3 dex is typically a 3$\sigma$ star-to-star error in
spectroscopic abundance analysis.  The remainder of stars, with chemical
composition altered by FG polluters (whatever they were) are dubbed second-generation (SG) stars and divided into two subgroups: those with intermediate
composition I (if the ratio [O/Na]$>-0.9$ dex, in blue in
Figure~\ref{f:solonao}) and those with extreme E composition ([O/Na]$<-0.9$ dex,
green points). 

Although more sophisticated methods of statistical cluster analysis (Valle
et al. 2021) only retrieve the main blocks of FG and SG stars, our finer
separation into I and E stars is not arbitrary, but motivated by the existence
of long tails in [O/Na] distributions observed in a number of GCs, such as NGC~2808
and NGC~5904 (M~5) (see Carretta et al. 2009b). In turn, the fractions of stars
within each group are correlated to physical properties of GCs, including the total
mass (see Table 4 and Figure 14 in Carretta et al. 2010a). We also double checked
the PIE attribution using a k-means algorithm. We retrieved three groups almost
identical to the ones adopted here, with minor differences at the group edges,
corroborating the adoption of the same procedure of all papers of our series on
Na-O anti-correlations for the sake of homogeneity. 

\begin{figure*}
\centering
\includegraphics[scale=0.62]{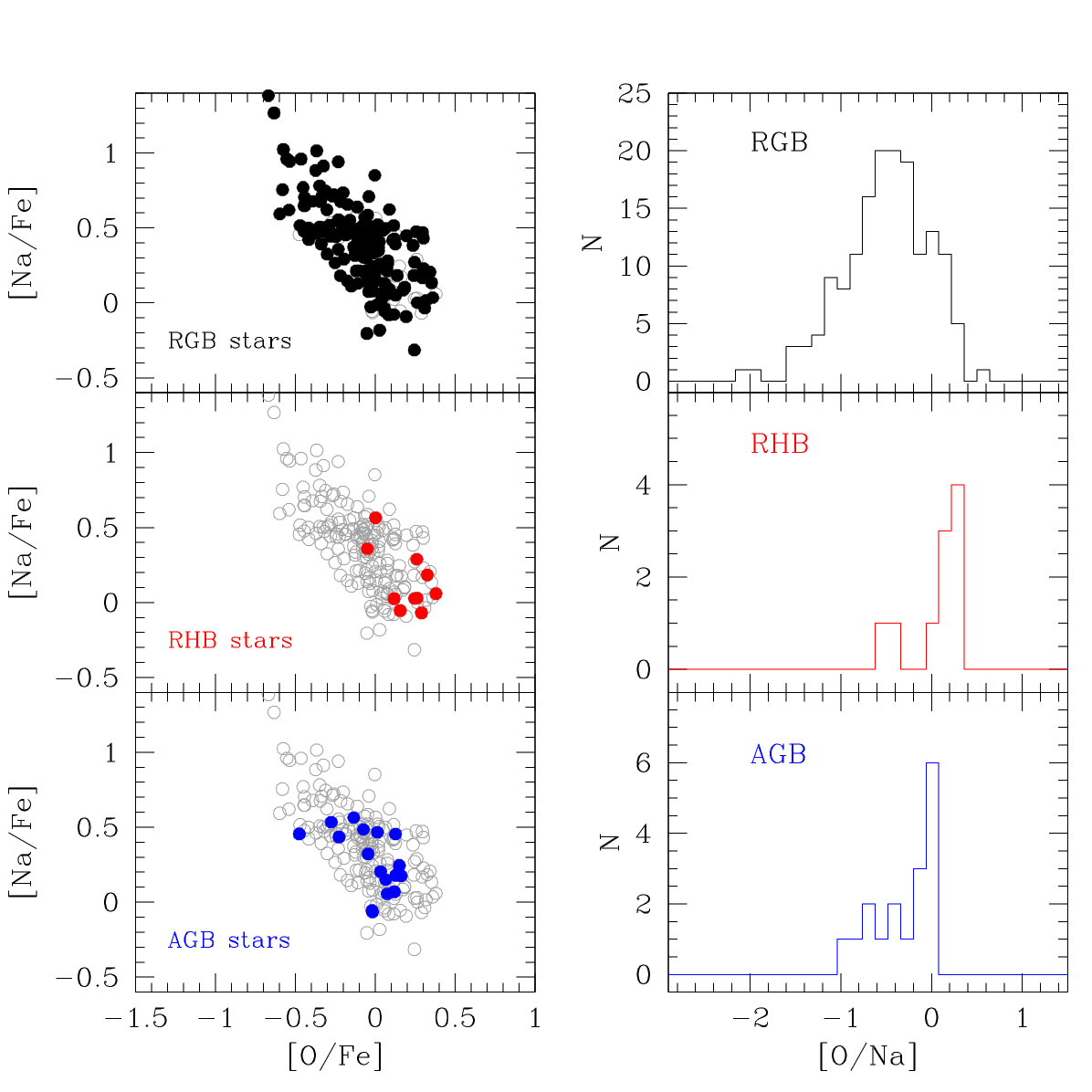}
\caption{Anti-correlation of Na and O abundances in stars of NGC~6388 in
different evolutionary phases. In all three panels on the left, open
circles are our total sample. In the top left panel, we plot
RGB stars as filled circles. In the middle and lower panels, we superimpose values for RHB and AGB
stars, in red and blue filled circles, respectively. In the right column, we show
the observed distributions of the [O/Na] abundance ratios for RGB, RHB, and AGB
stars, from top to bottom, respectively.}
\label{f:naononrgb}
\end{figure*}

In the present work, we confirm the fractions we found in Carretta
and Bragaglia (2018) for NGC~6388, reducing the associated Poisson errors by half thanks to the
much enlarged present sample. A third of stars show a composition with O and Na
levels identical to those of field Galactic stars of similar
metallicity (P=$29\pm 4\%$). The remaining SG stars are then split into two
components, the bulk consisting of stars with intermediate composition 
(I=$54\pm 5\%$). Finally, we confirm once again that in NGC~6388 there is a
noticeable fraction of stars showing an extremely modified composition 
(E=$16\pm 3\%$). These computations are made over 184 stars, because although
two stars are lacking O abundances, one of them (l63p307) has [Na/Fe]$=+0.102$ 
dex, a value that firmly places it in the P population. 

The fraction E also includes the two stars with the highest Na abundances
([Na/Fe]$>1.2$ dex). They are members of NGC~6388 according to both the RV and
metallicity measured on our spectra and $Gaia$ astrometry (Vasiliev and
Baumgardt 2021). They lie nicely on the locus of the Na-O anti-correlation
and are also the stars with the lowest O abundances in our sample ([O/Fe]$<-0.6$
dex, see  Table~\ref{t:protonUNIFIN}). We note that even if all the six 
stars flagged as possible non-members in APOGEE DR17 were excluded, our
results on multiple stellar populations and their respective composition in
NGC 6388 would not change. Their exclusion would only change the fraction of P, I, and E
stars within their associated Poissonian errors.

Formally, we observe a monotonic increase of the metal abundance along the
sequence PIE. The average [Fe/H] of the three groups raises from
$-0.493$ dex for 55 P stars to $-0.485$ dex and $-0.481$ dex for I (100 objects)
and E stars (30 objects), respectively. Although statistically not significant,
this progression is what is expected following an increasing helium abundance in SG
stars of the I and E groups. The effect is to increase the strength of metallic
lines (see B\"ohm-Vitense 1979), and that of neutral lines more than that of ionised
lines. This is clearly confirmed by the iron difference between P and E stars,
which we find to be 0.006 dex for Fe~{\sc ii}, half of that derived from Fe~{\sc
i} lines. Although the combination of tiny differences in abundance between the
groups and internal errors associated with iron prevents a stronger statement,
these results qualitatively conform to expectations.

\subsection{Horizontal and asymptotic giant branch stars}

Our sample in NGC~6388 includes a small number of stars that are not
on their first ascent on the red giant branch (RGB), since our selection
criterion for membership was exclusively based on RVs (see Sect.2). 
This provides the opportunity to homogeneously compare the behaviour of
multiple stellar populations in different evolutionary phases. The main
spectroscopic feature, the anti-correlation of Na-O abundances, is examined in
Fig.~\ref{f:naononrgb}. In the panels of the left column, we highlight the
abundances of ten red HB stars stars (RHB, red filled circles) and 17 asymptotic
giant branch stars (AGB, blue filled circles) on the overall Na-O
anti-correlation obtained for the whole sample. While there are no differences in
the metal abundances (the average [Fe/H] values agree within 0.003 dex), as
expected, the different patterns for proton-capture species are evident. 

Also this second finding is hardly surprising. 
At odds with the PIE
classification of RGB stars (fractions 25, 56, and 19\%, respectively), 
 when we only consider RHB stars we find that 
the vast majority (70\%) are in the P component, with only 30\% falling in the I group of SG stars (just the reverse proportion with
respect to RGB stars). Moreover, no RHB star is found with the highly modified composition
typical of the E component. 

Our findings agree with the well-known scenario where stars with
different He contents are located on distinct portions of the HB. Spectroscopic
detections of He in HB stars are hard to obtain, but since He variations
are correlated to star to star abundance variations in light
elements (see Gratton et al. 2004, 2012a, 2019; Bastian and Lardo 2018) it is
much easier to prove this scenario by analysing the pattern of other elements,
including Na and O (see Gratton et al. 2011, 2012b, 2013, 2014, 2015, Villanova
et al. 2009, 2012, Marino et al. 2011a). The almost complete segregation of
Na-poor/O-rich stars on the red extreme of the HB is a natural consequence of
the strong correlation linking the  chemical variations in proton-capture
species in GC stars and the highest temperature that may be reached on the ZAHB
in each GC, as found by  Carretta et al. (2007b). In turn, these observations agree with the prediction by D'Antona et al. (2002) that variations in
helium (or their proxies, variations in light elements) are a key ingredient to
explain the HB morphology. In NGC~6388, all the analysed HB stars belong to the
RHB, so they behave almost like a simple stellar population, with the [O/Na]
ratio peaking at a high value (see Fig.~\ref{f:naononrgb}, right column, middle
panel).

The lower panels in Fig.~\ref{f:naononrgb} show the situation for the 17 AGB 
stars in our sample. They are 11\% of the RGB stars in our sample, so we are
confident that we did not miss a significant fraction of objects in this evolutionary
phase, since AGB stars are rarer (see discussion in Gratton et al. 2010 and 
their Table 1). 
From the lower left panel in Fig.~\ref{f:naononrgb} and the number counts, we
see that the fraction of SG stars with intermediate composition (the I
component) is similar in RGB and AGB stars (56\%\ and 53\%, respectively). The
largest differences are for the primordial P component (25\% for RGB and 41\%
for AGB stars) and, in particular, for the extreme E component with only one AGB
star classified as such (18\% for RGB and 6\% for AGB stars). 
This agrees with previous findings.  Gratton et al. (2010) discussed the well-established lack of CN-strong objects among AGB stars in GCs (e.g. Norris et al. 1981,
Campbell et al. 2010) and correlated it with the HB properties, in particular
with the expectation that the less massive HB stars do not even begin their AGB
phase (AGB manqu\'e). That means that the most He-rich (Na-rich and O-poor)
stars in GCs, those we would classify as E, do not reach the AGB. More recent
results, sometimes contradictory, and further discussions on the different
fractions of FG and SG in RGB and AGB stars can be found, for instance, in Marino et
al. (2017), Wang et al. (2017), and MacLean et al. (2018).

Although the distributions of both RHB and AGB stars peak at higher values than
RGB stars (right column in Fig.~\ref{f:naononrgb}), the average [O/Na] for
AGB stars is shifted to a lower value than for RHB stars. A look at the panels in
the left column shows that the effect is not due to different Na abundances
(mean [Na/Fe]$=+0.141$ rms=0.205 dex in RHB and  +0.275 rms=0.206 dex in AGB),
but rather to an actual difference in the average O  content (mean
[O/Fe]$=+0.199$ rms=0.140 dex in RHB and $-0.023$ rms=0.173 dex in AGB). We used
Student's and Welch's tests with the null hypothesis that the groups of RHB and
AGB stars are extracted from a distribution with the same means. This
hypothesis cannot be rejected in the case of Na (two-tail probability $p=0.114$,
25 d.o.f.), whereas it can be safely rejected regarding the [O/Fe] average
content ($p=1.2\times 10^{-3}$), confirming that we are probably observing a real
difference. The above discussion also confirms well-known
behaviours in NGC~6388 due to the global multiple population phenomenon, without the need to
invoke systematic or model-dependent effects.

\begin{figure}
\centering
\includegraphics[scale=0.40]{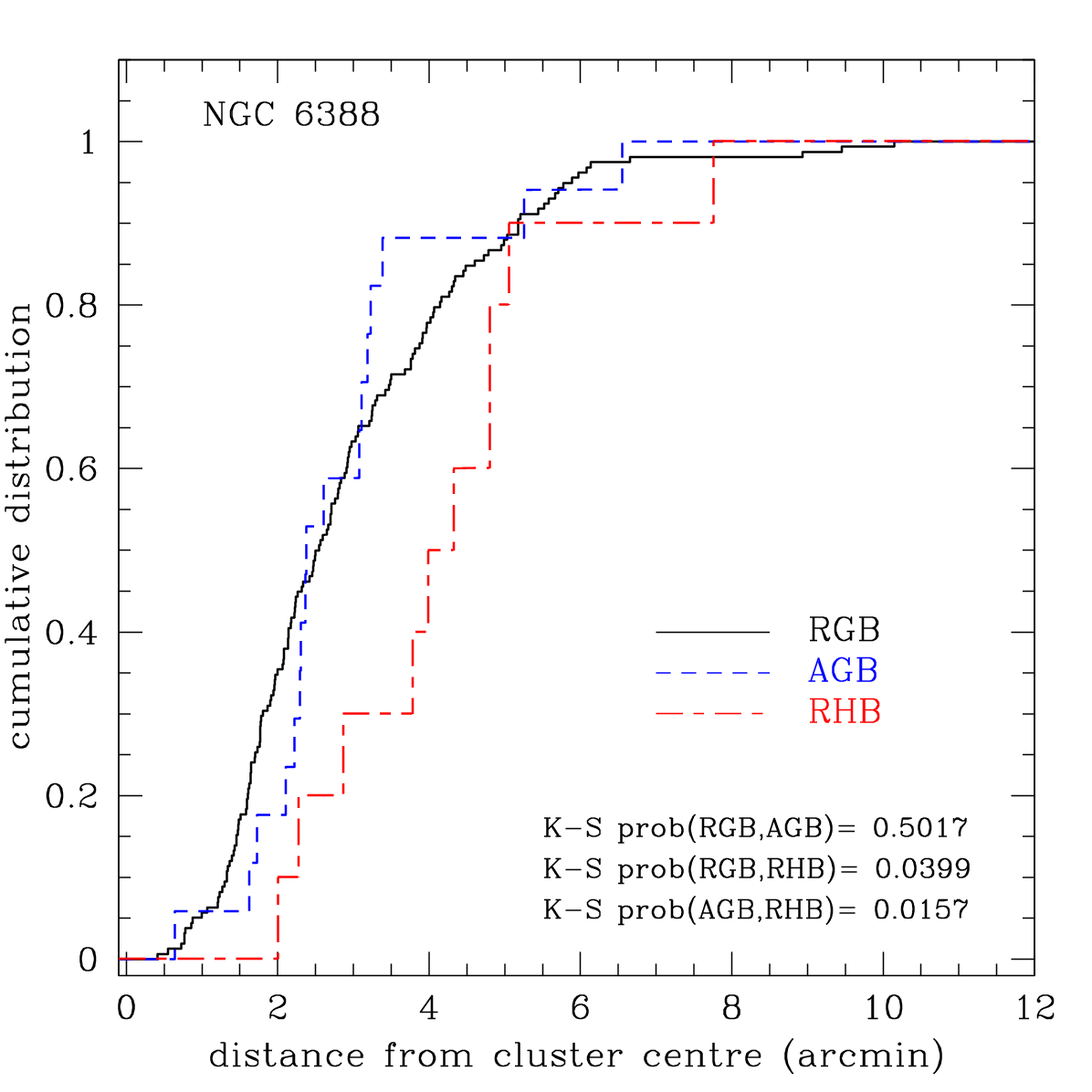}
\caption{Cumulative distributions of radial distances for stars in NGC~6388 in
different evolutionary phases: RGB (black solid line), RHB (red short-long
dashed line), and AGB (blue dashed line). The probabilities of the
Kolmogorov-Smirnov test are labelled in the panel.}
\label{f:ksRHB}
\end{figure}

A further difference concerns the radial distribution of stars in different
evolutionary phases. In Fig.~\ref{f:ksRHB}, we plot the cumulative distributions
of radial distances for stars in our sample, together with the results of a
Kolmogorov-Smirnov test to ascertain statistical differences between different
groups. We find that RHB stars are more externally distributed in the GC than
both RGB and AGB stars. Due to the limited size of the RHB and AGB samples, we
prefer not to expand further on this. On the other hand, we do not find any
relevant difference in the concentration of stars in the P, I, and E groups.

\begin{figure}
\centering
\includegraphics[scale=0.42]{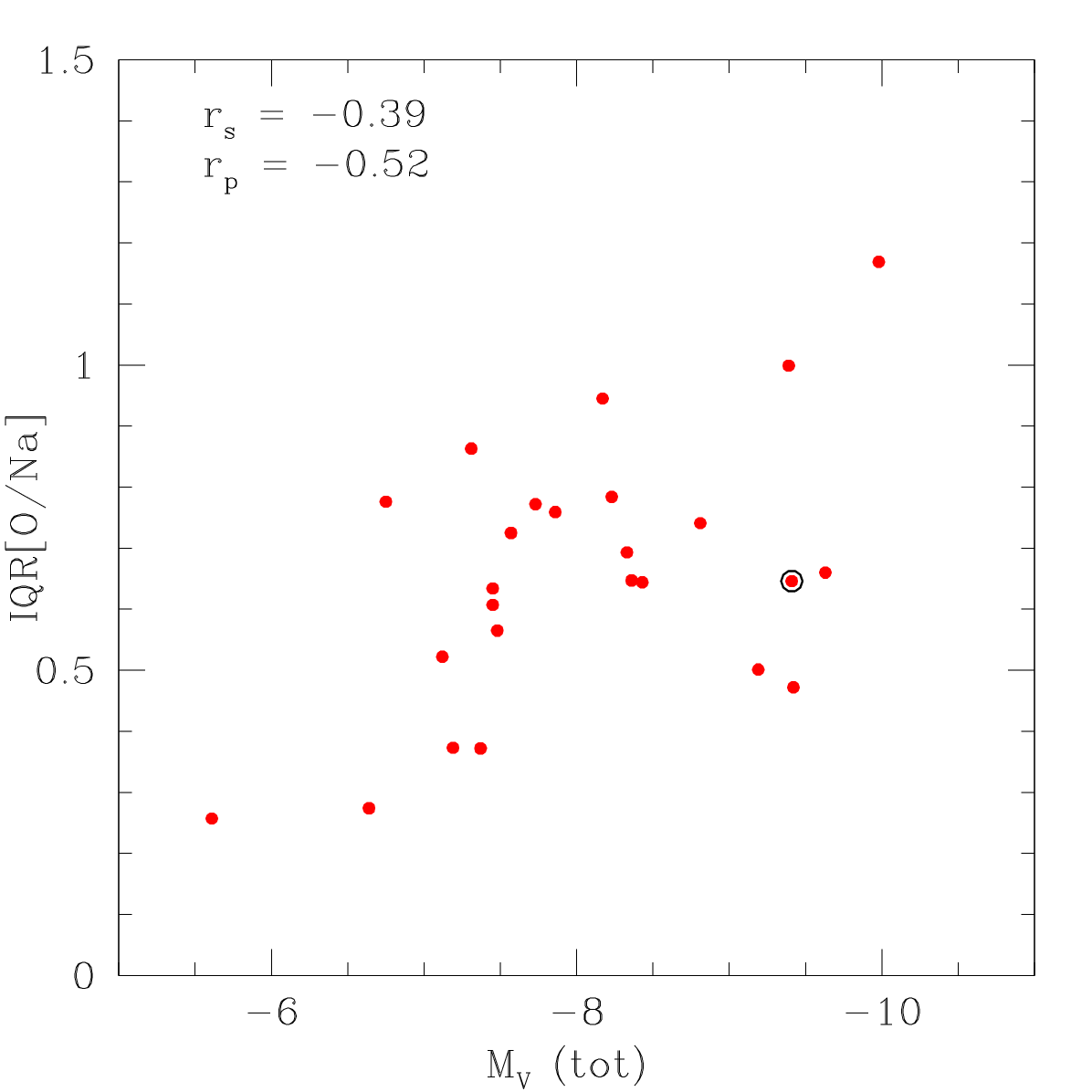}
\caption{Extension of Na-O anti-correlation in 25 GCs of our FLAMES survey
as measured by the inter-quartile range of the [O/Na] ratio, as a function of the
total mass (using the total cluster absolute magnitude as proxy). NGC~6388 is
indicated with an open circle. }
\label{f:iqr}
\end{figure}

\begin{figure*}
\centering
\includegraphics[bb=18 150 585 520, clip, scale=0.60]{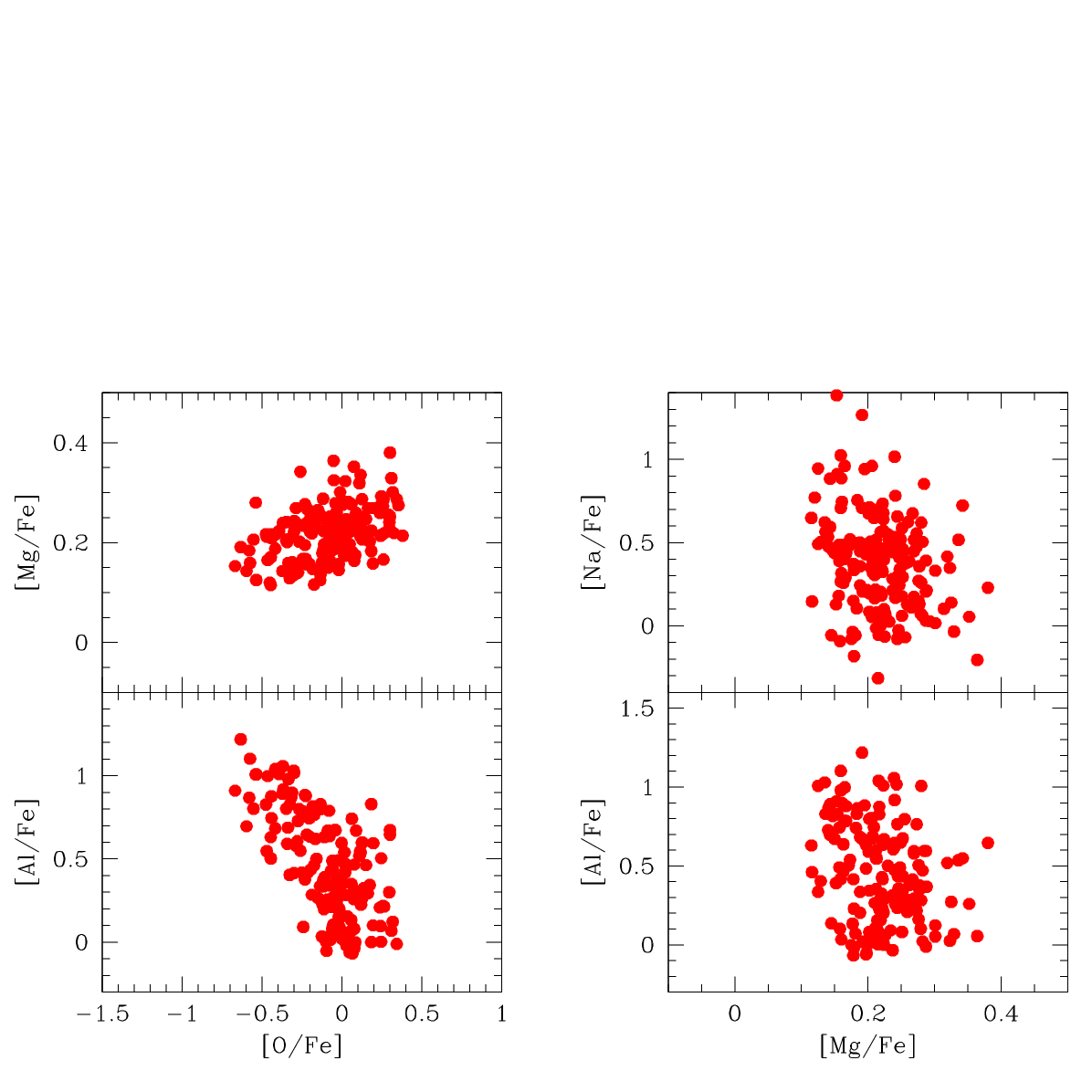}
\includegraphics[scale=0.60]{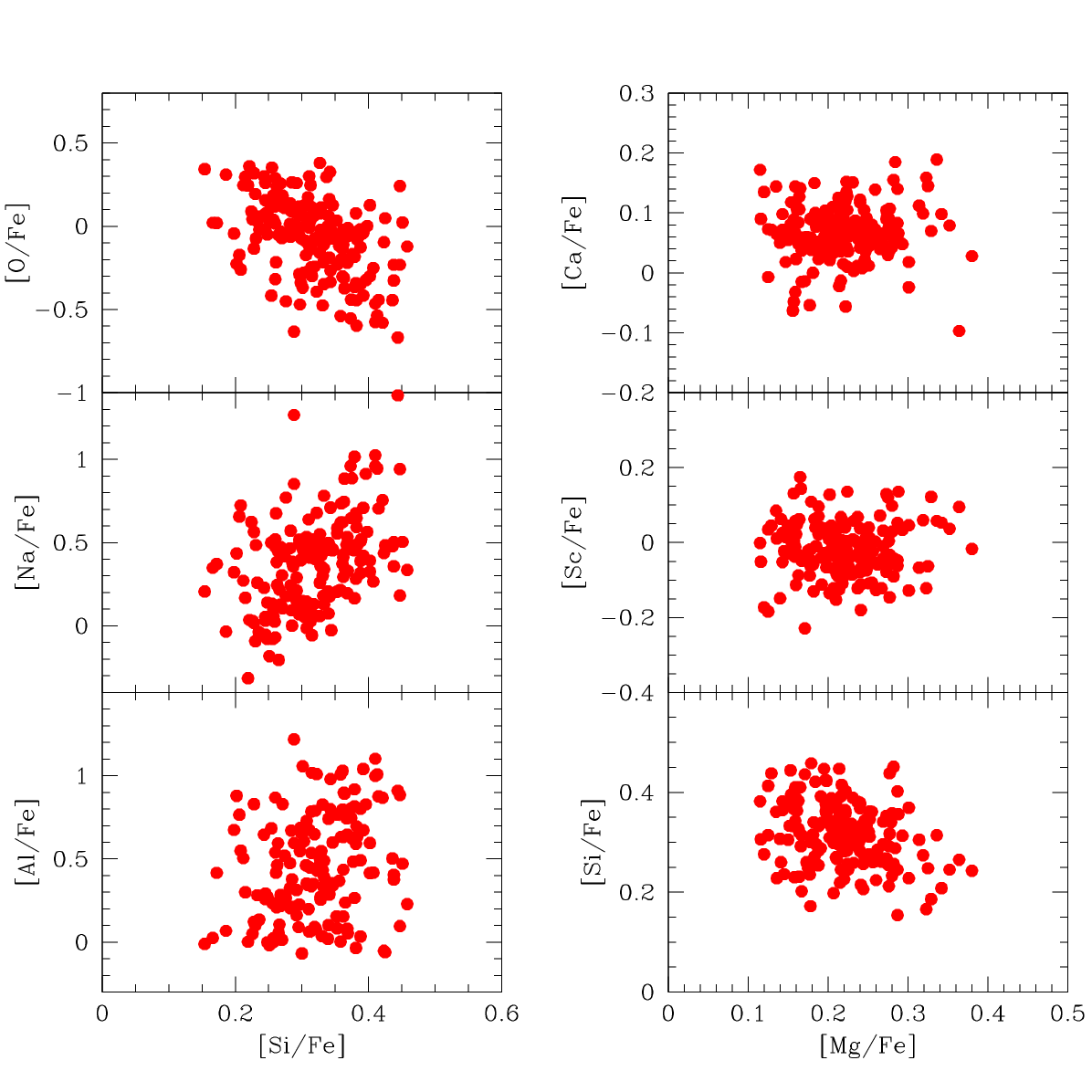}
\caption{Mutual correlations and anti-correlations among proton-capture elements
in stars of NGC~6388. Internal, star to star errors can be found in
Table~\ref{t:sensitivityuUNIFIN} and Table~\ref{t:sensitivitymUNIFIN}.}
\label{f:others}
\end{figure*}

\subsection{The peculiar Na-O anti-correlation of NGC~6388}

The inter-quartile range of the [O/Na] ratio was proposed by Carretta (2006) as a
robust estimate of the extension of the Na-O anti-correlation because it is
more insensitive to outliers than other traditional indicators (e.g. the intrinsic
spread as measured by the rms scatter). From the 183 stars in NGC~6388 with
measured O and Na abundances, we obtained IQR[O/Na]=0.638, which agrees
with the previous value (0.644) derived from a sample of only 49 stars in
Carretta and Bragaglia (2018). We then confirm that NGC~6388
belongs to the group of massive GCs whose extent of the Na-O anti-correlation is
too short with respect to what expected on the basis of their total mass. 
In Fig.~\ref{f:iqr}, we show the IQR[O/Na]-mass relation from our FLAMES survey
(see Carretta et al. 2010a), where we adopt the cluster total absolute magnitude
$M_V$ as a model-independent proxy of the present-day GC total mass. When we
plot the 25 GCs of our FLAMES survey homogeneously analysed, we note that
NGC~6388 falls in a group of four GCs standing out of the main trend, the other
ones being NGC~6441, NGC~104 (47~Tuc), and NGC~7078 (M~15).

This behaviour is not dependent on our abundance analysis, since it is confirmed
by literature data such as 104 giants for NGC~104 (Cordero et al. 2014) or 18
giants in M~15 (Sneden et al. 1997). 
At present, we have no explanation for  these four GCs having an extension
shorter than expected for the Na-O anti-correlation, given their rather large
total mass.

\subsection{The other proton-capture elements}

We determined the abundances of other elements (Mg, Al, Si, Ca, and Sc) involved
in the network of proton-capture reactions acting to modify the primordial
chemical composition of GC stars.  The mutual relations among these species in
NGC~6388 are illustrated in Fig.~\ref{f:others}. Internal errors can be found in
Table~\ref{t:sensitivityuUNIFIN} and Table~\ref{t:sensitivitymUNIFIN}.

\begin{figure}
\centering
\includegraphics[scale=0.42]{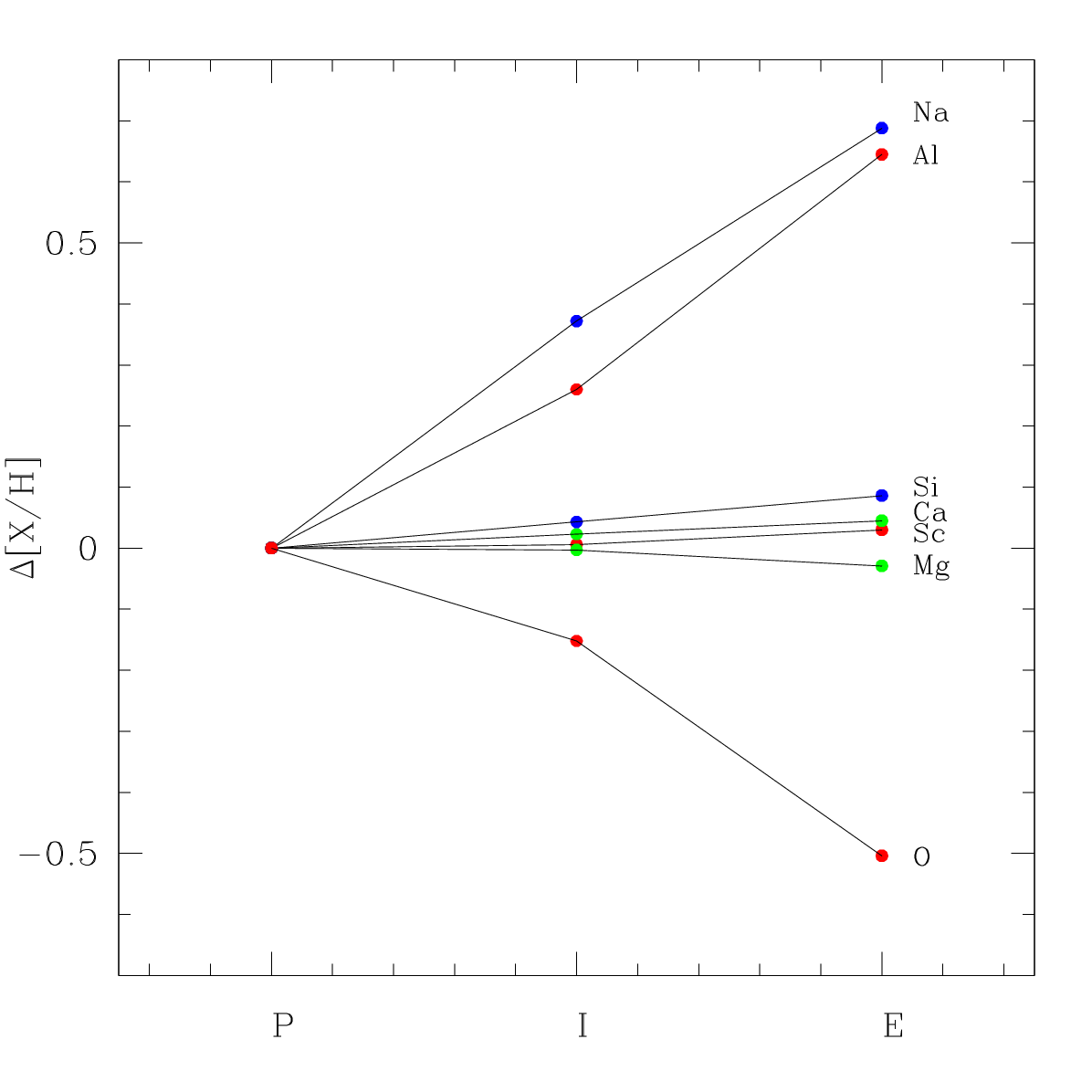}
\caption{Run of different element abundances among PIE groups in 
NGC~6388, indicated by the average differences with respect to the P group, 
whose composition is representative of the primordial abundances in the proto-GC.}
\label{f:trend}
\end{figure}

The trends visible in this figure point out once again that the whole
pattern we see among light elements in GC stars is clearly due to some form of
nucleosynthesis, because the variations in the chemical composition are not
randomly distributed, but follow well-known evolutions. There could be some
residual uncertainties related to the statistical grouping of stars, but the
three stellar populations are consistently ordered according to all abundances.
Enhancement in Al is accompanied by a depletion in Mg, Na is lower when O is
higher, and so on.

These trends are well summarised in Fig.~\ref{f:trend}, where we represent the
average abundances of different elements in the three components P, I, and E in
NGC~6388. In this figure, we plot the average differences with respect to the P
group, so that the trends are a direct representation of the changes in the
chemical composition due to the formation of multiple populations with respect
to the floor of primordial abundances in the proto-GC.

Not surprisingly, we retrieve the large enhancements in Na and Al contents,
paralleled by a noticeable O depletion, and a smaller decrease in Mg
abundance. Both the Si-Al correlation and the Mg-Si anti-correlation act as a
precision thermometer that probes the inner temperature reached by the FG
polluters. The small but steady increase in the [Si/Fe] ratio when progressing
from P to I and to E component (and the simultaneous decrease in [Mg/Fe]) has
been explained (e.g. Karakas and Lattanzio 2003; Yong et al. 2005) by
the leakage from the Mg-Al cycle on $^{28}$Si. This occurs when the two
reactions $^{27}$Al(p,$\gamma$)$^{28}$Si and $^{27}$Al(p,$\alpha$)$^{24}$Mg
switch as relevance at a well-defined temperature of $\sim 65$ MK (see Arnould et
al. 1999, their Figure 8). Concerning NGC~6388, this means that whatever the polluters
were, in the early evolution of the primordial population of the cluster they
were able to reach such a temperature.

In Carretta and Bragaglia (2019), we investigated the
high extreme of the temperature range possibly reached by the FG polluters
in order to reproduce the pattern of heavier proton-capture species, such as Sc and
Ca. The idea was to ascertain in NGC~6388 the presence (or lack) of the
anti-correlations found between the abundances of both Sc and Ca with those of
Mg in the massive cluster NGC~2808 (Carretta 2015). Their existence in
NGC~2808 was interpreted in the same framework of proton-capture reactions at
very high temperature used by Ventura et al. (2012) to explain the unique
pattern of the K-Mg anti-correlation in NGC~2419 (see Cohen and Kirby 2012,
Mucciarelli et al. 2012). The variations of Sc and Ca in NGC~2808 and the
anti-correlation K-Mg, later found by Mucciarelli et al. (2015), showed that this
extreme regime of H-burning can be traced even in more normal GCs than NGC~2419.
However, in Carretta and Bragaglia
(2019) we were able to show that the abundances of Sc and Ca in NGC~6388 cannot
be distinguished from those of field stars of similar metallicity.
The latter represent fairly well the unpolluted, primordial population of stars
where only the effects of supernova nucleosynthesis can be tracked. In turn,
these observations mean that the FG polluters in NGC~6388 were not able to
reach the temperature of about 150 MK, above which seed species including Ar and K
start to be affected, allowing the production of elements such as Sc and Ca (see
Prantzos et al. 2017). The constraints from the Si Al variations and the
evidence of Sc and Ca being largely unaffected would pinpoint a narrower
range of 100-120 MK for the temperature reached in the candidate
polluters if these can be identified with massive AGB stars (see D'Antona et
al. 2016). In Fig.~\ref{f:trend}, we note a slight enhancement in Sc and Ca in
the extreme E fraction of stars in NGC~6388, but it is not significant.

\begin{table}
\centering
\caption[]{Tests on (anti-)correlations.}
\begin{tabular}{lrrrrr}
\hline
rel.  & $r_s$  & $r_p$  &  Nr. &  t   &  prob.    \\
      & Spearm.& Pearson& stars& Stud.& two-tails \\
\hline
Al-O  & $-$0.602  &  $-$0.638 &  160  &  9.477  & $< 1.0 \times 10^{-6}$ \\
Na-O  & $-$0.676  &  $-$0.693 &  183  & 12.342  & $< 1.0 \times 10^{-6}$ \\
Na-Mg & $-$0.242  &  $-$0.267 &  184  &  3.738  & $  2.5 \times 10^{-4}$ \\
Na-Al &   +0.691  &    +0.704 &  160  & 12.460  & $< 1.0 \times 10^{-6}$ \\
Na-Si &   +0.443  &    +0.443 &  184  &  6.666  & $< 1.0 \times 10^{-6}$ \\
Al-Mg & $-$0.255  &  $-$0.264 &  160  &  3.440  & $  7.4 \times 10^{-4}$ \\
Si-Mg & $-$0.286  &  $-$0.305 &  183  &  4.309  & $  2.7 \times 10^{-5}$ \\
Si-Al &   +0.275  &    +0.266 &  159  &  3.458  & $  7.0 \times 10^{-4}$ \\
O-Si  & $-$0.479  &  $-$0.468 &  182  &  7.105  & $< 1.0 \times 10^{-6}$ \\
O-Mg  &   +0.397  &    +0.402 &  182  &  5.890  & $< 1.0 \times 10^{-6}$ \\
Ca-Mg &   +0.032  &    +0.041 &  184  &  0.554  & 0.581 \\
Sc-Mg & $-$0.067  &    +0.024 &  184  &  0.324  & 0.746 \\

\hline
\end{tabular}
\label{t:tabprob}
\end{table}

The scenario and the conclusions discussed above strongly support
statistics as shown in Table~\ref{t:tabprob}, where we list the parameters  of
linear regressions through the different relations among light elements shown in
Fig.~\ref{f:solonao} and Fig.~\ref{f:others}. We  tested the level of
significance for each regression, reporting the
results in the last two columns of Table~\ref{t:tabprob}. The two-tail probabilities listed 
show that all the relations involving O, Na, Mg, Al, and Si in NGC~6388 are real
with a high level of significance. On the contrary, the two anti-correlations
between Mg and Sc, and Mg and Ca are not significant. 
We can then confirm, based on robust
evidence, that in NGC~6388 the FG polluters, whatever they were, likely reached
a maximum inner temperature restricted to a narrow range between 100 and 120-150
MK, as found in Carretta and Bragaglia (2019).

\begin{figure*}
\centering
\includegraphics[scale=0.40]{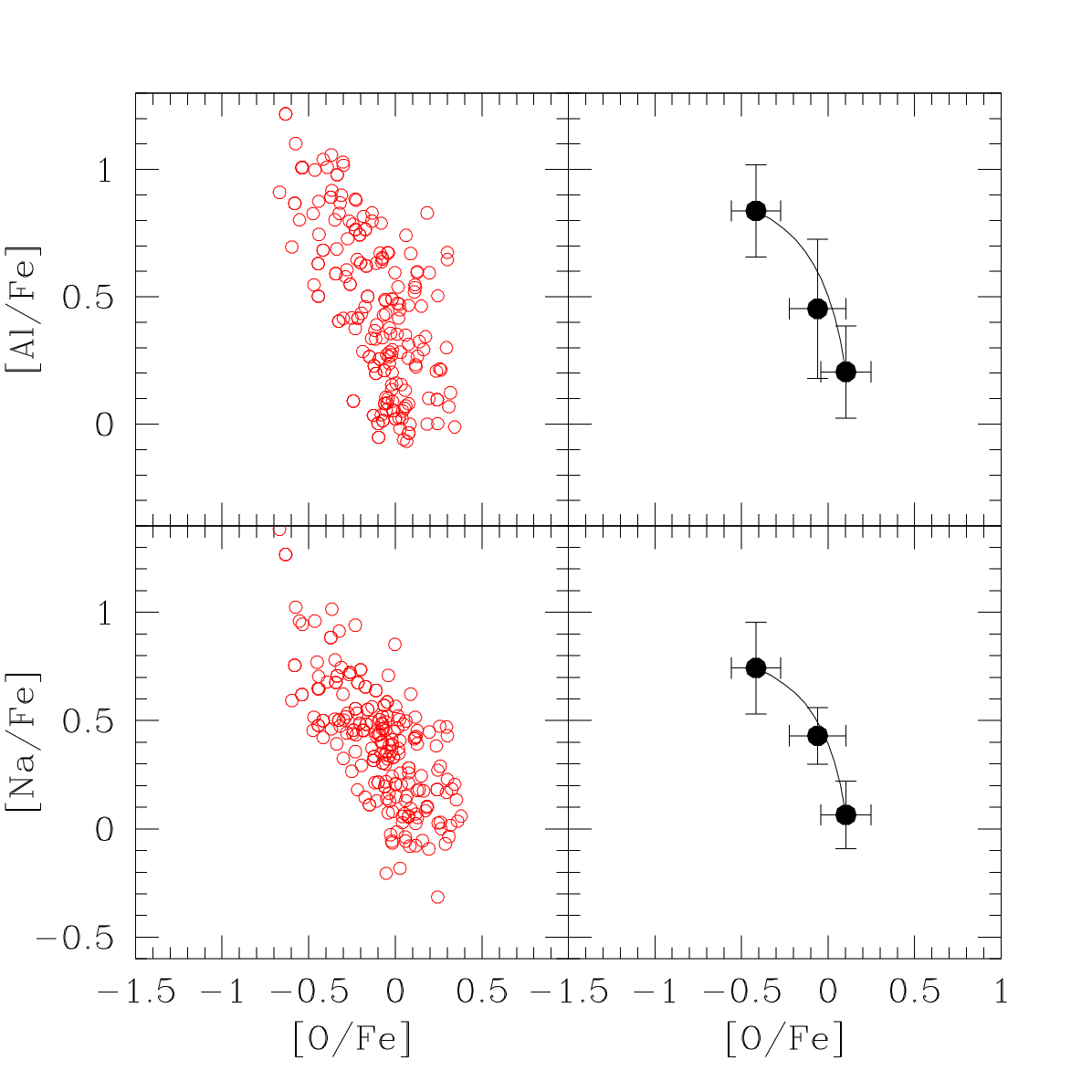}\includegraphics[scale=0.40]{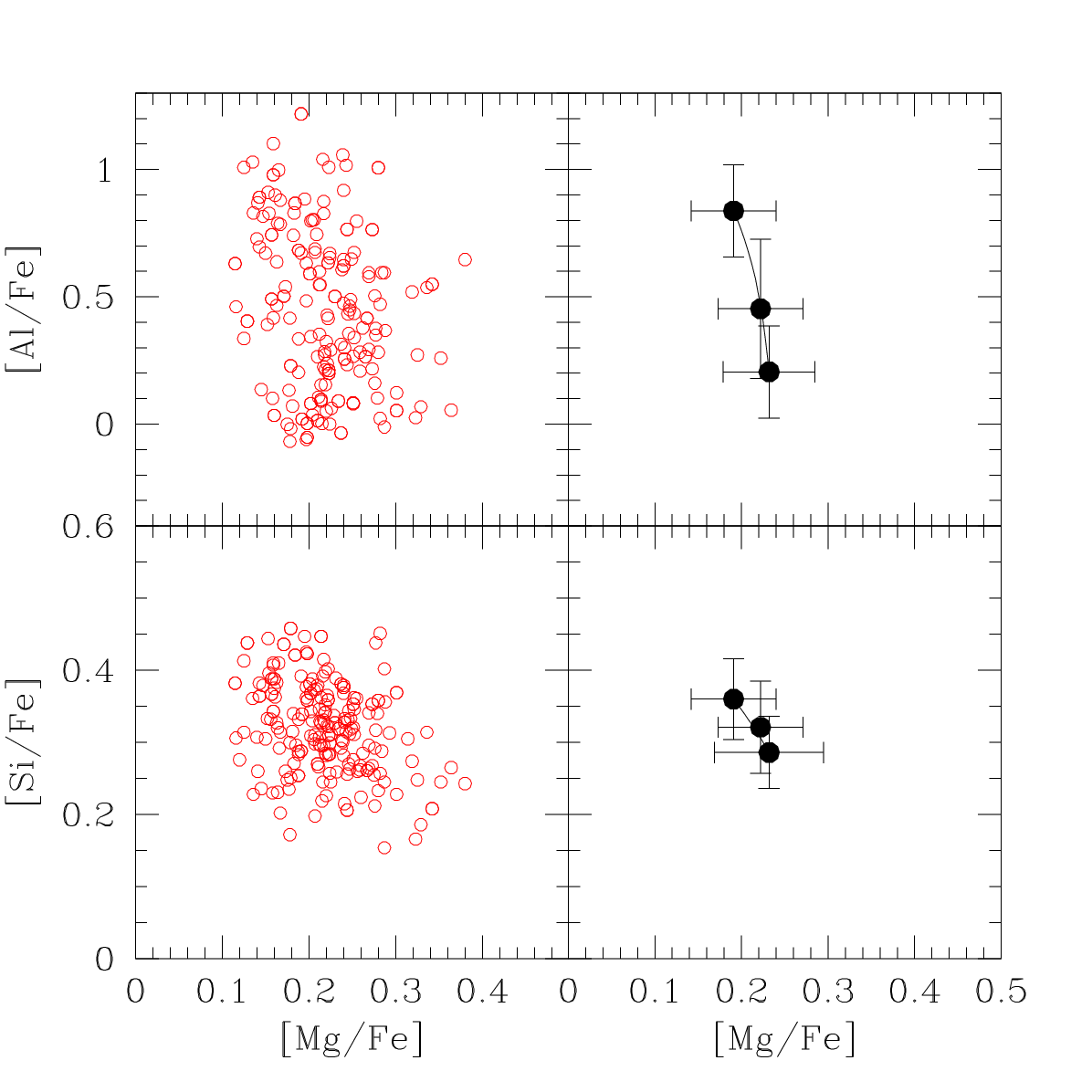}
\caption{Observed Al-O, Na-O, Al-Mg, and Si-Mg anti-correlations in NGC~6388
(empty red circles). Black filled points are the average values of each element
in the P, I, E groups. Solid lines are a simple dilution model anchored to the
primordial P and extreme E components.}
\label{f:dilunifin}
\end{figure*}

\subsection{Light elements, polluters, and dilution}

In Carretta and Bragaglia (2018), we tried to ascertain how many classes or kinds
of polluters may have been contributing to the chemical budget of NGC~6388. We
performed this exercise using the same approach adopted in Carretta et al.
(2012) to study the discrete components in NGC~6752.

We started with the simple dilution model (illustrated, e.g. in Carretta et al.
2009b), where we reproduce the chemical pattern of the various stellar
populations by mixing the composition of the E group with different fractions of
primordial gas, which is simply represented by the composition of the P
component. This means that the I population would be obtained by mixing a
fraction $dil$ of matter with E-like composition together with a fraction whose
composition is P-like. It follows that if only a class of polluters was in
action, then the value of $dil$ should be the same for all elements:

\begin{equation}
dil = { {[A({\rm X})_I-A({\rm X})_P]} \over {[A({\rm X})_E-A({\rm X})_P]} } = { {[A({\rm Y})_I-A({\rm Y})_P]} \over {[A({\rm Y})_E-A({\rm Y})_P]} }
,\end{equation}where $A(X)$ and $A(Y)$ are the abundances in number of atoms of elements X and
Y.

The results reached in Carretta and Bragaglia (2018) were not conclusive. The
main obstacle was the availability of Al abundances only for the 
limited sample of 24 stars with UVES spectra. In turn, this provided two
different possible divisions in the P, I, and E groups, depending on whether the groups
were based on the Al-O or the Na-O plane. As a consequence, we cannot not
entirely exclude the possibility that more than a single class of polluters
could be necessary to produce the observed composition of the I component.

With the present large sample of stars for
which homogeneous abundances of several proton-capture species are obtained in
NGC~6388, this problem is solved. For instance, our sample of stars with derived Al abundances is
increased by a factor 6.7, with respect to Carretta and Bragaglia
(2018). Thus, we repeated the exercise.

We found that in the chemical planes Al-O, Na-O, Al-Mg, and Si-Mg  a simple
dilution model, anchored to the mean values for P and E stars, now nicely passes
through the average value of the intermediate I population, as shown in
Fig.~\ref{f:dilunifin} and at variance with what was found in Carretta and
Bragaglia  (2018).
We thus conclude that the hints of multiple classes of polluters seen in
Carretta and Bragaglia (2018) were likely due to the limited size of the
available samples (especially Al abundances). Our present results are compatible
with the existence of  a single class of FG polluters able to reproduce the
observed characteristics of multiple stellar populations in NGC~6388, when their
ejecta were mixed with variable amount of pristine gas.

\begin{figure}
\centering
\includegraphics[scale=0.40]{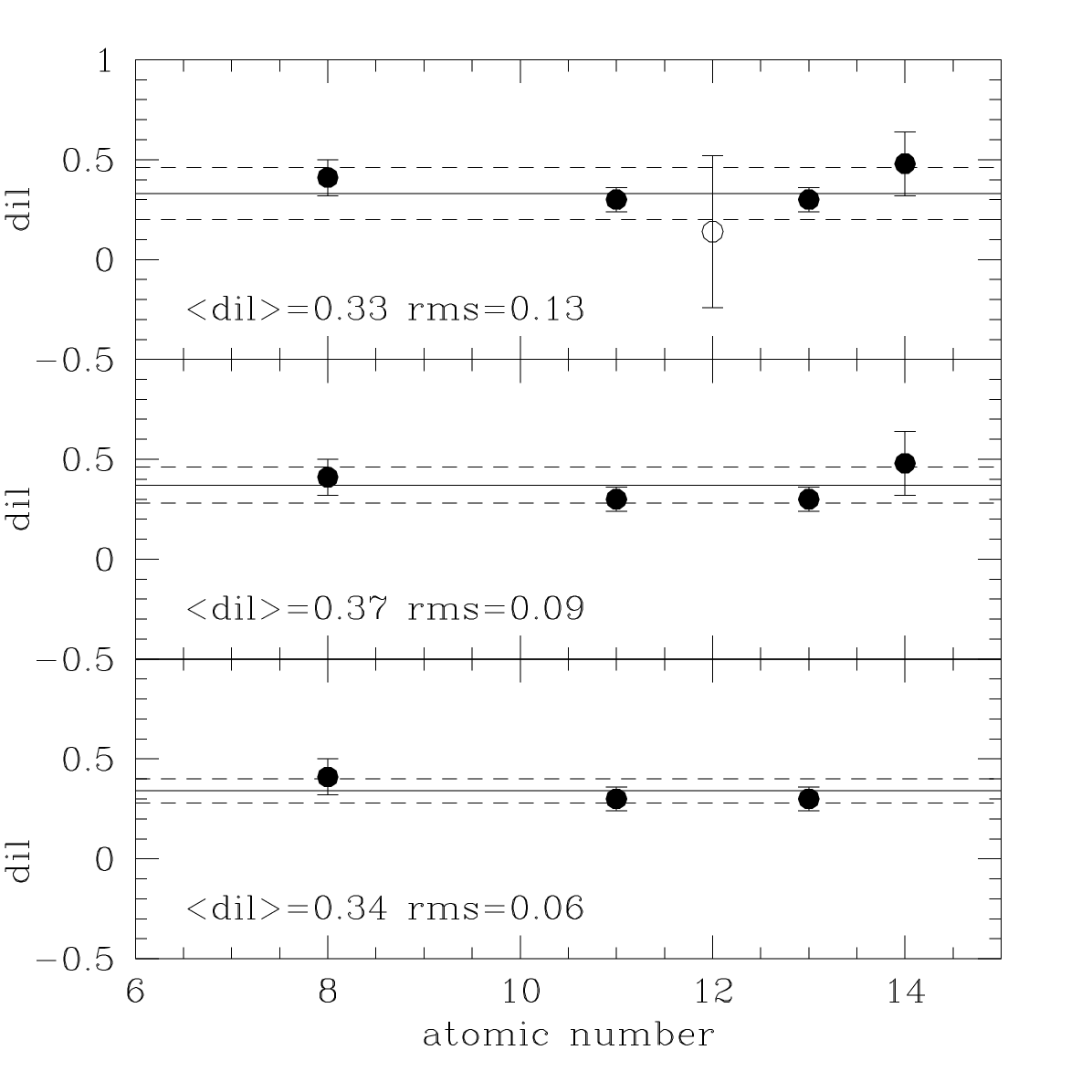}
\caption{Summary of the dilution factors for different elements (indicated by
the atomic number. In the three panels we indicate the average values (solid
lines) and the associated  $\pm 1\sigma$ (dashed line) resulting: i) from all the
five considered species, ii) neglecting Mg (empty symbol with large error bar), and iii)
considering only O, Na, and Al, (from top to bottom, respectively).}
\label{f:summarydil}
\end{figure}

A summary of the above exercise is given in Fig.~\ref{f:summarydil}, where we
plot the resulting dilution factors $dil$ for each considered species
(indicated by the atomic number). The value corresponding to Mg is indicated
with a different symbol (empty circle) to stress the large associated error. The
same behaviour was already noted in Carretta and Bragaglia (2018) and
tentatively attributed to the small variations in the abundance of Mg with
respect to the primordial value. However, we note that star-to-star variations
in the Si content are also small, yet its derived dilution factor is compatible
with the others species. At present, we do not have a satisfactory explanation
for the behaviour of the dilution of Mg.

The values we found for $dil$ are $0.41\pm 0.09$, $0.30\pm 0.06$, 
$0.14\pm 0.38$, $0.30\pm 0.06$, and $0.48\pm 0.16$ for O, Na, Mg, Al, and Si,
respectively. Apart from Mg, all these values lie within $\pm 1\sigma$ from the
average values labelled in Fig.~\ref{f:summarydil}, a strong indication for a unique
class of polluters acting in NGC~6388.

\section{The chemical composition of NGC~6388 in context}

\subsection{$\alpha$-capture elements}

\begin{figure}
\centering
\includegraphics[scale=0.45]{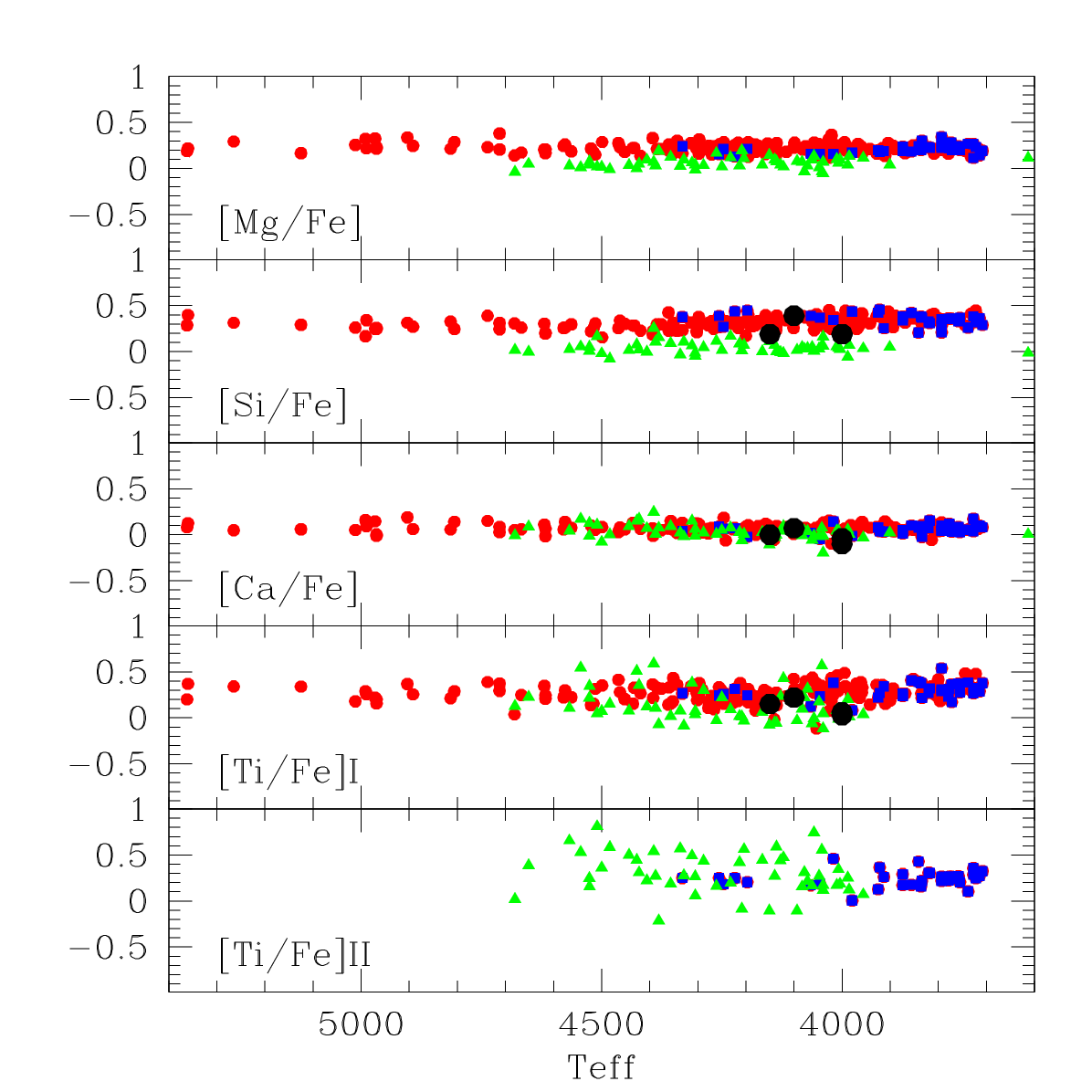}
\caption{Abundance ratios [Mg/Fe], [Si/Fe], [Ca/Fe], [Ti/Fe]{\sc i}, and 
[Ti/Fe]{\sc ii} for stars in NGC~6388 as a function  of the effective
temperature. For the present work, red filled circles indicate stars with
GIRAFFE spectra, whereas stars with UVES spectra are represented with blue
filled squares. Large black circles indicate the four stars analysed by Minelli
et al. (2021a), and green filled triangles are stars in APOGEE DR17 flagged as
members of NGC~6388.}
\label{f:alphateff5}
\end{figure}

Thanks to the spectral range of UVES spectra and to the large size of our
overall dataset, we were able to analyse the full set of $\alpha$-elements in
NGC~6388 for a large number of stars. Both products of hydrostatic burning (O,
Mg) and explosive nucleosynthesis (Si, Ca, Ti) were analysed. Their run as a
function of the effective temperature is summarised in Fig.~\ref{f:alphateff5}
(red and blue points for stars with GIRAFFE and UVES spectra, respectively).
Internal error bars for this figure are listed in 
Table~\ref{t:sensitivityuUNIFIN} and Table~\ref{t:sensitivitymUNIFIN} for
abundances derived from UVES and GIRAFFE spectra, respectively. 
No dependence of abundance on T$_{\rm eff}$ is present for any of these 
elements, over a range of luminosity of more than four magnitudes. 

Since the key work by Tinsley (1979), abundance ratios were
used as probe of the birth properties of stars. The interplay between star
formation and stellar lifetimes (driven by the mass) means that elements from
different nucleosynthetic sites can be combined into abundance ratios useful to
uncover the primordial environment where the observed stars were born.
In particular, the $\alpha$-Fe plane is a privileged indicator, since a
similarity in this plane for two stellar populations indicates a similar
chemical evolution of the systems. However, the metallicity at which the so-called knee due to the onset of Type Ia SNe occurs depends on the system's
stellar mass and on the efficiency of star formation in the progenitor galaxy
(see, e.g. the review by Tolstoy et al. 2009). Owing to the impact of these
abundance patterns on the clues for the origin of NGC~6388, our results on
$\alpha$-elements deserve a detailed discussion.

The average
amount of $\alpha$-elements in NGC~6388 is overabundant with respect to the
solar level, which is compatible with a chemical  composition dominated by the
contribution of type II SNe acting before a significant number of SN Ia started
to add increasing amounts of iron and lower the ratio [$\alpha$/Fe]. This
evidence, together with the time delay expected for the bulk explosions of SN
Ia, is also consistent with GCs in the bulge being old and 
nearly coeval to halo GCs, with similar levels of $\alpha$-element
overabundance.

The average [Si/Fe] ratio reflects the typical overabundance with respect to the
solar values found also  in the majority of GCs and for the other 
$\alpha$-elements in Fig.~\ref{f:alphateff5}.  We found a mean [Si/Fe] of 0.31 dex from
150 stars with GIRAFFE spectra and 0.35 dex from 34 stars observed with
UVES, after combining the results for the 12 new stars of the present study with
the sample analysed in Carretta and Bragaglia (2018). We thus confirm the high
value we found in our first analysis for NGC~6388 (Carretta et al. 2007a), which is at
odds with the low values derived by the APOGEE DR16 ($<$[Si/Fe]$>=-0.03\pm 0.1$
dex, as reported in Horta et al. 2020) and DR17 ($<$[Si/Fe]$>=+0.045\pm 0.061$
dex), from 53 stars flagged as members of NGC~6388 (Abdurro'uf et al. 2022).

Conversely, in NGC~6388 we found a value ($\sim 0.07$ dex, see 
Table~\ref{t:meanabuUNIFIN}) for the  average [Ca/Fe] ratio lower than for
other $\alpha$-elements.  However, this is fully consistent with
our previous analyses (Carretta et al. 2007a, Carretta and Bragaglia 2018).
Gratton et al. (2006) observed a similarly low value in the massive bulge
cluster NGC~6441, and they advanced the hypothesis that this deficiency in Ca
could be an artefact of the  analysis, due to using strong lines in bright and
cool giants. However, this explanation hardly applies to our sample, which
includes stars spanning a range  of about 1700 K in T$_{\rm eff}$. 

A comparison of the abundance of $\alpha$-elements in NGC~6388 from both optical
and infrared studies is provided in Fig.~\ref{f:alphateff5}. To our  derived
abundances we superimposed the four stars reanalysed by Minelli et al. (2021a)
(large black circles) and the about 50 stars flagged as members of NGC~6388 in
the APOGEE DR17 (filled green triangles). Looking at this Figure, clear offsets
are present between the abundances from infrared and optical spectra, with both
optical analyses giving consistent results (at least for the $\alpha$-elements,
see Carretta and Bragaglia 2022b and  the next section). In particular, Mg and Si
abundances from APOGEE are  lower than we find here, whereas a good agreement is
found for Ca abundances. For Ti, we are instead seeing a larger spread in APOGEE
abundances.

Part of the offsets may be explained by differences in the adopted solar
reference abundances. They are not given in the DR17 paper, but assuming that they are
not changed from DR16, we used the solar values provided by Smith et al. (2021)
for DR16 to compute the corrections needed to bring infrared data on our solar
scale (respectively +0.19, +0.11. +0.13, and -0.01 dex for Mg, Si, Ca, and Ti).
The overall agreement would improve for Mg and Si, but it would worsen for Ca.

\begin{table*}
\begin{center}
\caption{Averages and rms scatters for $\alpha$-element ratios in NGC~6388 and NGC~6441.}
\setlength{\tabcolsep}{1.5mm}
\begin{tabular}{lccccccccc}
\hline\hline
abund.           & this work    & this work    &  M21          & WKA07         &   APOGEE     & NGC~6441   & NGC~6441      & M20 & M20     \\
ratio               &   UVES       &      GIRAFFE &               &               &     DR17     & G06  & G07  & all & S/N$>$70\\
\hline
$[$Mg/Fe$]$         & +0.212 0.049 & +0.221 0.052 &               &  +0.259 0.175 & +0.059 0.056 & +0.34 0.09 & +0.38 0.14 & +0.069 0.167 & +0.152 0.140\\
$[$Si/Fe$]$         & +0.350 0.063 & +0.309 0.061 &  +0.240 0.100 &  +0.368 0.158 & +0.045 0.061 & +0.33 0.11 & +0.41 0.19 & +0.118 0.190 & +0.223 0.219\\
$[$Ca/Fe$]$         & +0.062 0.049 & +0.069 0.046 &$-$0.018 0.071 &$-$0.185 0.223 & +0.032 0.074 & +0.03 0.04 & +0.21 0.19 & +0.070 0.193 & +0.130 0.243\\
$[$Ti/Fe$]${\sc i} & +0.295 0.086 & +0.270 0.098 &  +0.115 0.087 &$-$0.106 0.120 & +0.132 0.177 & +0.29 0.10 & +0.33 0.20  &              &                \\
$[$Ti/Fe$]${\sc ii}& +0.244 0.086 &              &               &  +0.294 0.128 & +0.304 0.221 & +0.33 0.14 &             &              &                \\
\hline
\end{tabular}
\label{t:meanalpha}
\begin{list}{}{}
\item[M21:] Minelli et al. (2021a).
\item[WKA07:] Wallerstein, Kovtyukh, Andrievsky (2007).
\item[G06,G07:] Gratton et al. (2006: UVES); Gratton et al. (2007: GIRAFFE). 
\item[M20:] M\'esz\'aros et al. (2020).
\end{list}
\end{center}
\end{table*}

For a more quantitative comparison,  in Table~\ref{t:meanalpha} we report the averages
and rms scatters for $\alpha$-elements from the present and other studies:
Minelli et al. (2021a); Wallerstein et al. (2007), who studied eight cool giants
in NGC~6388 (we report the values from their analysis with photometric
gravities); and the mean values from APOGEE DR17. As a further comparison, we
added the average values derived from UVES and GIRAFFE spectra by Gratton
et al. (2006, 2007, respectively) in NGC~6441, often considered a twin GC of
NGC~6388, with similar origin and characteristics. Finally, the last two
columns provide the values for NGC~6388 derived by M20, both from all available
stars and by selecting stars with high S/Ns according to their criteria.

The values of [Mg/Fe] and [Si/Fe] from APOGEE DR17 are
lower than those of the analyses based on optical spectra. They are also
lower than the values given by M20, in particular when only high-S/N stars are
used in their reanalysis with BACCHUS of an earlier SDSS/APOGEE release.
On the contrary, the Ca level seems to be actually low in these
GCs, with the notable exception of NGC~6441 from GIRAFFE spectra (see, however,
the warning in Carretta and Bragaglia 2021, 2022b about the  time on targets for
NGC~6441 and the relative low S/N in that cluster).

\begin{figure}
\centering
\includegraphics[scale=0.45]{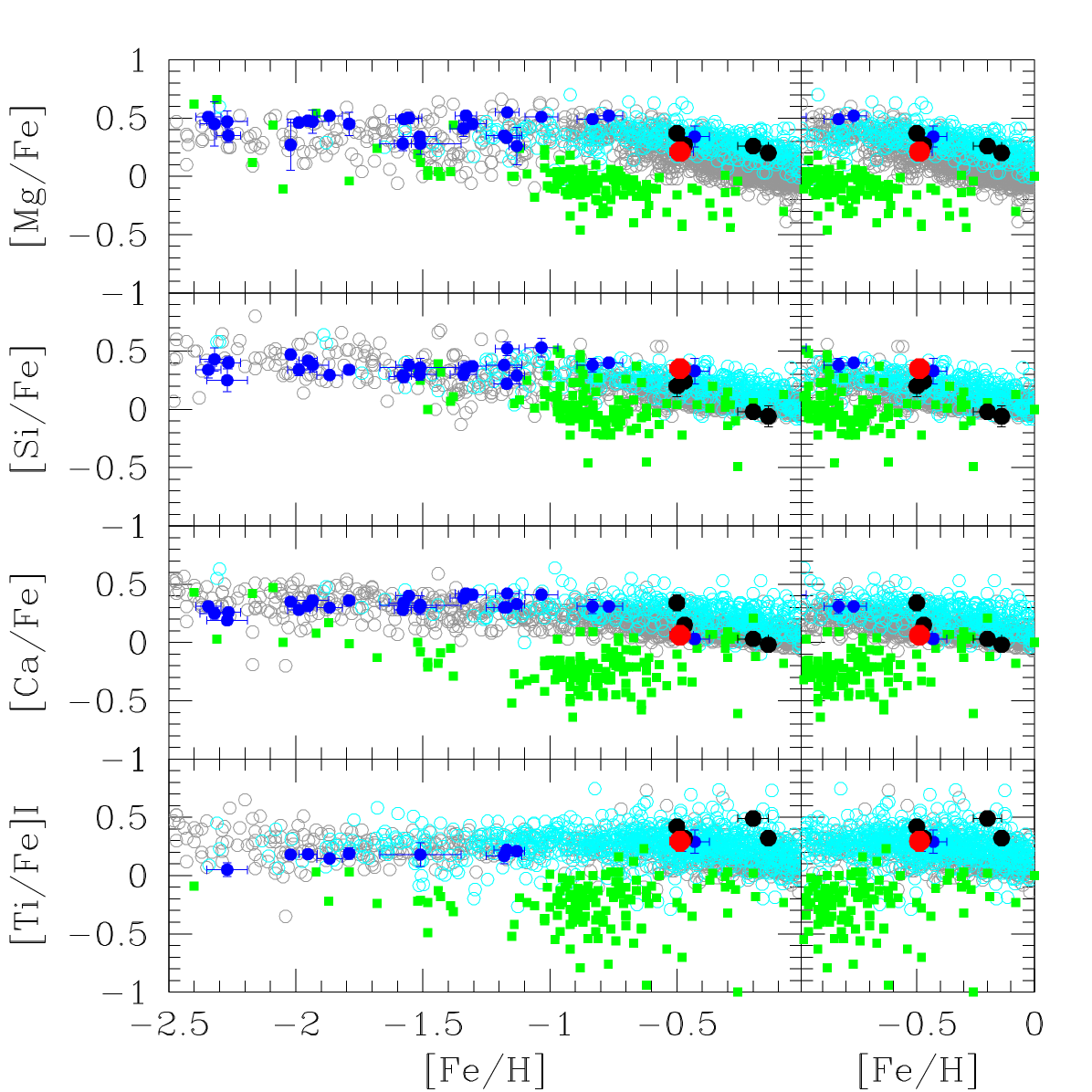}
\caption{Comparison of abundances of $\alpha$-elements with literature data.
Left panels: Mean abundance ratios [Mg/Fe], [Si/Fe], [Ca/Fe], and [Ti/Fe]{\sc i}
(from top to bottom) as a function of metallicity from the homogeneous analysis
of GCs in our FLAMES survey (filled blue points). Black filled circles are the
average ratios in four metal-rich GCs from Mu\~{n}oz and collaborators. The
larger red point indicates NGC~6388 (present  work). Grey and cyan open circles
are field stars in the disc and bulge of the Milky Way, respectively, from
different studies.  References for all the data used in this figure are reported
in Appendix B, Tables B.1 and B.2. Green filled squares are field stars in the
dSph galaxies Fornax (Letarte et al. 2010, Lemasle et al. 2014) and Sagittarius
(Minelli et al. 2021b). Right panels: Enlargement around the metallicity of
NGC~6388.}
\label{f:alphaGC}
\end{figure}

In Fig.~\ref{f:alphaGC}, we also compare  the average abundances of Mg, Si, Ca,
and Ti from  the present work with the abundances in the GCs homogeneously
analysed in our FLAMES survey (see Carretta et al. 2006 and following studies,
referenced in Table B.1), represented by blue filled circles with error bars.
These data have been complemented by four metal-rich GCs (three bulge clusters and one
disc cluster) from Mu\~{n}oz and collaborators.  
As reference, in the same figure we also plot field stars of the Galaxy, both in
the halo and disc components (open grey circles) and in the bulge (cyan open
circles), from several literature studies. 

For the sake of clarity, all references to the data used are provided in
Appendix B (Tables B.1 and B.2), where we also give the identification of each
GC and its [Fe/H] value. Finally, we also plot abundances for the two more
massive dwarf spheroidal galaxies associated with the MW, one still distinct
(Fornax: Letarte et al. 2010, Lemasle et al. 2014) and the other already
accreted and disrupting in the MW (Sagittarius, Minelli et al. 2021b). These
stars are used as a robust benchmark for the pattern of $\alpha$-elements in
external systems of lower total mass than the Galaxy. The combination of
lower star formation and chemical evolution shows up in a lower metallicity knee
and resulting underabundance of these elements, once such systems were possibly
accreted into the main Galaxy, as is currently happening to Sgr.

\begin{figure}
\centering
\includegraphics[scale=0.45]{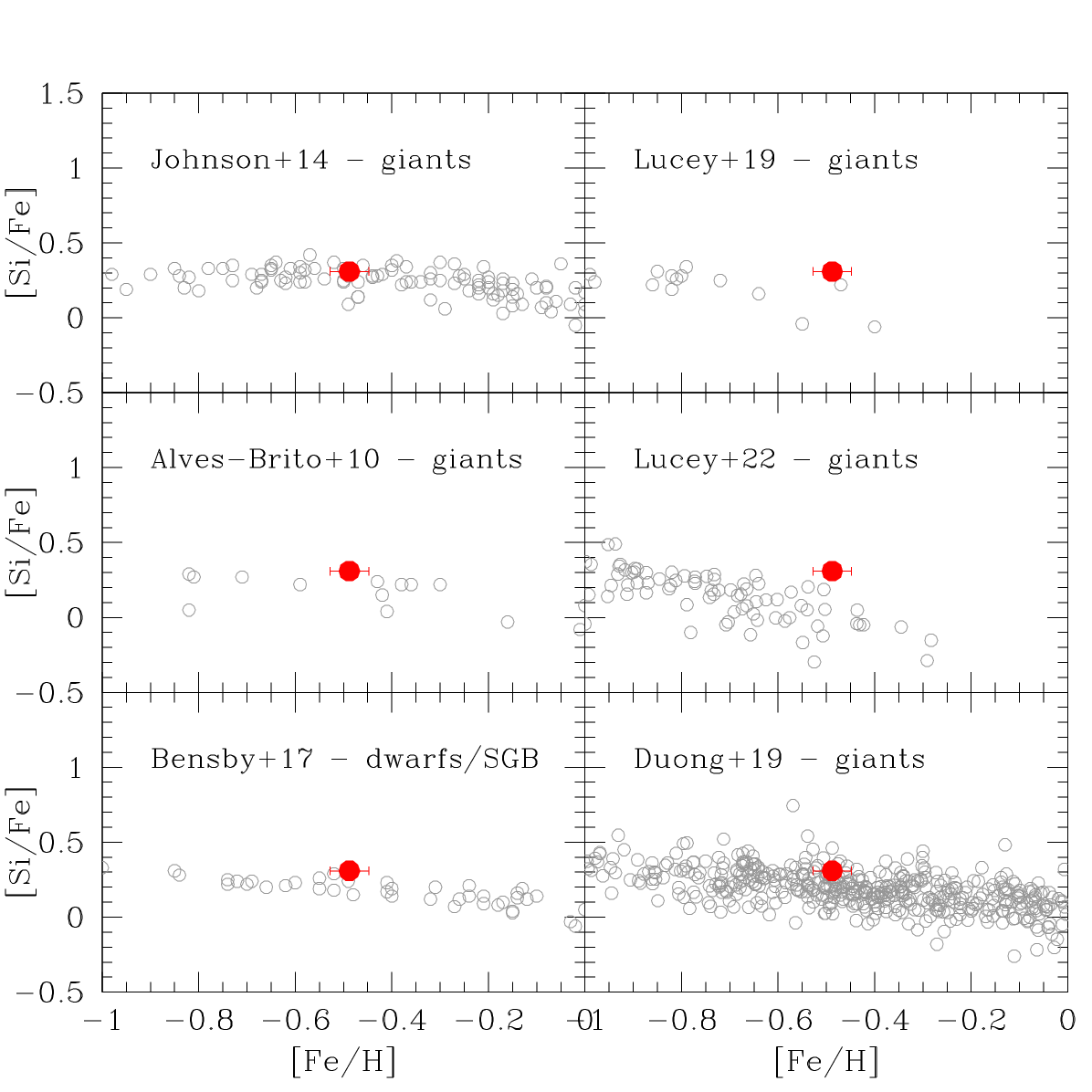}
\caption{Comparison of the average [Si/Fe] ratio in NGC~6388 (red point with
error bar) to six different studies of bulge field stars in the metallicity
range centred on the cluster: Johnson et al. (2014), Alves-Brito et al. (2010),
Bensby et al. (2017), Lucey et al. (2019, 2022), and Duong et al. (2019). The
evolutionary stage of each sample is reported in the panels.}
\label{f:solosi6}
\end{figure}

The average abundance of GCs nicely follows the pattern of $\alpha$-elements of
field MW stars, a plateau at low metallicity followed by a decrease when the
metallicity increases after the `knee', signalling the major onset of SN Ia.
As other metal-rich GCs ([Fe/H]$\gsim -0.7$ dex), NGC~6388 participates in
this trend well. 
In particular, the average [Si/Fe] ratios in NGC~6388 and in NGC~6441 are in 
very good agreement with the mean level shown by halo and disc GCs in the MW and
with  the abundance of bulge field stars, both giants (as in GCs) and dwarfs, as
shown in detail in Fig.~\ref{f:solosi6}, where we display some of the studies
used in  Fig.~\ref{f:alphaGC}, in a restricted range around the mean metallicity
of NGC~6388.

\begin{figure}
\centering
\includegraphics[scale=0.45]{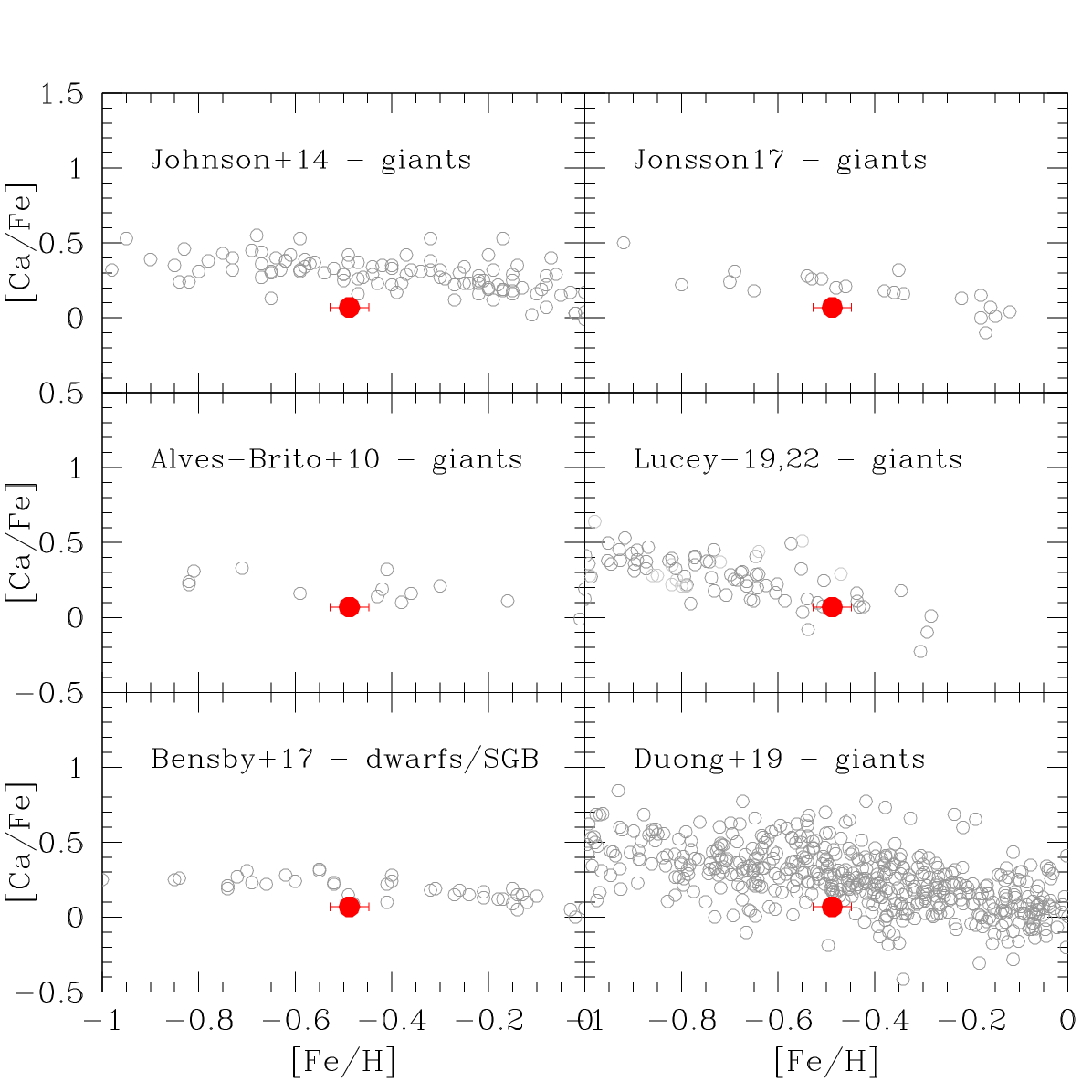}
\caption{Comparison of average [Ca/Fe] ratio in NGC~6388 (red point with
error bar) to seven different studies of bulge field stars in the metallicity
range centred on the cluster. To the studies listed in the previous figure, we
added the work by J\"onsson et al. (2017).}
\label{f:soloca6}
\end{figure}

Calcium abundances in NGC~6388 seem to lie at the lower envelope of the field
bulge star distribution (Fig.~\ref{f:alphaGC}). Overall, NGC~6388 and 
NGC~6441,  together with the other bulge GCs (with the exception of NGC~6440,
Mu\~{n}oz et al. 2017) seem to be more compatible with the disc stars.
However, the mean ratio [Ca/Fe] in NGC~6388 is still consistent with several
studies focusing on bulge stars (Fig.~\ref{f:soloca6}), again taking into
account small offsets related to the abundance analysis. 
Furthermore, we remind the reader that in NGC~6388, Ca  is not involved into the network of
proton-capture reactions (Carretta and Bragaglia 2021).
This result is quantified in Fig.~\ref{f:trend} and Table~\ref{t:tabprob};
among all relations concerning proton-capture elements, the only ones resulting
as statistically not significant are those involving Ca and Sc. 

Among the species examined in Fig.~\ref{f:alphaGC}, only Ti does not show the
typical decline at high metallicity. We remark that this element is not a
classical $\alpha$-element 
and is located at the boundary between $\alpha$
and Fe-peak species. Moreover, Ti abundances are only available for less than a
half of the GC sample, our FLAMES survey being focused on species (such as Mg, Si,
and sometimes Ca) involved in the multiple population phenomenon.
For the stars observed with UVES, both neutral and singly ionised lines were
available. We found, on average, a difference of [Ti/Fe]{\sc ii} - [Ti/Fe]{\sc i}
= $-$0.051 dex, with rms=0.096 dex, which is not significant.

From this section, our results and the similarity between NGC 6388 and both the
bulge and disc field of the MW suggest a high-mass environment for the progenitor of
this cluster. Coupled with the chemo-dynamical evidence presented in a previous
paper (CB22b), this is indicative of a formation within a massive component of
the Galaxy, likely the bulge itself. The high metallicity of NGC~6388 clearly
excludes this GC from being associated with a  metal-poor structure recently reported
from APOGEE data by Horta et al. (2021) and considered possibly accreted in the
early MW before the presently observed main Galactic bulge (including its
GC population and NGC~6388) was  formed.

\subsection{Fe-peak elements}

Together with iron, we derived the abundances of seven species of
the iron-peak group in NGC~6388: [Sc/Fe]{\sc ii}, [V/Fe], [Cr/Fe]{\sc i}, [Mn/Fe],
[Ni/Fe], [Co/Fe], and [Zn/Fe]. Most abundances are derived from UVES spectra of
the 12 stars newly observed in this work, integrated by the sample of Carretta
and Bragaglia (2018). However, abundances of Sc and Ni are also measured for the
larger sample of stars with GIRAFFE HR13 spectra, using on average 2 and
11 lines for Sc and Ni, respectively. 

\begin{figure}
\centering
\includegraphics[scale=0.45]{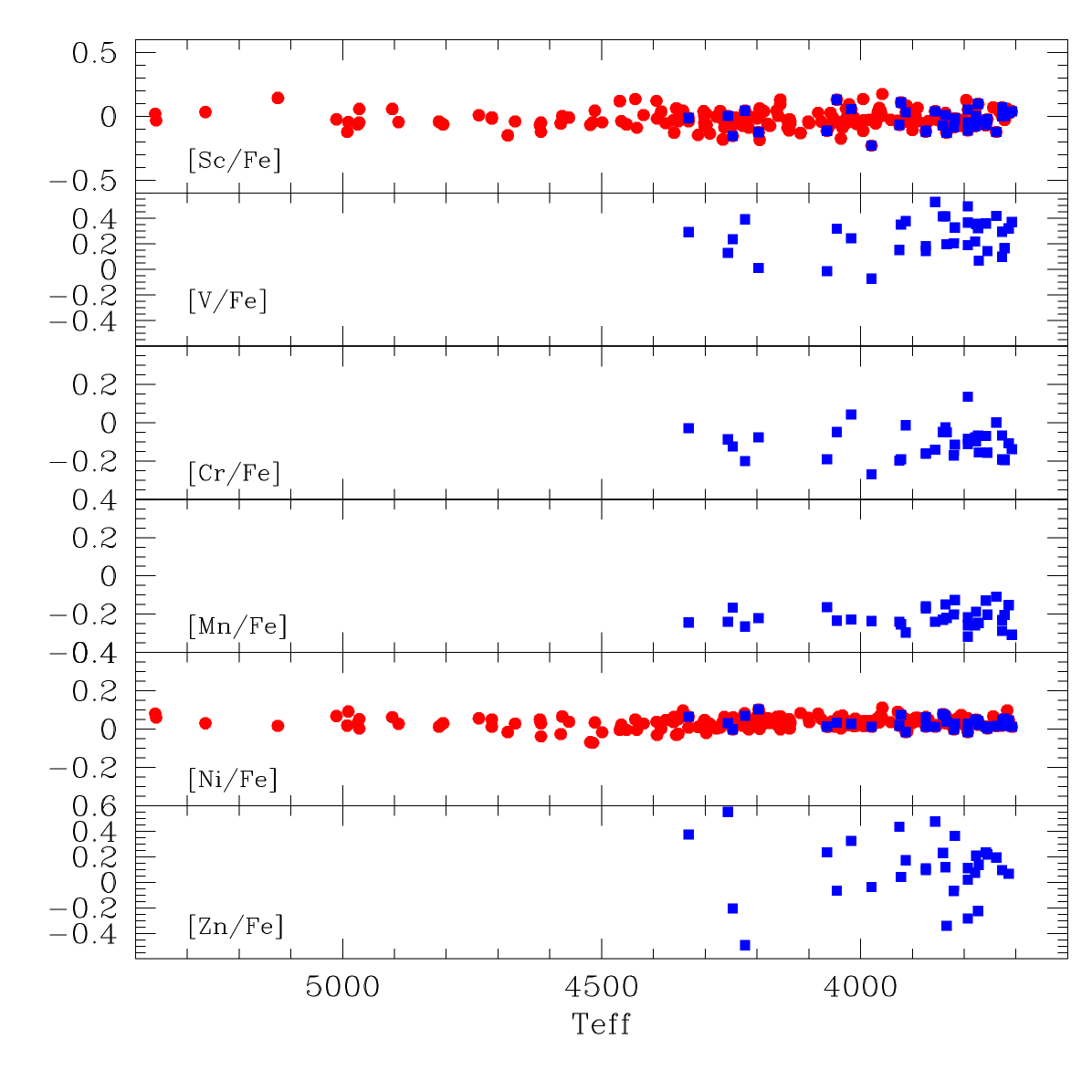}
\caption{Abundance ratios [Sc/Fe]{\sc ii}, [V/Fe], [Cr/Fe]{\sc i}, [Mn/Fe],
[Ni/Fe], and [Zn/Fe] (from top to bottom) as a function  of the effective
temperature. Symbols are as in Fig.~\ref{f:alphateff5}.}
\label{f:ironteff}
\end{figure}

The derived abundances present no dependence on the temperature of stars over a
range of about 1700 K (see Fig.~\ref{f:ironteff}), particularly evident in the
cases of Sc and Ni, and are roughly clustered within $\pm 0.10$ dex from the
solar value, apart from V, which shows a slight overabundance, and Mn, which
shows a clear deficiency, also reflected by the pattern of field stars at
similar metallicity (see below).

In Carretta and Bragaglia (2022b) the average abundances of Sc, V, and Zn from
the present work were used to check whether the level of these iron-peak elements
could be used for chemically tagging NGC~6388 (accreted or formed in situ).
Comparing the abundance of stars in NGC~6388 to a large ensemble of field stars
(both disc and bulge stars) at similar metallicity, we were able to exclude a
significant difference between cluster and field stars for all the three species
under scrutiny, thus rejecting the accretion origin.

\begin{figure}
\centering
\includegraphics[scale=0.45]{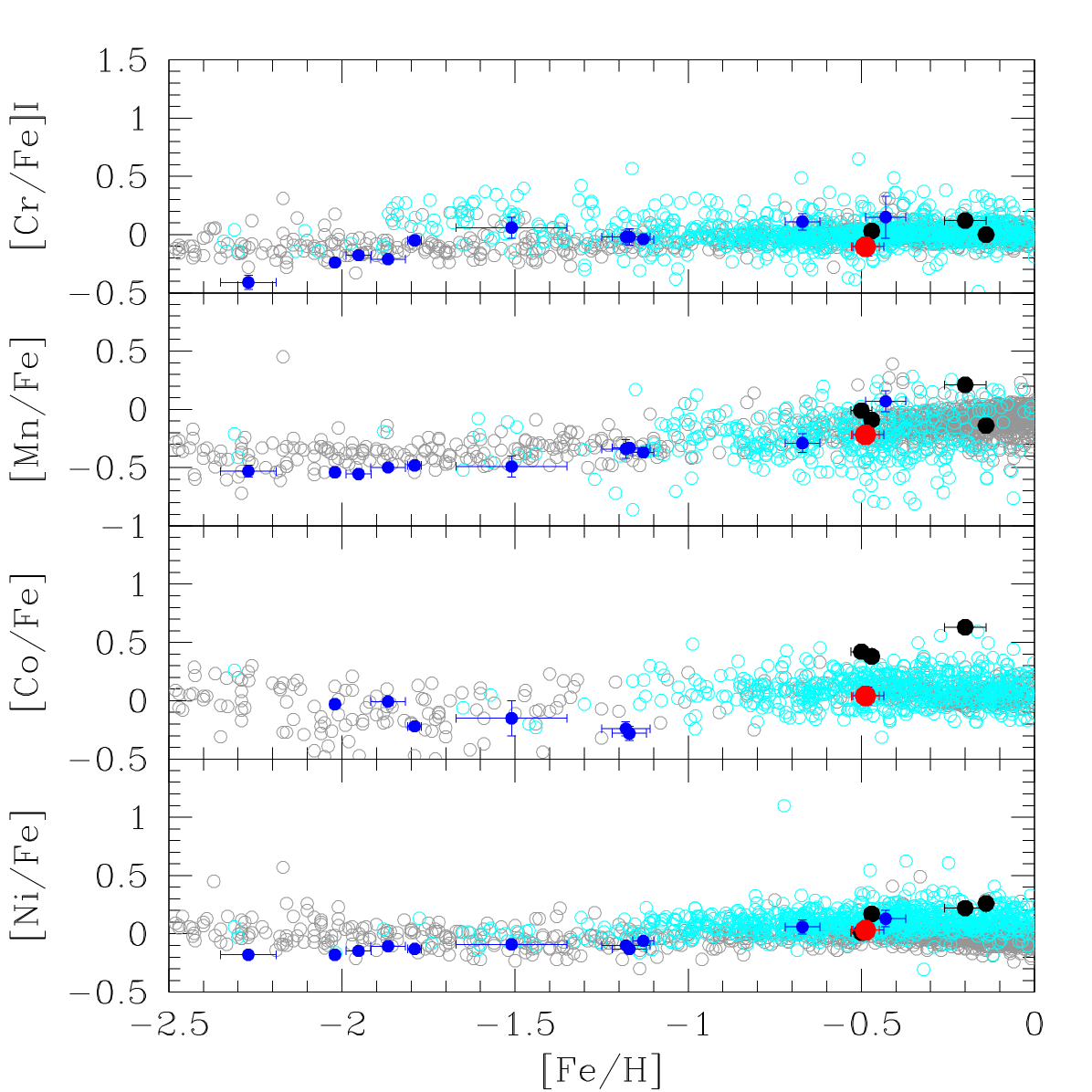}
\caption{Comparison of average abundance ratios [Cr/Fe], [Mn/Fe], [Co/Fe],
and [Ni/Fe] as a function of metallicity for NGC~6388, the GCs in our FLAMES survey, some
metal-rich GCs from Mu\~{n}oz and collaborators, and field stars in the disc and
bulge of the Milky Way from several studies. Symbols are as in
Fig.~\ref{f:alphaGC}, and references are given in Appendix B.}
\label{f:iron4oriplus}
\end{figure}

The in situ nature of NGC~6388 is supported and strengthened by the abundance of
the other elements of the iron group  derived here. In
Fig.~\ref{f:iron4oriplus}, we compare the mean abundances of Cr, Mn,  Co, and Ni
obtained for NGC~6388 to several samples of field stars in the Milky Way,
together with the average abundances for a number of GCs from our FLAMES survey
and from Mu\~{n}oz and collaborators.  No significant difference is found
between cluster and field MW stars. 

The same pattern is followed well by other classical, less massive bulge GCs,
with the possible exception of Co, whose level seems to exceed the locus defined
by field bulge and disc stars. This is particularly evident for NGC~6528
(Mu\~{n}oz et al. 2018).
However, when NGC~6441 (Gratton et al. 2006) is also considered among the bulge
populations, we may stress again that in general the iron-peak elements
perform poorly, in picking up objects presumably of accreted origin. 

\subsection{Neutron-capture elements: Zr and Ba}

We obtained the abundances of the neutron-capture elements Y, Zr, Ba, La, Ce,
and Nd, sampling the first and second peak of species produced mainly by the
$s$-process. We also measured Eu, typically produced by the $r$-process in
the presence of higher neutron densities. Most abundances were obtained from UVES
spectra, due to their larger spectral coverage. However, we were also able to derive
Ba and Zr abundances from GIRAFFE HR13 spectra by measuring the Ba~{\sc ii}
6141~\AA\  line in all of the 150 stars and up to a maximum of five Zr~{\sc i}
lines in 138 stars. We first concentrate on these two elements and defer the
analysis of all others to the next sub-section. 

\begin{figure}
\centering
\includegraphics[scale=0.45]{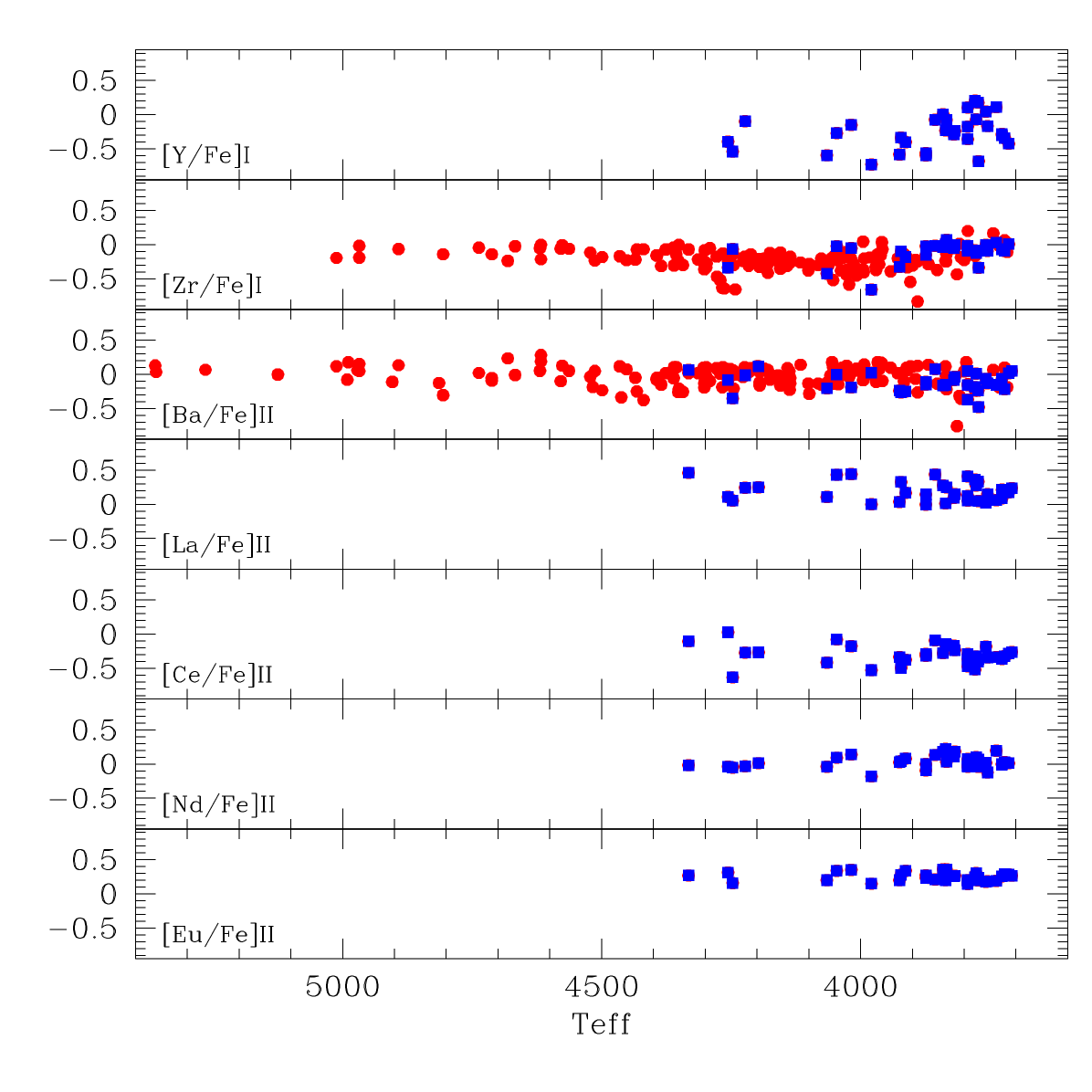}
\caption{Abundance ratios of neutron-capture elements in NGC~6388 as a 
function of the effective temperature. Red filled circles indicate stars with
GIRAFFE spectra, whereas stars with UVES spectra are represented by blue
filled squares.}
\label{f:ncapteff}
\end{figure}

Sources of atomic parameters for these Ba and Zr lines can be found in Table~8
of Gratton et al.  (2007), with the exception of the $\log gf$ value for the
Ba~{\sc ii}  line, taken instead   from Sneden et al. (2003). Due to the
strength of this line and the consequent dependence on microturbulent velocity,
abundances of Ba were derived using the relation as a function of surface
gravity for $v_t$ (Worley et al. 2013) and a constant metallicity value ($-0.48$
dex) for all stars. This expedient allows us to avoid spurious trend of
abundances as a function of the  microturbulent velocity (see e.g. Carretta et
al. 2015). The absence of trends with effective temperature for all derived
elements is shown in Fig.~\ref{f:ncapteff}.

Large samples of stars with abundances of $s$-process elements are necessary to
explore another defining property of the type II GCs.
Together with an enhancement in metallicity [Fe/H], the metal-rich
component is also proposed to be enriched in  neutron-capture elements. This
characteristic may manifest in a range of variations, going from the small
amounts observed in NGC~1851 (Carretta et al. 2011) up to the large excesses
detected in $\omega$ Cen (Johnson and Pilachowski 2010; Marino et al. 2011b). 

In Carretta and Bragaglia (2022a), we demonstrated that NGC~6388 is a typical
mono-metallic GC, whose intrinsic spread in [Fe/H] is fully compatible with
uncertainties derived from abundance analysis. Hence,  finding here that neither
Ba (measured in 185 stars) nor Zr (in 168 stars)  show  evidence of enhancement
in part of the stars of this GC does not come as a surprise.

\begin{figure} 
\centering 
\includegraphics[scale=0.50]{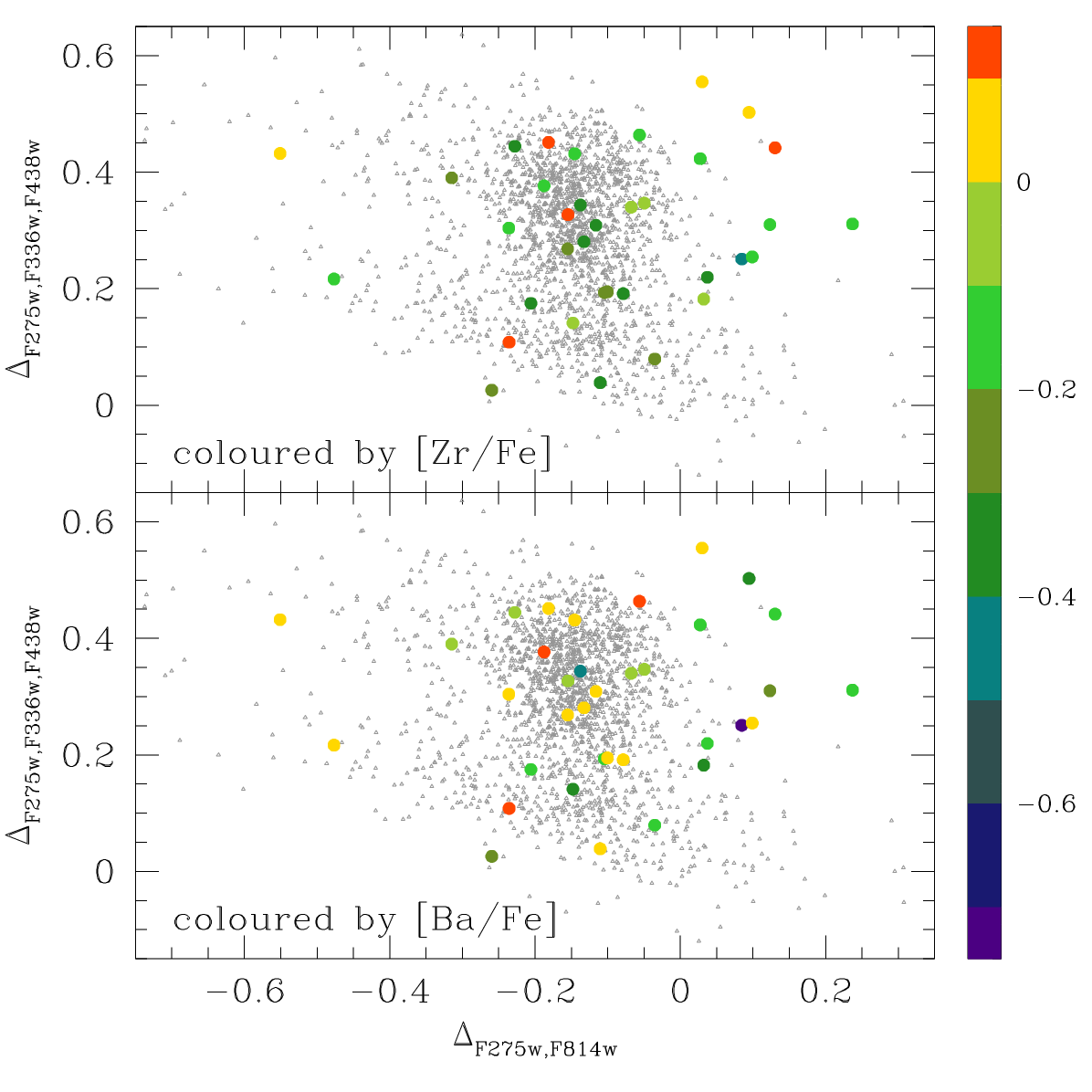}
\caption{Pseudo-colour maps
of NGC~6388 from Carretta and Bragaglia (2022a). Larger symbols indicate stars
of the present work cross-identified with the HST photometry in Nardiello et al.
(2018). The colour-coding in the upper and bottom panels is done according to the
abundances of Zr~{\sc i} and Ba, respectively, obtained in the present work.}
\label{f:cromobazr} 
\end{figure}

As a first test, we split our dataset at [Fe/H]$=-0.488$ dex (the mean
metallicity of the GIRAFFE sample). Then, we compared the  cumulative
distribution of Ba (and Zr) abundances for all stars more metal rich and more
metal poor than this value using a Kolmogorov-Smirnov test. For both elements we
obtain a K-S probability $p \sim 0.30$. This means that it is statistically not
possible to safely reject the null hypothesis that the two distributions are
extracted from the same parent population. 
A second test involves the pseudo-colour maps constructed using HST photometry
(Milone et al. 2017). Figure~\ref{f:cromobazr} shows the chromosome map derived
by us using the public data available from the HST archive (Nardiello et al.
2018; see Carretta and Bragaglia 2022a for details). In the figure, the stars in
our sample falling in the small central region covered by the HST are indicated by
larger symbols, coloured according to the derived Ba (upper panel) and Zr~{\sc
i} (lower panel) abundances derived in the present work. There is no significant
difference  between stars scattered to the red in the pseudo-colour map and the
other stars in this photometric plane. 
The direct implication of these tests is that NGC~6388 does not qualify as a
type II GC either for some enhancement in metallicity or in the amount of
$s$-process elements. In turn, the red RGB stars scattered in the pseudo-colour
maps obtained from HST photometry are simply not yet explained by any observable
change or alterations in their chemical composition.

\begin{figure} 
\centering 
\includegraphics[scale=0.40]{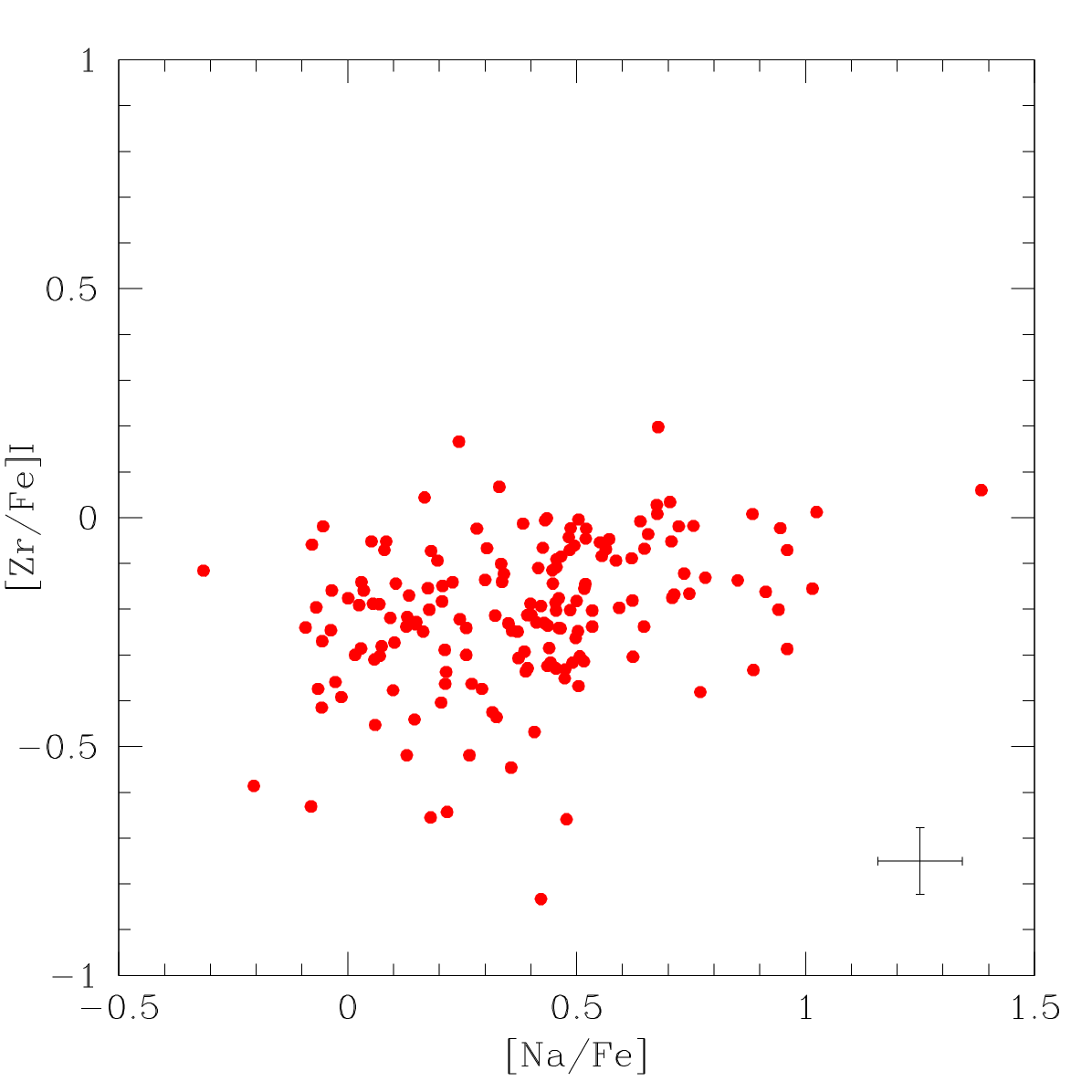} 
\caption{Abundance ratio [Zr/Fe]~{\sc i} as a function of [Na/Fe] in 168
stars of NGC~6388.} 
\label{f:zrna} 
\end{figure}

\begin{figure} 
\centering 
\includegraphics[scale=0.40]{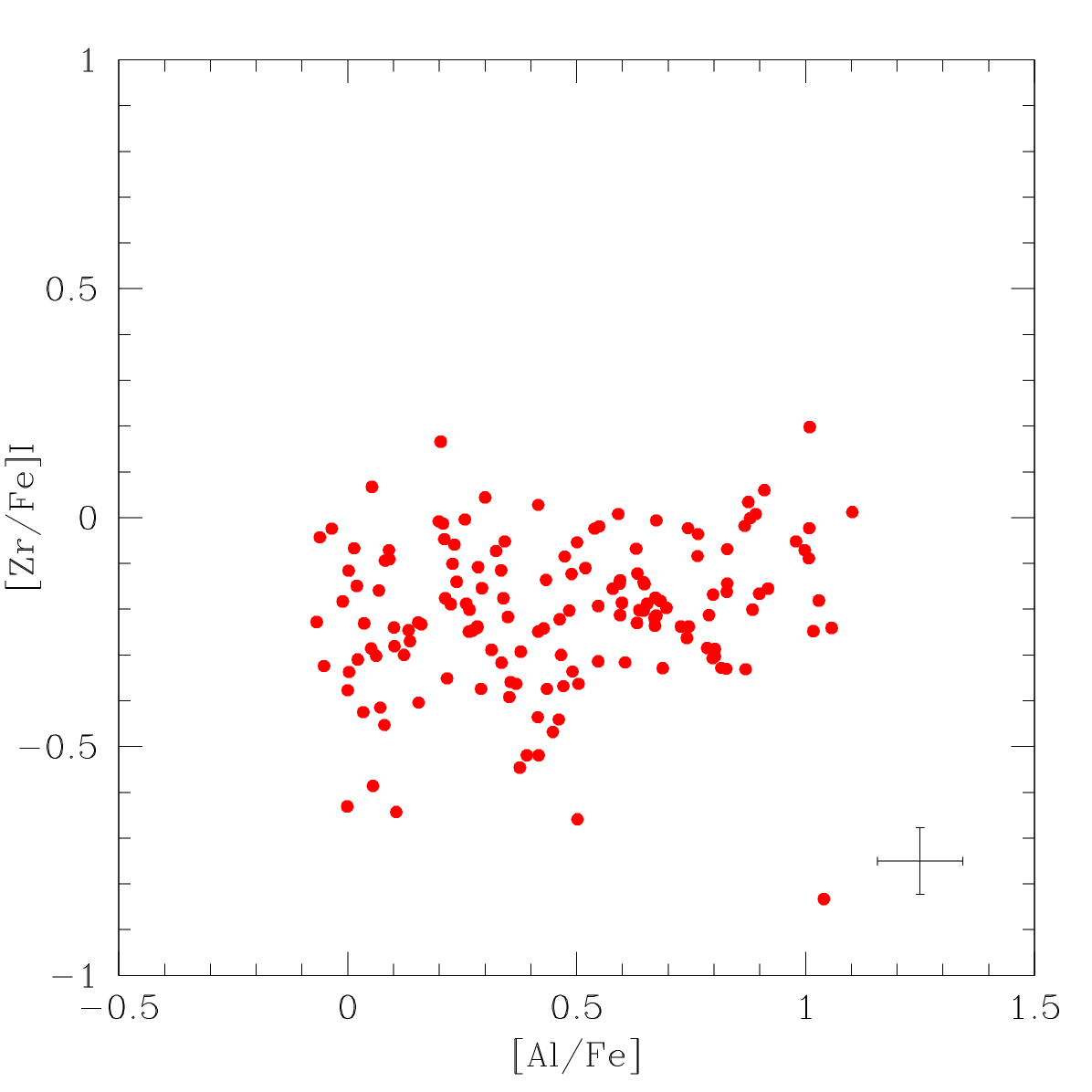} 
\caption{As in Fig.~\ref{f:zrna}, but for 151 stars with both Zr and Al
abundances.}
\label{f:zral} 
\end{figure}

On the other hand, we found  
statistically significant correlations between the Zr abundance and a few
light elements that are enhanced in the proton-capture reactions typical of the
FG polluters. In Fig.~\ref{f:zrna} and Fig.~\ref{f:zral}, we show the ratio 
[Zr/Fe]{\sc i} as a function of [Na/Fe] and [Al/Fe]. 
We tested the level of significance for a linear regression between Zr and both
Na and Al: the two-tail probabilities ($p=2.4\times 10^{-5}$ and $p=0.033$  for
Na and Al, respectively) allow us to conclude that the observed relations are real,
with a high level of significance.

This is not the first time a similar correlation has been found. To our
knowledge, the first paper to discuss correlations between elements enhanced in
SG stars (such as Na) and others based on a large enough number of stars was Yong
et al. (2013) on NGC~6752. They used differential analysis to derive very
precise abundances (at 0.01 dex level) and found positive correlations, generally
statistically significant, between many elements and Na. This could indicate
that the same stars that enhanced Na over primordial values were also
responsible for the increase in the other elements, among them some
neutron-capture ones (unfortunately, they did not measure Zr). Their
conclusion was that the abundance trends are real and  discussed three
potential mechanisms to explain them (besides the possibility of systematic
errors in stellar parameters, which were regarded very unlikely): a star-to-star
variation in  CNO abundance, or in He (which is strictly connected to Na in
multiple populations), or inhomogeneous early chemical evolution (i.e.
metallicity variations). They favoured a combination of the last two, but
encouraged similar studies on other clusters  to try and clarify this issue.

At odds with those findings, Schiappacasse-Ulloa \& Lucatello (2023) did not
find correlation between Na and neutron-capture elements in the same cluster.
They analysed about 160 stars in NGC~6752, from the main-sequence turn-off up to
the RGB bump, deriving abundances of elements from different nucleosynthetic
chains, among which we find Na, Y, and Ba. While they saw a mild Na-Y correlation, this
is not statistically significant. Also, the Ba abundance does not correlate with Na
abundance, and  they concluded that the stars that enhanced the Na level did not
contribute Y and Ba.

Finally,  Kolomiecas et al. (2022) derived the Na and Zr abundances of about 240
RGB stars in 47~Tuc, finding a statistically significant positive correlation
between them. Their conclusion was  that some amount of Zr should have been
produced by the same primordial stars which were enriched in Na the SG stars.
Unfortunately, this cannot be attributed unequivocally to a single class of 
polluters, either AGBs or  massive stars, or a combination of the two.   We follow
Yong et al. (2013) and Kolomiecas et al. (2022) in encouraging the  extention of  this
kind of analysis to more elements and more clusters.

\begin{figure}
\centering
\includegraphics[scale=0.40]{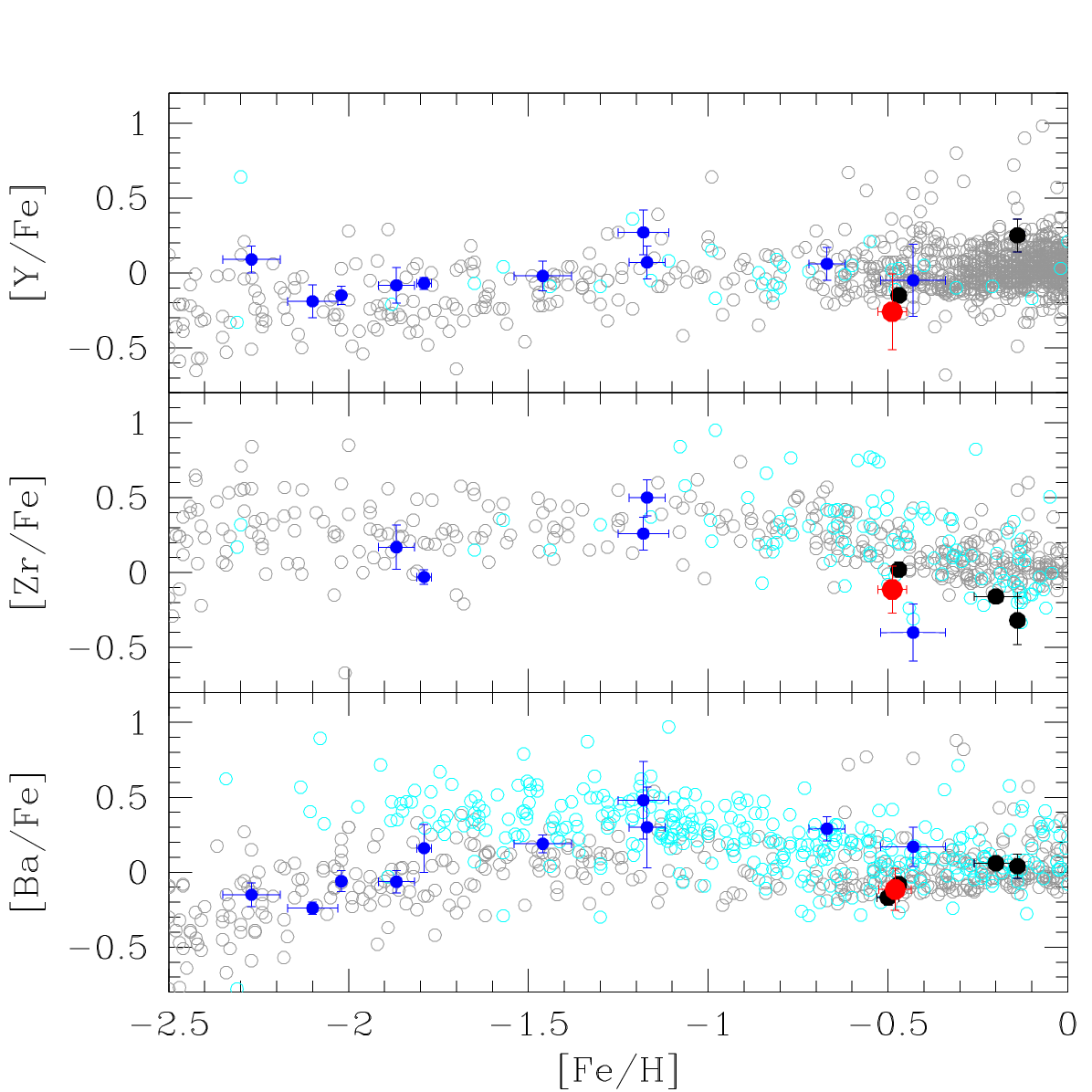}
\caption{As in Fig.~\ref{f:alphaGC}, but for average abundance ratios
[Y/Fe], [Zr/Fe], and [Ba/Fe]. References for the samples are in Appendix B.}
\label{f:ncapt3oriplus}
\end{figure}

\begin{figure}
\centering
\includegraphics[scale=0.40]{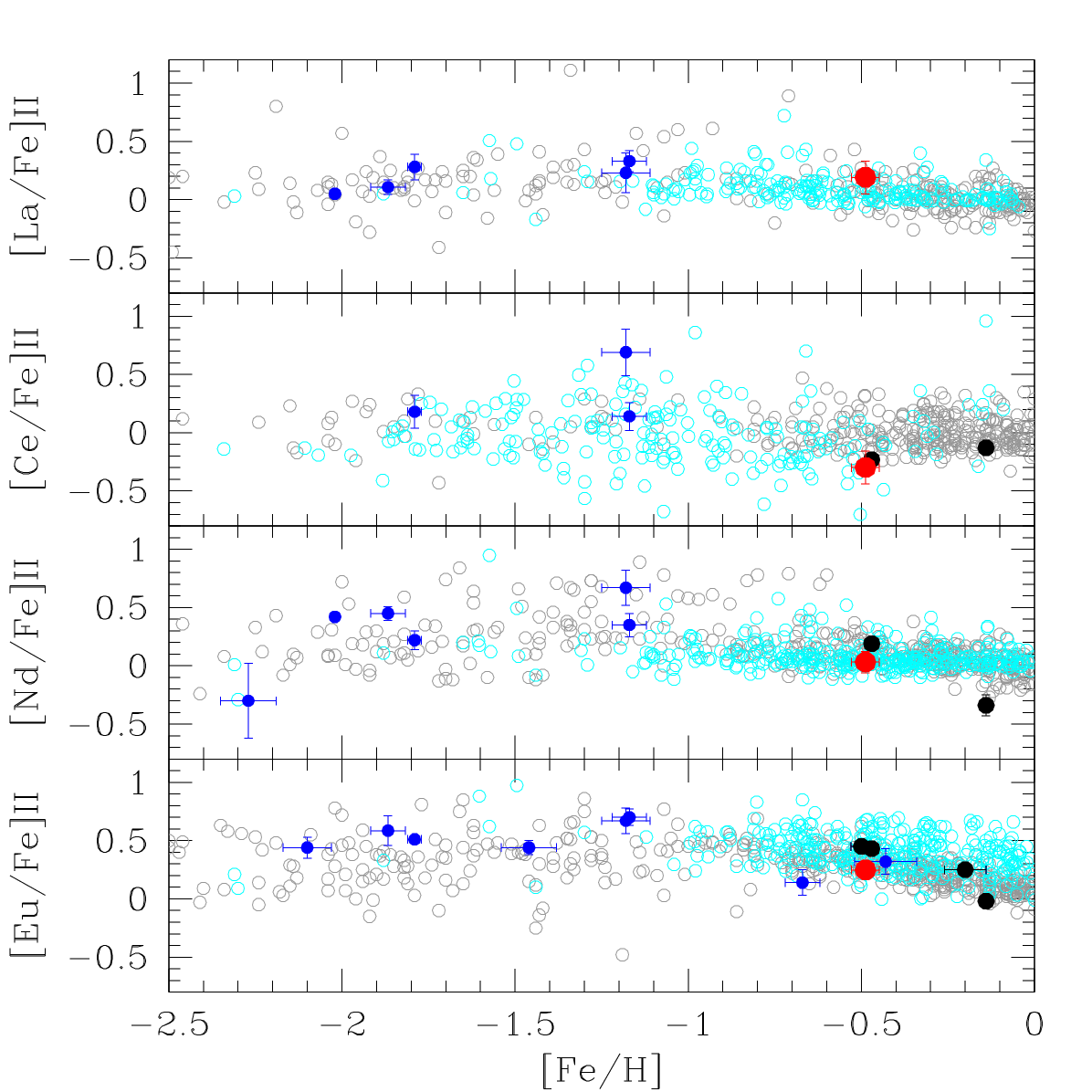}
\caption{As in Fig.~\ref{f:alphaGC}, but for average abundance ratios
[La/Fe], [Ce/Fe], [Nd/Fe], and [Eu/Fe]. References for the samples are in
Appendix B.}
\label{f:ncapt4oriplus}
\end{figure}

\subsection{Neutron-capture elements: Overal pattern in NGC~6388}

The average abundances in NGC~6388 are compared to those of field disc and bulge
stars, as well as to the mean abundances of previous analyses of GCs, in
Fig.~\ref{f:ncapt3oriplus} and Fig.~\ref{f:ncapt4oriplus} (symbols are as in
Fig.~\ref{f:alphaGC}). The  neutron-capture elements in NGC~6388 seem to be consistent
with those of bulge field stars of similar metallicity. Some offsets seem to
exist with respect to the disc component around [Fe/H]$\sim -0.5$ dex, but not
enough to be very significant, maybe with the exception of Zr. However, the
deficiency in [Zr/Fe] (middle panel of Fig. ~\ref{f:ncapt3oriplus}) is also
shared by NGC~6441 and two other more metal-rich GCs. Together with the mean
abundances of the four metal-poor GCs, the overall pattern could be explained
by the decreasing abundances of Zr (produced in the main $s$-process component
in AGB stars) as the metallicity increases due to the continuous injection of
fresh iron from SN Ia (e.g. Tinsley 1979).

Another possibility is the one highlighted by Kobayashi et al. (2020); the 
contribution from nucleosynthetic sources with an intrinsically long time delay
is more affected by the star formation timescale. Inefficient star formation (as
in the halo) may give a higher level of neutron-capture elements relatively to iron.
On the other hand, rapid star formation implies a smaller contribution of
ejecta from longer lived, lower mass AGB stars, which are producers of elements from the
main $s$-process component. However, this second alternative would apply to both
the light and heavy $s$-process elements (Zr, La, Ce) produced in the 
main $s$-process, whereas there is some evidence that NGC~6388 and 
other bulge GCs show the same level of neutron-capture elements as seen in halo
metal-poor GCs, at least for La, Ba, and Nd.

Concerning the $r$-process, in the bottom panel of Fig. ~\ref{f:ncapt4oriplus} 
we trace the run of [Eu/Fe] as a function of metallicity. The pattern of
constant values in the halo phase, followed by a gradual decrease as [Fe/H]
increases, is explained well by the scenario pioneered by Tinsley (1979),
pointing to the origin of Eu in the same sites (massive stars) where the 
$\alpha$-elements are produced. After the knee in [Fe/H]  is reached in the
main progenitor galaxy, the Eu production would remain essentially flat, but the
[Eu/Fe] ratio is lowered by the increasing contribution of SN Ia. When our
sample of clusters, mainly constituted by metal-poor GCs, is complemented by
metal-rich GCs, it is easy to appreciate that the above scenario is satisfied
by both field and GC stars. The interplay between enrichment from core-collapse
and thermonuclear SNe was essentially the same, regardless of the star formation
occurring in GCs or in the general field.

\begin{figure}
\centering
\includegraphics[scale=0.40]{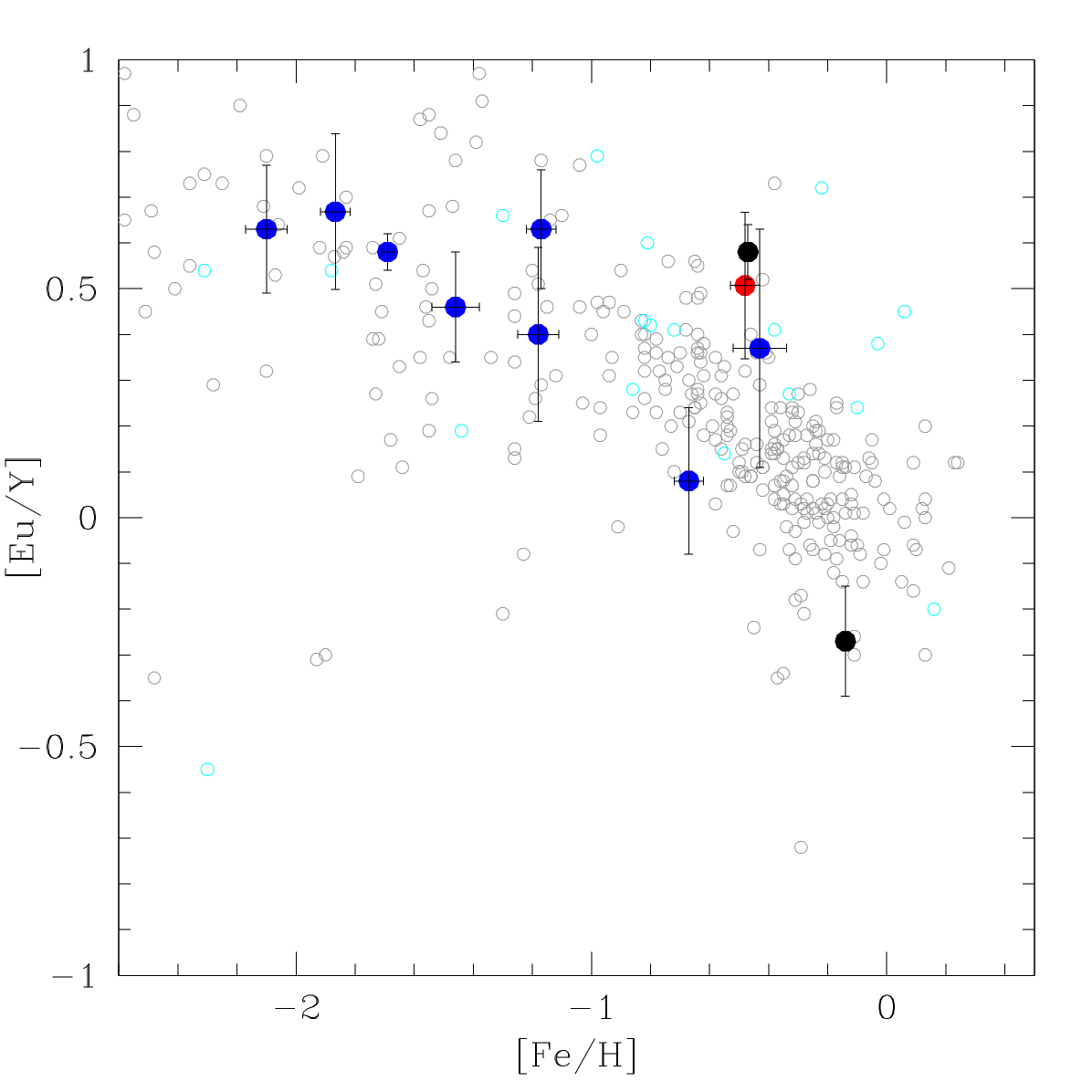}
\caption{As in Fig.~\ref{f:alphaGC}, but for the average abundance ratio [Eu/Y]
of the Eu and Y species, which are reference elements for the $r$ and $s$-processes,
respectively. Field bulge stars (cyan) are from Lucey et al. (2019) and Gratton
et al. (2006). Halo and disc stars (grey circles) are from James et al. (2004)
and the compilation by Venn et al. (2004).}
\label{f:euy}
\end{figure}

Finally, in Fig.~\ref{f:euy} we use the relative strengths of the $r$-process
and $s$-process to probe the relative contribution from high-mass stars (mainly
responsible for yields of the weak component of the $s$-process and the
$r$-process) and from low- or intermediate-mass AGB stars (main $s$-process).
In this figure, the pure $r$-process element Eu is compared to Y, chosen as
reference element for the $s$-process. Again, by plotting the ratio [Eu/Y] as a
function of the metallicity (liberally used as a chemical `chronometer') we are
exploiting the different mass ranges, and therefore different evolutionary 
timescales, of the involved stars to give a general picture of the enrichment
process in field and GC stars.

At low metallicities, both GC and field stars show high [Eu/Y] ratios,
approaching the scaled Solar System pure $r$-process level, with scarce or
no contribution  from the $s$-process in AGB stars. As lower mass stars with longer
evolutionary timescales appear on the enrichment scene, an increase in the
production of  $s$-process elements lowers the [Eu/Y] ratio. The decrease seems
to happen in lockstep both in field and GC stars until about [Fe/H]$\sim -0.5$.
At this metallicity, our present analysis does confirm the earlier results
presented in Carretta et al. (2007a) for NGC~6388 and in Gratton et al. (2006) for NGC~6441.  In
particular, NGC~6388 seems to have a [Eu/Y] higher than observed in field stars
of similar metallicities.  Although the larger errors associated with NGC~6441 make
its ratio still compatible with the field stars, the ratio in NGC~6388 is in
better agreement with the ratio in metal-poor GCs, where the contribution of AGB
stars to the $s$-process was not yet relevant.

In Carretta et al. (2007a), we put forward the hypothesis that this excess in [Eu/Y]
could be explained by an enhanced contribution of massive stars to the
enrichment in the bulge, since light $s$-process elements such as Y can be also
produced in the weak $s$-process component within the He-burning core of massive
stars (Couch et al. 1974, Travaglio et al. 2004). This idea would also explain
the high [Eu/Y] ratios measured in metal-rich bulge field stars (Gratton et al.
2006, cyan circles in Fig.~\ref{f:euy}). However, we note that the disc GC
NGC~5927 (Mura-Guzm\`{a}n et al. 2018) also shares this excess ratio; therefore, a larger sample of GCs at high metallicity is required to answer the question.

\begin{figure}
\centering
\includegraphics[bb=39 167 521 465, clip, scale=0.52]{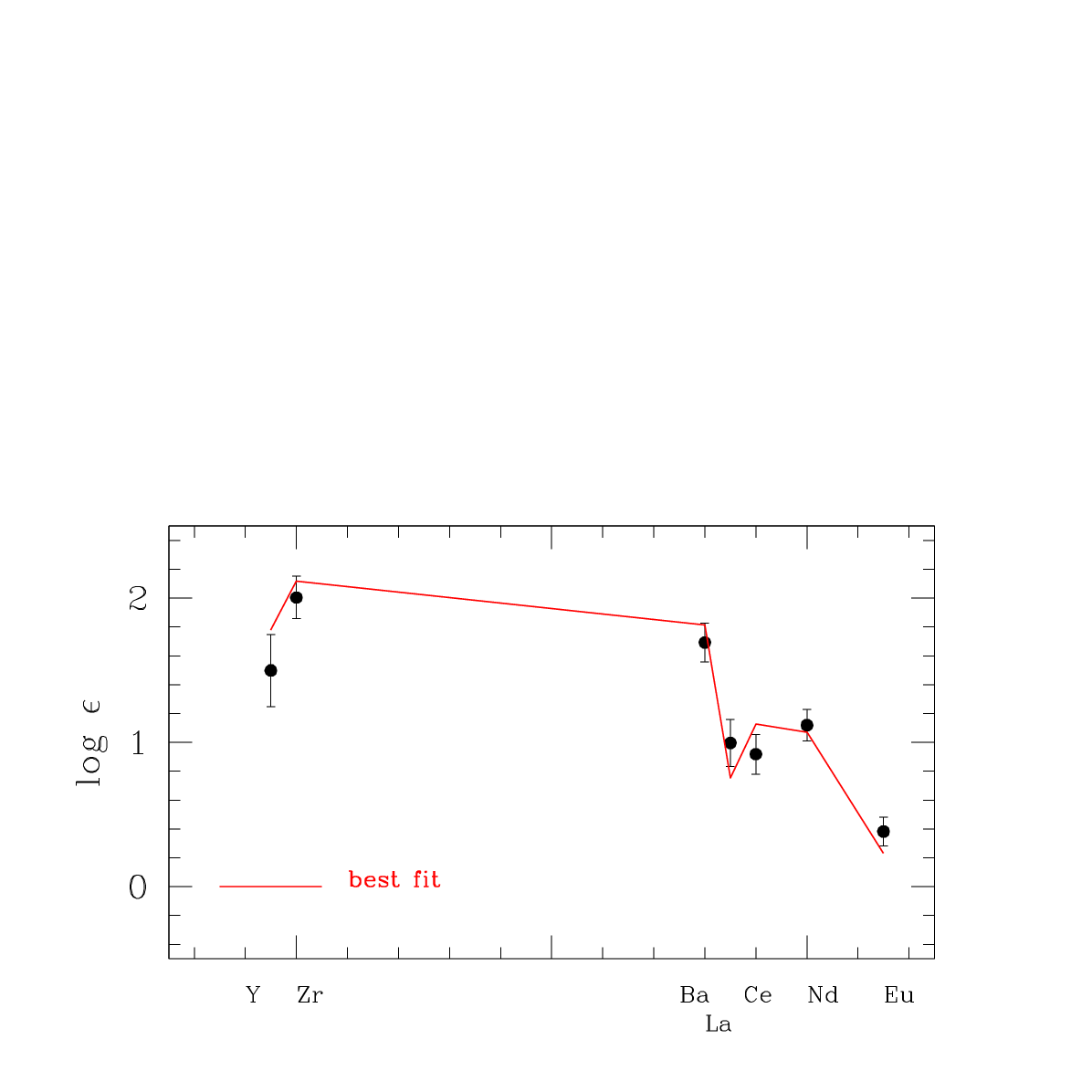}
\caption{Average abundances of Y, Zr, Ba, La, Ce, Nd, and Eu in NGC~6388 from
UVES spectra. The respective rms scatters are also represented as associated 
error bars. Our best-fit estimate using solar-scaled pure $r$-process and
$s$-process abundances is shown as a red line (see text for the adopted
procedure).}
\label{f:ncaptrs}
\end{figure}

A more quantitative approach is summarised in Fig.~\ref{f:ncaptrs}, where
we compare the average abundances of neutron-capture elements derived from UVES
spectra in NGC~6388 to the ratios of $r-$ and $s-$ elements estimated by
Simmerer et al. (2004) in the Solar System. For this comparison we followed the
approach suggested by Raffaele Gratton and employed in Carretta et al. (2015)
for NGC~6093. To reproduce the pattern of neutron-capture elements in NGC~6388,
our best fit must consider the sum of two contributions: a solar-scaled
$r$-process contribution, with a scaling by -0.27 dex, and a solar-scaled
$s$-process contribution, with a scaling by -0.51 dex.
Taking into account the derived metallicity for NGC~6388 ([Fe/H]$=-0.48$ dex),
these scaling factors imply abundance ratios of [$r$/Fe]$=+0.21$ dex and 
[$s$/Fe]$=-0.03$ dex. 

The excess of elements produced by the $r$-process is very similar to the one we
obtain from the $\alpha$-elements. From GIRAFFE and UVES spectra, we derive mean
values of [$\alpha$/Fe]$=+0.22$ dex and +0.23 dex (regardless of inclusion  or
exclusion of Mg in the average). The excess in $r$-process elements with respect
to the solar value may be interpreted as an iron deficiency due to the fact that
we see almost exclusively the contribution of massive stars, whereas the one from SN Ia
is missing. On the other side, our data show that it is necessary to also
consider a significant part of the $s$-process, which scale almost exactly as Fe, to explain the observations in NGC~6388 well.

\section{Summary and conclusions}

We present the homogeneous spectroscopic analysis of a large sample of stars in
NGC~6388. Concerning the proton-capture elements, we find that all
stars observed in NGC~6388 nicely trace the typical correlations and
anti-correlations that are the unique trademark of GCs. The exceptions are the
heaviest species (Ca, Sc) involved in the network of proton-capture reactions in
H-burning at high temperature. No statistically significant variation is found for these two
species between FG and SG stars, confirming qualitative results shown in Carretta and
Bragaglia (2019) for NGC~6388.

Star-to-star variations in Si, correlated to abundance changes in Al and
anti-correlated to Mg depletions, support leakage from the Mg-Al cycle on Si.
In turn, this requires temperatures as high as about 65 MK in the FG polluters. 

A simple dilution model is compatible with a single class of polluters 
injecting processed matter in the intra-cluster medium at early time in
the proro-GC. Mixing this polluted material (typical of the composition of the
SG group E) with different amounts of pristine gas (whose composition is
represented by the P group of stars), we can obtain a good agreement for the
composition of the intermediate SG group for all the involved species.

The extent of the Na-O anti-correlation, the privileged, unambiguous signature of multiple
stellar populations in GCs, seems to be too short in NGC~6388 with respect
to its large total mass. Together with a few other massive GCs (47~Tuc, M~15,
NGC~6441), NGC~6388 lies slightly below the main trend in the IQR[O/Na] versus $M_V$
relation describing the dependence of the extent of Na-O anti-correlation as a
function of cluster total mass (Carretta 2006; Carretta et al. 2010a).

The inventory of $\alpha$-element abundances in NGC~6388 is very compatible with
the chemical pattern of field stars in the Milky Way, both in the disc and
bulge. The abundance ratios in NGC~6388 participate in the classical trend
defined by the interplay between star formation and lifetimes of the main
stellar nucleosynthesis sites, namely type II and type Ia SNe. The resulting 
plateau at low metallicity, followed by a knee and the decrease of $\alpha$/Fe]
ratio at increasing [Fe/H] is followed by both field and GC stars, including
NGC~6388.

We found no evidence of a low level of Si, as derived by infrared APOGEE data.
Therefore, we cannot support an extragalactic origin for NGC~6388, as suggested
by Horta et al. (2020). We note that all the studies using optical spectroscopy
converge on finding normal, high values of [Si/Fe] for NGC~6388. Its seems that
there could be some offsets between optical and infrared analyses due to still 
poorly understood effects concerning Si. A low value of Ca is instead derived
for NGC~6388, as well as for bulge stars in a similar metallicity regime,
regardless of whether optical or infrared spectra are used. 

The average abundances of elements of the iron group in NGC~6388 closely follow
the pattern of chemical enrichment typical of field stars in the Milky Way. We
then confirm and strengthen the results by Carretta and Bragaglia (2022b):
NGC~6388 is clearly not of extragalactic origin, but likely formed in situ in
the Galactic bulge, and the iron-peak species can only be used to trace the GCs
of the Sagittarius dwarf, whose content in such elements is typically lower than
in the autochthonous stars of our Galaxy. Consistently, both these elements and
the high Si level, normal for old GCs, point toward the in situ origin of
NGC~6388.

We do not detect any signature of enhancement in neutron-capture elements in a
fraction of stars of NCG~6388. In particular, there is no significant difference
in the abundance of Ba and Zr between FGs, SG stars, and the stars scattered to
the red of the RGB. This evidence corroborates the fact that NGC~6388 is not a GC
of a distinct (type II) class and also leaves the red RGB stars unexplained, at
least from a chemical perspective.

Statistically significant correlations are found between Zr abundance and both
Na and Al abundances. Similar results are also found in a few other cases and
warrant being extended to a larger sample of GCs.

The excess of the $r$-process element Eu in NGC~6388 is consistent with the values
of more metal-poor old GCs and similar to the ratio of $\alpha$-elements,
showing the contribution of massive stars coupled to the small injection of Fe
from thermonuclear SNe at the epoch of cluster formation. The overall pattern of
neutron-capture elements from high-resolution UVES spectra shows, however, that
it is necessary to also consider a contribution from the $s$-process in this GC.

\begin{acknowledgements}
This research made use of the products of the Cosmic-Lab project funded by
the European Research Council. We thank E. Dalessandro for helpful discussions.
This research has made use of the SIMBAD database (in particular Vizier),
operated at CDS, Strasbourg, France, of the NASA's Astrophysical Data System,
and of TOPCAT (http://www.starlink.ac.uk/topcat/). 
This paper makes use of the data collected by the HST Treasury Program GO 13297.
We acknowledge funding from PRIN INAF 2019 “Building up the halo: chemo-dynamical tagging in the age of large surveys”, PI Lucatello.
\end{acknowledgements}

\newpage
\begin{appendix}
\section{Abundances of individual stars}

Abundances for individual stars in NGC~6388 are listed in
Table~\ref{t:protonUNIFIN} for proton-capture elements,
Table~\ref{t:alphaUNIFIN} for $\alpha$-elements, Table~\ref{t:fepeakUNIFIN} for
elements of the Fe-group, and Table~\ref{t:neutronUNIFIN} for species from
neutron-capture processes measured on UVES spectra. Finally, in
Table~\ref{t:zr1baUNIFIN} we list the abundances of the neutron-capture elements
Zr~{\sc i} and Ba~{\sc ii} that could be derived also from GIRAFFE spectra. For
this table, as well as for  tables relative to light elements, $\alpha$-elements
and Fe-peak elements (Tables A.1, A.2, A.3, and A.5), only an excerpt is
provided here as a guidance of the content. Complete tables can be found at CDS
Strasbourg. 

\begin{table*}[h]
\centering
\caption[]{Light element abundances.}
\begin{tabular}{rcccccrcccrcrcc}
\hline
Star   &   nr &  [O/Fe]{\sc i} & rms&lim&nr &  [Na/Fe]{\sc i} &  rms &  nr &  [Mg/Fe]{\sc i} &  rms  &  nr &  [Al/Fe]{\sc i} &  rms  & phase\\
\hline
n63a    &   2  &$-$0.099     & 0.110&1  &  4  &   +0.215        & 0.066&  4  &       +0.198    &  0.133&  2  &    +0.003       & 0.031 & RGB \\ 
n63b    &   2  &$-$0.149     & 0.137&1  &  3  &   +0.111        & 0.041&  3  &       +0.265    &  0.067&  2  &    +0.265       & 0.027 & RGB \\ 
n63c    &   2  &$-$0.372     & 0.157&1  &  4  &   +0.884        & 0.090&  4  &       +0.143    &  0.097&  2  &    +0.891       & 0.151 & RGB \\ 
n63d    &   2  &$-$0.242     & 0.075&1  &  4  &   +0.497        & 0.060&  4  &       +0.234    &  0.099&  2  &    +0.091       & 0.083 & RGB \\ 
n63e    &   2  &$-$0.159     & 0.053&1  &  4  &   +0.551        & 0.050&  4  &       +0.230    &  0.173&  2  &    +0.501       & 0.071 & RGB \\ 
n63f    &   2  &$-$0.112     & 0.002&1  &  4  &   +0.639        & 0.085&  4  &       +0.223    &  0.116&  2  &    +0.199       & 0.126 & RGB \\ 
n63g    &   1  &$-$0.633     &      &1  &  3  &   +1.267        & 0.120&  4  &       +0.191    &  0.063&  2  &    +1.218       & 0.088 & RGB \\ 
n63h    &   2  &$-$0.061     & 0.013&1  &  4  &   +0.196        & 0.044&  4  &       +0.202    &  0.104&  2  &    +0.081       & 0.071 & RGB \\ 
n63i    &   2  &$-$0.047     & 0.008&1  &  4  &   +0.586        & 0.179&  4  &       +0.251    &  0.078&  2  &    +0.083       & 0.018 & RGB \\ 
n63l    &   2  &  +0.001     & 0.023&1  &  4  &   +0.207        & 0.126&  4  &       +0.192    &  0.063&  2  &    +0.020       & 0.108 & RGB \\ 
n63m    &   2  &$-$0.200     & 0.083&1  &  4  &   +0.735        & 0.112&  4  &       +0.222    &  0.146&  2  &    +0.633       & 0.159 & RGB \\ 
n63n    &   2  &$-$0.445     & 0.043&1  &  4  &   +0.648        & 0.085&  4  &       +0.115    &  0.148&  2  &    +0.630       & 0.163 & RGB \\ 
l63p001 &   2  &  +0.237     & 0.001&1  &  2  &   +0.383        & 0.082&  2  &       +0.259    &  0.031&  2  &    +0.208       & 0.018 & RGB \\ 
l63p002 &   2  &  +0.301     & 0.105&1  &  2  &   +0.229        & 0.001&  2  &       +0.380    &  0.030&  2  &    +0.646       & 0.218 & RGB \\ 
l63p003 &   1  &  +0.021     &      &0  &  2  &   +0.348        & 0.075&  2  &       +0.323    &  0.035&  1  &    +0.025       &       & RGB \\ 
l63p004 &   2  &  +0.245     & 0.113&1  &  2  & $-$0.315        & 0.142&  2  &       +0.215    &  0.083&  2  &    +0.002       & 0.080 & RGB \\ 
l63p006 &   1  &  +0.120     &      &1  &  2  &   +0.426        & 0.063&  2  &       +0.246    &  0.104&  0  &                 &       & RGB \\ 

\hline
\end{tabular}
\label{t:protonUNIFIN}
\end{table*}

\begin{table*}[h]
\centering
\caption[]{$\alpha$-element abundances.}
\begin{tabular}{rccccrcccrccc}
\hline
Star   &   nr &  [Si/Fe]{\sc i} &  rms  & nr &  [Ca/Fe]{\sc i} &  rms  &  nr &  [Ti/Fe]{\sc i} &  rms   &  nr&  [Ti/Fe]{\sc ii} &  rms  \\
\hline
n63a    &   9  &    +0.358       & 0.166 & 16 &       +0.034    & 0.140 &  25 &   +0.171  &  0.211 & 11 &      +0.264      & 0.254 \\ 
n63b    &   0  &                 &       & 11 &       +0.040    & 0.225 &  25 &   +0.325  &  0.189 & 8  &      +0.357      & 0.201 \\ 
n63c    &   8  &    +0.364       & 0.167 & 18 &       +0.098    & 0.222 &  29 &   +0.288  &  0.271 & 11 &      +0.274      & 0.143 \\ 
n63d    &   6  &    +0.319       & 0.148 & 16 &       +0.087    & 0.206 &  28 &   +0.322  &  0.170 & 12 &      +0.246      & 0.163 \\ 
n63e    &   8  &    +0.327       & 0.180 & 19 &       +0.072    & 0.177 &  26 &   +0.272  &  0.151 & 12 &      +0.307      & 0.194 \\ 
n63f    &   8  &    +0.310       & 0.172 & 19 &       +0.152    & 0.198 &  27 &   +0.320  &  0.182 & 12 &      +0.303      & 0.174 \\ 
n63g    &   8  &    +0.288       & 0.278 & 9  &       +0.081    & 0.217 &  21 &   +0.378  &  0.220 & 10 &      +0.324      & 0.230 \\ 
n63h    &   9  &    +0.368       & 0.194 & 15 &       +0.026    & 0.114 &  27 &   +0.277  &  0.164 & 10 &      +0.205      & 0.236 \\ 
n63i    &   8  &    +0.353       & 0.151 & 19 &       +0.052    & 0.164 &  27 &   +0.309  &  0.153 & 12 &      +0.263      & 0.149 \\ 
n63l    &   8  &    +0.339       & 0.158 & 17 &       +0.051    & 0.149 &  28 &   +0.263  &  0.179 & 11 &      +0.292      & 0.176 \\ 
n63m    &   7  &    +0.359       & 0.198 & 17 &       +0.121    & 0.174 &  27 &   +0.314  &  0.261 & 10 &      +0.265      & 0.176 \\ 
n63n    &   9  &    +0.382       & 0.259 & 13 &       +0.172    & 0.202 &  27 &   +0.267  &  0.225 & 11 &      +0.288      & 0.289 \\ 
l63p001 &   3  &    +0.262       & 0.060 & 6  &       +0.139    & 0.088 &  3  &   +0.298  &  0.074 & 0  &                  &       \\ 
l63p002 &   3  &    +0.243       & 0.031 & 6  &       +0.028    & 0.112 &  3  &   +0.294  &  0.084 & 0  &                  &       \\ 
l63p003 &   3  &    +0.166       & 0.006 & 6  &       +0.159    & 0.116 &  2  &   +0.288  &  0.098 & 0  &                  &       \\ 
l63p004 &   3  &    +0.219       & 0.084 & 5  &       +0.101    & 0.054 &  3  &   +0.139  &  0.130 & 0  &                  &       \\ 
l63p006 &   3  &    +0.270       & 0.062 & 6  &       +0.061    & 0.160 &  4  &   +0.256  &  0.172 & 0  &                  &       \\ 

\hline
\end{tabular}
\label{t:alphaUNIFIN}
\end{table*}

\begin{table*}[h]
\centering
\setlength{\tabcolsep}{1mm}
\caption[]{Iron-peak abundances.}
\tiny
\begin{tabular}{rccccrcccrcccrcccrcccc}
\hline
Star   
&   nr &  [Sc/Fe]{\sc ii} &  rms &  nr &  [V/Fe]{\sc i} &rms &  nr   &[Cr/Fe]{\sc i} &  rms & nr  & [Mn/Fe]{\sc  i} &rms &  nr  &[Co/Fe]{\sc i}&rms &  nr   & [Ni/Fe]{\sc i} &rms&  nr &[Zn/Fe]{\sc i} &  rms \\
\hline
n63a    & 8  &  $-$0.035  &  0.134 & 13 &  +0.069   & 0.261  &   17  & $-$0.155 &   0.230   &  8  & $-$0.247  &   0.244  &    5 &$-$0.027 &   0.283 &   41  &   +0.023 &   0.209 &  1  &  +0.137  &             \\ 
n63b    & 3  &    +0.072  &  0.239 & 8  &  +0.100   & 0.266  &   15  & $-$0.066 &   0.211   &  8  & $-$0.231  &   0.295  &    5 &  +0.083 &   0.334 &   26  &   +0.018 &   0.358 &  0  &     &          \\ 
n63c    & 8  &    +0.026  &  0.155 & 13 &  +0.321   & 0.236  &   21  & $-$0.107 &   0.284   &  7  & $-$0.153  &   0.095  &    5 &  +0.047 &   0.175 &   40  &   +0.045 &   0.280 &  1  &  +0.066  &             \\ 
n63d    & 7  &    +0.004  &  0.220 & 13 &  +0.167   & 0.246  &   17  & $-$0.194 &   0.194   &  8  & $-$0.204  &   0.182  &    5 &  +0.111 &   0.178 &   31  &   +0.054 &   0.171 &  0  &     &          \\ 
n63e    & 8  &  $-$0.086  &  0.125 & 11 &  +0.205   & 0.205  &   25  & $-$0.170 &   0.253   &  6  & $-$0.202  &   0.204  &    5 &  +0.077 &   0.143 &   43  & $-$0.001 &   0.185 &  1  &$-$0.067  &             \\ 
n63f    & 8  &  $-$0.016  &  0.151 & 13 &  +0.328   & 0.204  &   25  & $-$0.114 &   0.238   &  6  & $-$0.126  &   0.196  &    5 &  +0.134 &   0.175 &   41  &   +0.029 &   0.169 &  1  &  +0.363  &             \\ 
n63g    & 8  &    +0.040  &  0.242 & 13 &  +0.371   & 0.312  &   13  & $-$0.138 &   0.259   &  6  & $-$0.308  &   0.297  &    4 &  +0.073 &   0.291 &   21  &   +0.012 &   0.286 &  0  &     &          \\ 
n63h    & 8  &  $-$0.023  &  0.191 & 12 &  +0.145   & 0.261  &   22  & $-$0.156 &   0.251   &  7  & $-$0.203  &   0.260  &    4 &$-$0.026 &   0.157 &   37  &   +0.004 &   0.177 &  1  &  +0.218  &             \\ 
n63i    & 8  &  $-$0.109  &  0.186 & 13 &  +0.191   & 0.175  &   25  & $-$0.112 &   0.256   &  6  & $-$0.317  &   0.119  &    5 &$-$0.031 &   0.174 &   41  & $-$0.015 &   0.160 &  1  &  +0.020  &             \\ 
n63l    & 8  &  $-$0.112  &  0.178 & 12 &  +0.184   & 0.194  &   22  & $-$0.159 &   0.235   &  7  & $-$0.169  &   0.194  &    5 &  +0.137 &   0.106 &   39  &   +0.013 &   0.155 &  1  &  +0.109  &             \\ 
n63m    & 8  &    +0.004  &  0.179 & 13 &  +0.357   & 0.168  &   13  & $-$0.098 &   0.172   &  6  & $-$0.187  &   0.182  &    4 &$-$0.044 &   0.174 &   40  &   +0.049 &   0.231 &  1  &  +0.208  &             \\ 
n63n    & 8  &  $-$0.001  &  0.096 & 13 &  +0.297   & 0.201  &   21  & $-$0.191 &   0.232   &  7  & $-$0.288  &   0.183  &    5 &$-$0.019 &   0.163 &   37  &   +0.047 &   0.219 &  1  &  +0.096  &             \\ 
l63p001 & 2  &    +0.003  &  0.012 &    &           &        &   1   & $-$0.037 &           &  0  &           &          &    0 &         &         &   12  &   +0.066 &   0.104 &  0  &          &             \\ 
l63p002 & 2  &  $-$0.017  &  0.017 &    &           &        &   1   & $-$0.126 &           &  0  &           &          &    0 &         &         &   12  &   +0.051 &   0.057 &  0  &          &             \\ 
l63p003 & 2  &  $-$0.122  &  0.071 &    &           &        &   1   &   +0.256 &           &  0  &           &          &    0 &         &         &     8   &   +0.019 &   0.100 &  0  &          &          \\ 
l63p004 & 2  &  $-$0.066  &  0.007 &    &           &        &   1   & $-$0.487 &           &  0  &           &          &    0 &         &         &   10  & $-$0.068 &   0.093 &  0  &          &             \\ 
l63p006 & 2  &  $-$0.044  &  0.014 &    &           &        &   1   &   +0.080 &           &  0  &           &          &    0 &         &         &     10  &   +0.028 &   0.152 &  0  &          &          \\ 

\hline
\end{tabular}
\label{t:fepeakUNIFIN}
\end{table*}

\begin{table*}[h]
\centering
\setlength{\tabcolsep}{1mm}
\caption[]{Neutron-capture abundances.}
\small
\begin{tabular}{lccccccccccrccrccrc}
\hline
Star   
&   nr &[Y/Fe]{\sc i}  &  rms    & nr & [Zr/Fe]{\sc ii}&  rms  &  nr &  [La/Fe]{\sc ii} &  rms  & nr & [Ce/Fe]{\sc ii} &  rms  &  nr &  [Nd/Fe]{\sc ii}&  rms &  nr &  [Eu/Fe]{\sc ii} &  rms  \\
\hline
n63a   & 2 &  $-$0.680 &   0.134 &  0 &                &       &  2  &    +0.047 &  0.040&  1 &  $-$0.365       &       &  3  &  +0.070         & 0.082 & 2  &    +0.238         &   0.064 \\ 
n63b   & 1 &  $-$0.300 &         &  1 &                &       &  2  &    +0.214 &  0.022&  0 &                 &       &  3  &  +0.003         & 0.066 & 0  &                   &         \\ 
n63c   & 2 &  $-$0.424 &   0.018 &  1 &$-$0.280        &       &  2  &    +0.173 &  0.044&  1 &  $-$0.285       &       &  3  &  +0.012         & 0.157 & 2  &    +0.283         &   0.020 \\ 
n63d   & 2 &  $-$0.340 &   0.221 &  1 &$-$0.205        &       &  2  &    +0.174 &  0.018&  1 &  $-$0.323       &       &  3  &  +0.029         & 0.203 & 2  &    +0.288         &   0.168 \\ 
n63e   & 2 &  $-$0.287 &   0.202 &  1 &  +0.133        &       &  2  &    +0.094 &  0.175&  1 &  $-$0.169       &       &  3  &  +0.111         & 0.013 & 2  &    +0.260         &   0.035 \\ 
n63f   & 2 &  $-$0.237 &   0.129 &  1 &  +0.091        &       &  2  &    +0.147 &  0.066&  1 &  $-$0.233       &       &  3  &  +0.183         & 0.059 & 2  &    +0.265         &   0.000 \\ 
n63g   & 0 &           &         &  1 &$-$0.371        &       &  2  &    +0.238 &  0.279&  1 &  $-$0.260       &       &  0  &                 &       & 2  &    +0.267         &   0.161 \\ 
n63h   & 2 &  $-$0.166 &   0.035 &  1 &$-$0.066        &       &  2  &    +0.149 &  0.027&  1 &  $-$0.341       &       &  3  &$-$0.123         & 0.099 & 2  &    +0.183         &   0.013 \\ 
n63i   & 2 &  $-$0.173 &   0.216 &  1 &$-$0.228        &       &  2  &    +0.122 &  0.155&  1 &  $-$0.290       &       &  3  &$-$0.024         & 0.210 & 2  &    +0.202         &   0.071 \\ 
n63l   & 2 &  $-$0.594 &   0.071 &  1 &$-$0.071        &       &  2  &    +0.145 &  0.053&  1 &  $-$0.289       &       &  3  &  +0.000         & 0.200 & 2  &    +0.233         &   0.108 \\ 
n63m   & 2 &  $-$0.070 &   0.115 &  1 &$-$0.395        &       &  2  &    +0.273 &  0.115&  1 &  $-$0.325       &       &  3  &  +0.102         & 0.179 & 2  &    +0.305         &   0.025 \\ 
n63n   & 2 &  $-$0.281 &   0.257 &  1 &$-$0.464        &       &  2  &    +0.088 &  0.009&  1 &  $-$0.364       &       &  3  &$-$0.010         & 0.133 & 2  &    +0.257         &   0.084 \\ 

\hline
\end{tabular}
\label{t:neutronUNIFIN}
\end{table*}

\begin{table*}[h]
\centering
\setlength{\tabcolsep}{1mm}
\caption[]{[Zr/Fe]{\sc i} and [Ba/Fe]{\sc ii} abundances.}
\small
\begin{tabular}{rcrccrc}
\hline
Star     &nr   &[Zr/Fe]{\sc i} &rms & nr &  [Ba/Fe]{\sc ii} &  rms   \\
\hline
n63a     &  4  &  $-$0.337 &  0.113 &  3 &$-$0.478  &  0.025   \\ 
n63b     &  0  &           &        &  0 &          &          \\ 
n63c     &  4  &    +0.008 &  0.252 &  3 &  +0.016  &  0.149   \\ 
n63d     &  6  &  $-$0.091 &  0.275 &  3 &$-$0.214  &  0.071   \\ 
n63e     &  4  &  $-$0.054 &  0.222 &  3 &$-$0.083  &  0.069   \\ 
n63f     &  3  &  $-$0.008 &  0.029 &  3 &$-$0.036  &  0.102   \\ 
n63g     &  0  &           &        &  3 &  +0.049  &  0.221   \\ 
n63h     &  4  &  $-$0.094 &  0.228 &  3 &$-$0.031  &  0.098   \\ 
n63i     &  3  &  $-$0.094 &  0.063 &  3 &  +0.049  &  0.059   \\ 
n63l     &  3  &  $-$0.149 &  0.038 &  3 &$-$0.148  &  0.045   \\ 
n63m     &  3  &  $-$0.122 &  0.280 &  3 &  +0.011  &  0.082   \\ 
n63n     &  4  &  $-$0.068 &  0.162 &  3 &$-$0.061  &  0.322   \\ 
l63p001  &  4  &  $-$0.013 &  0.127 &  1 &  +0.129  &          \\ 
l63p002  &  4  &  $-$0.141 &  0.170 &  1 &$-$0.091  &          \\ 
l63p003  &  0  &           &        &  1 &$-$0.077  &          \\ 
l63p004  &  3  &  $-$0.116 &  0.197 &  1 &$-$0.035  &          \\ 
l63p006  &  1  &  $-$0.066 &        &  1 &  +0.136  &          \\ 

\hline
\end{tabular}
\label{t:zr1baUNIFIN}
\end{table*}

\clearpage
\section{References for studies listed in Figures~\ref{f:alphaGC},
 \ref{f:iron4oriplus}, \ref{f:ncapt3oriplus}, and \ref{f:ncapt4oriplus}}

In Figures~\ref{f:alphaGC}, \ref{f:iron4oriplus}, \ref{f:ncapt3oriplus}, and
\ref{f:ncapt4oriplus} we compare the average abundance ratios
derived in NGC~6388 for some species to a sample of GCs and field stars, both in the disc and in
the bulge, from a number of studies.
Not all the elements were available for GC stars, whereas all species were
sampled in the abundance analyses of comparison field stars in the Milky Way.\\

The GCs in our FLAMES survey, ordered by increasing metallicity, are listed in
Table~\ref{t:appB_noi} (we give the [Fe/H] value from UVES spectra used
in the plots, from Carretta et al. 2009c or from individual papers, in parentheses).
We note that some elements were analysed in the same study for many GCs, whereas 
other elements were obtained in the individual papers.

To GCs homogeneously analysed in our FLAMES survey (or with similar procedures),
we added four metal-rich GCs from the same group of investigators (see
Table~\ref{t:appB_other}). For field Milky Way stars (in halo, disc, or bulge components), we used many studies, as detailed
in the same table.

\begin{landscape}
\begin{table}    \setlength{\tabcolsep}{0.7mm}
\caption{Globular cluster in our FLAMES survey: Elements and references.}
\begin{tabular}{llllll}
\hline \hline
GC and [Fe/H] & Ref for [Fe/H] & Ref for Mg, Si & Ref for Ca & Ref for other elem. & Other elements \\
\hline
NGC~7099 (M~30, $-2.344$) &Carretta et al. 2009c &Carretta et al. 2009a &Carretta et al. 2010b &  & \\
NGC~7078 (M~15, $-2.320$) &Carretta et al. 2009c &Carretta et al. 2009a &Carretta et al. 2010b &  & \\
Terzan~8 ($-2.270$)       &Carretta et al. 2014a &Carretta et al. 2014a &Carretta et al. 2014a &Carretta et al. 2014a &Ti~{\sc i}, Cr, Mn, Ni, Y~{\sc ii}, Ba, Nd \\                      
NGC~4590 (M~68, $-2.265$) &Carretta et al. 2009c &Carretta et al. 2009a &Carretta et al. 2010b &  & \\
NGC~4833 ($-2.020$)       &Carretta et al. 2014b &Carretta et al. 2014b &Carretta et al. 2014b &Carretta et al. 2014b &Ti~{\sc i}, Cr, Mn, Ni, Y~{\sc ii}, Y~{\sc ii},Ba, Nd \\
NGC~6397 ($-1.988$)       &Carretta et al. 2009c &Carretta et al. 2009a &Carretta et al. 2010b &James et al. 2004a & Y~{\sc ii}, Ba, Eu \\
NGC~6535 ($-1.952$)       &Bragaglia et al. 2017 &Bragaglia et al. 2017 &Bragaglia et al. 2017 &Bragaglia et al. 2017 &Ti~{\sc i}, Cr, Mn, Ni\\ 
NGC~6809 (M~55, $-1.934$) &Carretta et al. 2009c &Carretta et al. 2009a &Carretta et al. 2010b &  & \\
NGC~5634 ($-1.867$)       &Carretta et al. 2017  &Carretta et al. 2017  &Carretta et al. 2017  &Carretta et al. 2017  &Ti~{\sc i}, Cr, Mn, Co, Ni, Y~{\sc ii}, Ba, La, Nd, Eu\\
NGC~6093 (M~80, $-1.790$) &Carretta et al. 2015  &Carretta et al. 2015  &Carretta et al. 2015  &Carretta et al. 2015  &Ti~{\sc i}, Cr, Mn, Co, Ni, Y~{\sc ii}, Ba, La, Ce, Nd, Eu\\
NGC~1904 (M~79, $-1.579$) &Carretta et al. 2009c &Carretta et al. 2009a &Carretta et al. 2010b &  & \\
NGC~6254 (M~10, $-1.575$) &Carretta et al. 2009c &Carretta et al. 2009a &Carretta et al. 2010b &  & \\
NGC~6752  ($-1.555$)      &Carretta et al. 2009c &Carretta et al. 2009a &Carretta et al. 2010b &James et al. 2004b & Y~{\sc ii}, Ba, Eu \\
NGC~3201 ($-1.512$)       &Carretta et al. 2009c &Carretta et al. 2009a &Carretta et al. 2010b &  & \\
NGC~6715 (M~54, $-1.510$) &Carretta et al. 2010c &Carretta et al. 2010c &Carretta et al. 2010c &Carretta et al. 2010c &Ti~{\sc i}, Cr, Mn, Co, Ni\\
NGC~5904 (M~5, $-1.340$)  &Carretta et al. 2009c &Carretta et al. 2009a &Carretta et al. 2010b &  & \\
NGC~6218 (M~12, $-1.330$) &Carretta et al. 2009c &Carretta et al. 2009a &Carretta et al. 2010b &  & \\
NGC~288 ($-1.305$)        &Carretta et al. 2009c &Carretta et al. 2009a &Carretta et al. 2010b &  & \\
NGC~1851 ($-1.180$)       &Carretta et al. 2011  &Carretta et al. 2011  &Carretta et al. 2011  &Carretta et al. 2011  &Ti~{\sc i}, Cr, Mn, Co, Ni, Y~{\sc ii}, Zr~{\sc ii}, Ba, La, Nd, Eu\\
NGC~362 ($-1.170$)        &Carretta et al. 2013  &Carretta et al. 2013  &Carretta et al. 2013  &Carretta et al. 2013  &Ti~{\sc i}, Cr, Mn, Co, Ni, Y~{\sc ii}, Zr~{\sc ii}, Ba, La, Nd, Eu\\
NGC~6121 (M~4, ($-1.168$) &Carretta et al. 2009c &Carretta et al. 2009a &Carretta et al. 2010b &  & \\
NGC~2808 ($-1.130$)       &Carretta  2015        &Carretta 2015   &Carretta 2015         &Carretta 2015         &Ti~{\sc i}, Cr, Mn, Ni\\
NGC~6171 (M~107, $-1.033$)&Carretta et al. 2009c &Carretta et al. 2009a &Carretta et al. 2010b &  & \\
NGC~6838 (M~71, $-0.832$) &Carretta et al. 2009c &Carretta et al. 2009a &Carretta et al. 2010b &  & \\
NGC~104 (47~Tuc, $-0.768$)&Carretta et al. 2009c &Carretta et al. 2009a &Carretta et al. 2010b & Carretta et al. 2004                     &Cr, Mn, Ni; Y~{\sc ii}, Ba, Eu\\ 
                           &                      &                      &                       & James et al. 2004a                       &                                 \\ 
NGC~6388 ($-0.480$)       &Carretta\&Bragaglia 2022a& this work       & this work                 & this work & Ti~{\sc i}, Cr~{\sc i}, Mn, Co, Ni, Y~{\sc i}, Zr~{sc i}, Ba, La, Ce, Nd, Eu \\
NGC~6441 ($-0.430$)       &Carretta et al. 2009c &Gratton et al. 2006   &Gratton et al. 2006   &Gratton et al. 2006   &Ti~{\sc i}, Cr, Mn, Ni, Y~{\sc i}, Zr~{\sc i}, Ba, Eu\\ 
\hline
\label{t:appB_noi}
\end{tabular}
\end{table}
\end{landscape}

\begin{table*}
    \centering
    \caption{Literature data: Elements and references.}
    \begin{tabular}{lll}
    \hline \hline
GC or system & Reference & elements \\
\hline
NGC~6440 ([Fe/H]$=-0.50$) &Mu\~{n}oz et al. 2017       &  Mg, Si, Ca, Sc, Ti, Mn, Ni,
Co, Ba, Eu\\
NGC~5927 ([Fe/H]$=-0.47$) &Mura-Guzm\'{a}n et al. 2018 &  Mg, Si, Ca, Sc, Ti, V,
Cr, Mn, Co, Ni, Y, Zr, Ce, Ba, Eu\\
NGC~6528 ([Fe/H]$=-0.20$) &Mu\~{n}oz et al. 2018       &  Mg, Si, Ca, Sc, Ti, V, Cr,
Mn, Co, Ni, Zr, Ba, Eu\\
NGC~6553 [Fe/H]$=-0.14$) &Mu\~{n}oz et al. 2020 & Mg, Si, Ca, Sc, Ti, V,
Cr, Mn, Ni, Y, Zr, Ce, Ba, Eu\\
\hline
halo/disc & Adibekyan et al. 2012        & Ca  \\
          & Battistini and Bensby 2016   & Zr, La, Ce, Nd \\
          & Beir\~{a}o et al. 2005       & Mg \\
          & Bensby et al. 2014           & Mg, Ca, Cr, Ni, Y, Ba\\
          & Brewer et al. 2016           & Mg, Mn, Ni, Y\\
          & Delgado Mena et al. 2017     & Si \\  
          & Gratton et al. 2003          & Mg, Si, Ca, Ti, Cr, Mn, Ni\\
          & Ishigaki et al. 2012, 2013   & Mg, Si, Ca, Ti, Cr, Mn, Co, Ni, Y, Zr, Ba, La, Nd, Eu\\
          & Lai et al. 2008              & Si\\
          & Neves et al. 2009            & Mg, Si. Ti, Cr, Mn, Co, Ni\\
          & Reddy et al. 2003            & Si \\
          & Reddy et al. 2006            & Ca \\
          & Reggiani et al. 2017         & Mg, Cr, Mn, Co, Ni, Y, Zr, Ba\\
          & Roederer et al. 2014.        & Si, Ca, Ti, Cr, Mn, Co, Ni, Y, Zr, Ba, La, Ce, Nd, Eu\\
\hline
bulge     & Alves-Brito et al. 2010      & Mg, Ca, Ti\\
          & Barbuy et al. 2013           & Mn \\
          & Bensby et al. 2017           & Mg, Si, Ca, Ti, Cr, Ni, Y, Ba\\
          & da Silveira et al. 2018      & Mg, Si, Ca \\
          & Duong et al. 2019a           & Ti \\
          & Duong et al. 2019b           &Cr, Mn, Co, Ni, Zr, La, Nd, Eu \\
          & Forsberg et al. 2019         & Zr, La, Ce, Eu \\
          & Johnson et al. 2014          & Mg, Si, Ca, Cr, Co, Ni \\
          & J{\"o}nsson et al. 2017      & Mg, Ca, Ti\\
          & Lomaeva et al. 2019          & Cr, Mn, Co, Ni\\
          & Lucey et al. 2019            & Mg, Si, Ca, Ti, Cr, Mn, Co, Ni, Y, Zr, Ba, La, Nd, Eu \\
          & Lucey et al. 2022            & Ti\\
\hline
\end{tabular}
\label{t:appB_other}
\end{table*}

\end{appendix}

\begin{thebibliography}{}

\bibitem[]{} Abdurro'uf, Accetta, K., Aerts, C., et al. 2022, ApJS, 259, 35 
\bibitem[]{} Adibekyan, V.Zh., Sousa, S.G., Santos, N.C. et al. 2012, A\&A, 545,
  A32
\bibitem[]{} Alonso, A., Arribas, S., Martinez-Roger, C. 1999, A\&AS, 140, 261 
\bibitem[]{} Alonso, A., Arribas, S., Martinez-Roger, C. 2001, A\&A, 376, 1039 
\bibitem[]{} Alves-Brito, A., Mel\`{e}ndez, J., Asplund, M., Ram\`{i}rez, I.,
  Yong, D. 2010, A\&A, 513, A35
\bibitem[]{} Anders, E., Grevesse, N. 1989, GeCoA, 53, 197
\bibitem[]{} Arnould, M., Goriely, S., Jorissen, A. 1999, A\&A, 347, 572 
\bibitem[]{} Barbuy, B., Hill, V., Zoccali, M. et al. 2013, A\&A, 559, A5
\bibitem[]{} Bastian, N., Lardo, C. 2018, ARA\&A, 56, 83
\bibitem[]{} Battistini, C., Bensby, T. 2016, A\&A, 586, A49
\bibitem[]{} Beir\~{a}o, P., Santos, N.C., Israelian, G., Mayor, M. 2005, A\&A,
  438, 251
\bibitem[]{} Bensby, T., Feltzing, S., Lundstr\"om, I., Ilyin, I. 2005, A\&A,
  433, 185  
\bibitem[]{} Bensby, T., Feltzing, S., Oey, M.S. 2014, A\&A, 562, A71
\bibitem[]{} Bensby, T., Feltzing, S., Gould, A. et al. 2017, A\&A, 605, A89
\bibitem[]{} B\"ohm-Vitense, E. 1979, ApJ, 234, 521
\bibitem[]{} Bragaglia, A., Carretta, E., Gratton, R.G. et al. 2001, AJ, 121, 327 
\bibitem[]{} Bragaglia, A., Carretta, E., D'Orazi, V. et al. 2017, A\&A, 607, A44 
\bibitem[]{} Brewer, J.M., Fischer, D.A., Valenti, J.A., Piskunov, N. 2016,
  ApJS, 225, 32
\bibitem[]{} Carretta, E. 2006, AJ, 131, 1766 
\bibitem[]{} Carretta, E. 2015, ApJ, 810, 148 
\bibitem[]{} Carretta, E., Bragaglia, A. 2018, A\&A, 614, A109 
\bibitem[]{} Carretta, E., Bragaglia, A. 2019, A\&A, 627, L7 
\bibitem[]{} Carretta, E., Bragaglia, A. 2021, A\&A, 646, A9 
\bibitem[]{} Carretta, E., Bragaglia, A. 2022a, A\&A, 659, A122 
\bibitem[]{} Carretta, E., Bragaglia, A. 2022b, A\&A, 660, L1 
\bibitem[]{} Carretta, E., Gratton R.G., Bragaglia, A., Bonifacio, P., 
  Pasquini, L. 2004, A\&A, 416, 925 
\bibitem[]{} Carretta, E., Bragaglia, A., Gratton, R.G. et al. 2006, A\&A, 450, 523 
\bibitem[]{} Carretta, E., Bragaglia, A., Gratton, R.G. et al. 2007a, A\&A, 464, 967 
\bibitem[]{} Carretta, E., Recio-Blanco, A., Gratton, R.G., Piotto, G.,
  Bragaglia, A. 2007b, ApJ, 671, L125 
\bibitem[]{} Carretta, E., Bragaglia, A., Gratton, R.G., Lucatello, S. 2009a, 
 A\&A, 505, 139 
\bibitem[]{}  Carretta, E., Bragaglia, A., Gratton, R.G. et al. 2009b, 
  A\&A, 505, 117  
\bibitem[]{} Carretta, E., Bragaglia, A., Gratton, R.G., D'Orazi, V., Lucatello,
 S. 2009c, A\&A, 508, 695 
\bibitem[]{} Carretta, E., Bragaglia, A., Gratton, R.G. et al. 2010a, A\&A, 516, 55 
\bibitem[]{} Carretta, E., Bragaglia, A., Gratton, R.G. et al. 2010b, ApJ, 712, L21 
\bibitem[]{} Carretta, E., Bragaglia, A., Gratton, R.G. et al. 2010c, A\&A, 520, 95 
\bibitem[]{} Carretta, E., Lucatello, S., Gratton, R.G., Bragaglia, A., D'Orazi,
  V. 2011, A\&A, 533, 69 
\bibitem[]{} Carretta, E., Bragaglia, A., Gratton, R.G., Lucatello, S., 
  D'Orazi, V. 2012, ApJ, 750, L14 
\bibitem[]{} Carretta, E., Bragaglia, A., Gratton, R.G. et al. 2013, A\&A, 557, A138 
\bibitem[]{} Carretta, E., Bragaglia, A., Gratton, R.G. et al. 2014a, A\&A, 561,
  A87 
\bibitem[]{} Carretta, E., Bragaglia, A., Gratton, R.G. et al. 2014b, A\&A, 564,
  A60 
\bibitem[]{} Carretta, E., Bragaglia, A., Gratton, R.G. et al. 2015, A\&A, 578, 
  A116 
\bibitem[]{} Carretta, E., Bragaglia, A., Lucatello, S. et al. 2017, A\&A, 600, A118 
\bibitem[]{} Chen, J., Ferraro, F.~R., Cadelano, M., et al. 2021, Nature Astronomy, 5, 1170
\bibitem[]{} Cohen, J.G., Kirby, E.N. 2012, ApJ, 760, 86 
\bibitem[]{} Couch, R.G., Schmiedekamp, A.B., Arnett, W.D. 1974, ApJ, 190, 95
\bibitem[]{} Cordero, M.J., Pilachowski, C.A., Johnson C.I. et al. 2014, ApJ, 780, 94 
\bibitem[]{} D'Antona, F., Caloi, V., Montalb\'an, J., Ventura, P., Gratton, R.
  2002, A\&A, 395, 69 
\bibitem[]{} D'Antona, F., Vesperini, E., D'Ercole, A. et al. 2016, MNRAS, 458, 2122 
\bibitem[]{} da Silveira, C.R., Barbuy, B., Fria\c ca, A.C.S. et al. 2018,
  A\&A, 614, A149
\bibitem[]{} Delgado Mena, E., Tsantaki, M., Adibekyan, V. Zh. et al. 2017,
  A\&A, 606, A94
\bibitem[]{} Duong, L., Asplund, M., Nataf, D.M. et al. 2019, MNRAS, 486, 3586
\bibitem[]{} Duong, L., Asplund, M., Nataf, D.M., Freeman, K.C., Ness, M. 2019,
  MNRAS, 486, 5349
\bibitem[]{} Ferraro, F.~R., Carretta, E., Corsi, C.~E., et al. 1997, A\&A, 320, 757
\bibitem[]{} Forsberg, R., J{\"o}nsson, H., Ryde, N., Mattucci, F. 2019, A\&A,
  631, A113
\bibitem[]{} Gratton, R.G. 1988, Rome Obs. Preprint Ser., 29 
\bibitem[]{} Gratton, R.G., Carretta, E., Eriksson, K., Gustafsson, B. 1999,
  A\&A, 350, 955 
\bibitem[]{} Gratton, R.G., Bonifacio, P., Bragaglia, A., et al. 2001, 
 A\&A, 369, 87 
\bibitem[]{} Gratton, R.G., Carretta, E., Claudi, R., Lucatello, S.,
  Barbieri, M. 2003, A\&A, 404, 187 
\bibitem[]{} Gratton, R.G., Sneden, C., Carretta, E. 2004, ARA\&A, 42, 385
\bibitem[]{} Gratton, R.G., Lucatello, S., Bragaglia, A. et al. 2006, A\&A, 
  455, 271 
\bibitem[]{} Gratton, R.G., Lucatello, S., Bragaglia, A. et al. 2007, A\&A, 
  464, 953 
\bibitem[]{} Gratton, R.G., Lucatello, S., Carretta, E. et al. 2011, A\&A, 534, 123 
\bibitem[]{} Gratton, R.G., Carretta, E., Bragaglia, A. 2012a, A\&ARv, 20, 50 
\bibitem[]{} Gratton, R.G., Lucatello, S., Carretta, E. et al. 2012b, A\&A, 539, 19 
\bibitem[]{} Gratton, R.G., Lucatello, S., Sollima, A. et al. 2013, A\&A, 549, A41 
\bibitem[]{} Gratton, R.G., Lucatello, S., Sollima, A. et al. 2014, 563, A13   
\bibitem[]{} Gratton, R.G., Lucatello, S., Sollima, A. et al. 2015, A\&A, 573, A92 
\bibitem[]{} Gratton, R.G., Bragaglia, A., Carretta, E. et al. 2019, A\&ARv, 27, 8
\bibitem[]{} Harris, W.~E. 2010, arXiv:1012.3224
\bibitem[]{} Horta, D., Schiavon, R.P., Mackereth, J.T. et al. 2020, MNRAS, 493,
  3363 
\bibitem[]{} Horta, D., Schiavon, R.P., Mackereth, J.T. et al. 2021, MNRAS, 500, 1385
\bibitem[]{} Ishigaki, M.N., Chiba, M., Aoki, W. 2012, ApJ, 753, 64
\bibitem[]{} Ishigaki, M.N., Aoki, W., Chiba, M. 2013, ApJ, 771, 67
\bibitem[]{} James, G., Fran\c cois, P., Bonifacio, P. et al. 2004a, A\&A, 427, 825 
\bibitem[]{} James, G., Fran\c cois, P., Bonifacio, P. et al. 2004b, A\&A, 414,
  1071   
\bibitem[]{} Johnson, C.I., Pilachowski, C.A. 2010, ApJ, 722, 1373 
\bibitem[]{} Johnson, C.I., Rich, R.M., Kobayashi, C., Kunder, A., Koch, A. et
  al. 2014, AJ, 148, 67
\bibitem[]{} J{\"o}nsson, H., Ryde, N., Schultheis, M., Zoccali, M. 2017, A\&A,
  598, A101
\bibitem[]{} Karakas, A.I., Lattanzio, J..C. 2003, PASA, 20, 279 
\bibitem[]{} Kobayashi, C., Karakas, A.I., Lugaro, M. 2020, ApJ, 900, 179
\bibitem[]{} Kolomiecas, E., Dobrovolskas, V., Ku$\breve{c}$inskas, A., Bonifacio,
  P., Korotin, S. 2022, A\&A, 660, A46 
\bibitem[]{} Kurucz, R.L. 1993, CD-ROM 13, Smithsonian Astrophysical
  Observatory, Cambridge
\bibitem[]{} Lai, D.K., Bolte, M., Johnson, J.A. et al. 2008, ApJ, 681, 1524
\bibitem[]{} Lanzoni, B., Mucciarelli, A., Origlia, L. et al. 2013, ApJ, 769,
  107 (L13) 
\bibitem[]{} Lemasle, B., de Boer, T.J.L, Hill, V. et al. 2014, 
  A\&A, 572, 88
\bibitem[]{} Letarte, B., Hill, V., Tolstoy, E. et al. 2010, A\&A, 523, 17
\bibitem[]{} Lomaeva, M., J{\"o}nsson, H., Ryde, N., Schulteis, M., Thorsbro, B.
  2019, A\&A, 625, A141
\bibitem[]{} Lucey, M., Hawkins, K., Ness, M. et al. 2019, MNRAS, 488, 2283
\bibitem[]{} Lucey, M., Hawkins, K., Ness, M. et al. 2022, MNRAS, 509, 122
\bibitem[]{} Magain, P. 1984, A\&A, 134, 189 
\bibitem[]{} Marino, A.F., Milone, A., Piotto, G., et al. 2011b, ApJ, 731, 64 
\bibitem[]{} Marino, A.F., Villanova, S., Milone, A.P. et al. 2011a, ApJ, 730,
  L16 
\bibitem[]{} Massari, D., Koppelman, H.H., Helmi, A. 2019, A\&A, 630, L4
\bibitem[]{} M\'esz\'aros, S., Masseron, T., Garc\'ia-Hern\'andez, D.A. et al.
  2020, MNRAS, 492, 1641 (M20) 
\bibitem[]{} Milone, A.P., Marino, A.F., Renzini, A. et al. 2018, MNRAS, 481,
  5098 
\bibitem[]{} Minelli, A., Mucciarelli, A., Massari, D., et al. 2021a, ApJL, 918, L32
\bibitem[]{} Minelli, A., Mucciarelli, A., Romano, D., et al. 
  2021b, ApJ, 910, 114
\bibitem[]{} Mucciarelli, A., Bellazzini, M., Ibata, R. et al. 2012, MNRAS, 426, 2889 
\bibitem[]{} Mucciarelli, A., Bellazzini, M., Merle, T. et al. 2015, ApJ, 801, 68   
\bibitem[]{} Mu\~{n}oz, C., Villanova, S., Geisler, D. et al. 2017, A\&A, 605, 
  A12 
\bibitem[]{} Mu\~{n}oz, C., Geisler, D., Villanova, S. et al. 2018, A\&A, 620, 
  A96 
\bibitem[]{} Mu\~{n}oz, C., Villanova, S., Geisler, D. et al. 2020, MNRAS, 492,
  3742 
\bibitem[]{} Mura-Guzm\'{a}n, A., Villanova, S., Mu\~{n}noz, C. 2018, MNRAS, 
  474, 4541 
\bibitem[]{} Myeong, G.C., Vasiliev, E., Iorio, G. 2019, MNRAS, 488, 1235
\bibitem[]{} Nardiello, D., Libralato, M., Piotto, G. et al. 2018, MNRAS, 481, 
3382
\bibitem[]{} Neves, V., Santos, N.C., Sousa, S.G., Correia, A.C.M., Israelian,
  G. 2009, A\&A, 497, 563
\bibitem[]{} Prantzos, N., Charbonnel, C., Iliadis, C. 2017, A\&A, 608, 
 A28 
\bibitem[]{} R Core Team (2022). R: A language and environment for statistical
computing. R Foundation for Statistical Computing, Vienna, Austria.
URL https://www.R-project.org/
\bibitem[]{} Reddy, B.E., Tomkin, J., Lambert, D.L., Allende Prieto, C. 2003,
  MNRAS, 340, 304
\bibitem[]{} Reddy, B.E., Lambert, D.L., Allende Prieto, C  2006, MNRAS, 367,
  1329
\bibitem[]{} Reggiani, H., Mel\`endez, J., Kobayashi, C., Karakas, A., Placco,
  V. 2017, A\&A, 608, A46
\bibitem[]{} Roederer, I.U., Presto, G.W., Thompson, I.B. et al. 2014, AJ, 147,
  136
\bibitem[]{} Schiappacasse-Ulloa, J., Lucatello, S. 2023, MNRAS, 520, 5938
\bibitem[]{} Simmerer, J., Sneden, C., Cowan, J.J. et al. 2004, ApJ, 617, 1091 
\bibitem[]{} Smith, V.V., Bizyaev, D., Cunha, K. et al. 2021, AJ, 161, 254 
\bibitem[]{} Sneden, C., Kraft, R.P., Shetrone, M.D. et al. 1997, AJ, 114, 1964 
\bibitem[]{} Sneden, C., Cowan, J.J., Lawler, J.E. et al. 2003, ApJ, 591, 936
\bibitem[]{} Tinsley, B.M. 1979, ApJ, 229, 1046
\bibitem[]{} Tolstoy, E., Hill, V., Tosi, M. 2009, ARAA\&A, 47, 
  371g
\bibitem[]{} Travaglio, C., Gallino, R., Arnone, E. et al. 2004, ApJ, 601, 864
\bibitem[]{} Valle, G., Dell'Omodarme, M., Tognelli, E. 2022, A\&A, 658, A141
\bibitem[]{} Vasiliev, E., Baumgardt, H. 2021, MNRAS, 505, 5978
\bibitem[]{} Ventura, P., D'Antona, F., Di Criscienzo, M. et al. 2012, ApJ, 761, L30 
\bibitem[]{} Villanova, S., Piotto, G., Gratton, R.G. 2009, A\&A, 499, 755
\bibitem[]{} Villanova, S., Geisler, D., Piotto, G., Gratton, R.G. 2012, ApJ,
  748, 62
\bibitem[]{} Wallerstein, G., Kovtyukh, V.V., Andrievsky, S.M. 2007, AJ, 133,
  1373 
\bibitem[]{} Worley, C.C., Hill, V., Sobeck, J., Carretta, E. 2013, A\&A, 553,
  A47 
\bibitem[]{} Yong, D., Grundahl, F.,  Nissen, P.E., Jensen, H.R., Lambert, D.L.
 2005, A\&A, 438, 875 
\bibitem[]{} Yong, D., M\'{e}lendez, J., Grundahl, F. et al. 2013, MNRAS, 434,
  3452  


\end{thebibliography}
\end{document}